\documentclass[aps,prx,reprint,superscriptaddress,nofootinbib]{revtex4-2}
\usepackage{blindtext}
\usepackage{lipsum}
\usepackage{graphics}
\usepackage{amsmath}
\usepackage{graphicx}
\usepackage{graphics}
\usepackage{amssymb}
\usepackage{verbatim}
\usepackage{physics}
\usepackage{float}

\usepackage[colorlinks=true,citecolor=blue,linkcolor=blue,urlcolor=blue,pdftitle={QSLakes}]{hyperref}
\makeatletter
\def\l@subsubsection#1#2{}
\makeatother

\usepackage[dvipsnames]{xcolor}
\usepackage[normalem]{ulem}
\usepackage{wasysym} 
\usepackage[export]{adjustbox}

\makeatletter
\DeclareFontFamily{OMX}{MnSymbolE}{}
\DeclareSymbolFont{MnLargeSymbols}{OMX}{MnSymbolE}{m}{n}
\SetSymbolFont{MnLargeSymbols}{bold}{OMX}{MnSymbolE}{b}{n}
\DeclareFontShape{OMX}{MnSymbolE}{m}{n}{
    <-6>  MnSymbolE5
   <6-7>  MnSymbolE6
   <7-8>  MnSymbolE7
   <8-9>  MnSymbolE8
   <9-10> MnSymbolE9
  <10-12> MnSymbolE10
  <12->   MnSymbolE12
}{}
\DeclareFontShape{OMX}{MnSymbolE}{b}{n}{
    <-6>  MnSymbolE-Bold5
   <6-7>  MnSymbolE-Bold6
   <7-8>  MnSymbolE-Bold7
   <8-9>  MnSymbolE-Bold8
   <9-10> MnSymbolE-Bold9
  <10-12> MnSymbolE-Bold10
  <12->   MnSymbolE-Bold12
}{}

\let\llangle\@undefined
\let\rrangle\@undefined
\DeclareMathDelimiter{\llangle}{\mathopen}%
                     {MnLargeSymbols}{'164}{MnLargeSymbols}{'164}
\DeclareMathDelimiter{\rrangle}{\mathclose}%
                     {MnLargeSymbols}{'171}{MnLargeSymbols}{'171}
\makeatother

\usepackage{tikz}
\usepackage{tikz-cd}
\usetikzlibrary{arrows}
\usetikzlibrary{snakes}
\usetikzlibrary{intersections}
\usetikzlibrary{shapes.geometric}
\usetikzlibrary{decorations.pathmorphing, patterns,shapes}
\usetikzlibrary{decorations.markings}

\tikzset{
	partial ellipse/.style args={#1:#2:#3}{
		insert path={+ (#1:#3) arc (#1:#2:#3)}
	}
}

\tikzset{
	mid arrow/.style={postaction={decorate,decoration={
				markings,
				mark=at position .575 with {\arrow[#1]{stealth}}
	}}},
	near arrow/.style={postaction={decorate,decoration={
				markings,
				mark=at position .275 with {\arrow[#1]{stealth}}
	}}},
	far arrow/.style={postaction={decorate,decoration={
				markings,
				mark=at position .800 with {\arrow[#1]{stealth}}
	}}},
}

\pgfdeclarepatternformonly{south west lines}{\pgfqpoint{-0pt}{-0pt}}{\pgfqpoint{3pt}{3pt}}{\pgfqpoint{3pt}{3pt}}{
	\pgfsetlinewidth{0.4pt}
	\pgfpathmoveto{\pgfqpoint{0pt}{0pt}}
	\pgfpathlineto{\pgfqpoint{3pt}{3pt}}
	\pgfpathmoveto{\pgfqpoint{2.8pt}{-.2pt}}
	\pgfpathlineto{\pgfqpoint{3.2pt}{.2pt}}
	\pgfpathmoveto{\pgfqpoint{-.2pt}{2.8pt}}
	\pgfpathlineto{\pgfqpoint{.2pt}{3.2pt}}
	\pgfusepath{stroke}}

\newcommand{\rv}[1]{{ \color{blue}[RV] #1 }}
\newcommand{\rs}[1]{{ \color{red}[RS] #1 }}

\usepackage{ulem}

\definecolor{orange(ryb)}{HTML}{FFA500}
\definecolor{dodgerblue}{HTML}{1E90FF}
\definecolor{forest}{HTML}{025839}

\definecolor{edit}{HTML}{000000}

\begin{document}

\title{Quantum Spin Puddles and Lakes:\\NISQ-Era Spin Liquids from Non-Equilibrium Dynamics}
\author{Rahul Sahay}
\affiliation{Department of Physics, Harvard University, Cambridge, Massachusetts 02138 USA}
\author{Ashvin Vishwanath}
\affiliation{Department of Physics, Harvard University, Cambridge, Massachusetts 02138 USA}
\author{Ruben Verresen}
\affiliation{Department of Physics, Harvard University, Cambridge, Massachusetts 02138 USA}

\date{\today}

\begin{abstract}
While many-body quantum systems can host long-ranged entangled quantum spin liquids (QSLs), the ingredients for realizing these as ground states can be prohibitively difficult.
In a broad range of circumstances, one requires (i) a constrained Hilbert space and (ii) an extensive quantum superposition of such states.
The paradigmatic example is the toric code state, or $\mathbb{Z}_2$ spin liquid, which is a superposition of all closed loop states.
We show how simple non-equilibrium Hamiltonian dynamics can provide a more streamlined route toward creating such QSLs.
In particular, rather than cooling into the ground state of a complicated Hamiltonian, we show how a simple parameter sweep can dynamically project a family of initial product states into the desired constrained space, giving rise to a QSL.
For the toric code case, this is naturally achieved in systems where there is a separation in energy scales between the $e$- and $m$-anyons, such that one can sweep in a way that is adiabatic (sudden) with respect to the former (latter).
Although such a separation of scales does not extend to the thermodynamic limit, we use analytic arguments and tensor network numerics to argue that this method efficiently and robustly prepares a spin liquid in finite-sized regions, which we brand ``quantum spin lakes''.
This mechanism sheds light on recent experimental and numerical observations of the dynamical state preparation of the ruby lattice spin liquid in Rydberg atom arrays. 
In fact, the slow dynamics with respect to $m$-anyons suggests we can capture such quantum spin lake preparation by simulating the dynamics on tree lattices, which we confirm with highly-efficient tensor network simulations.
Finally, we use this mechanism to propose new experimental protocols, e.g., for preparing a finite-sized $U(1)$ spin liquid as a honeycomb Rokhsar-Kivelson dimer model using Rydberg atoms---which is all the more remarkable given its equilibrium counterpart is unstable in $2 + 1$D.
Our work thus opens up a new avenue in the study of non-equilibrium physics, as well as the preparation and exploration of exotic states of finite extent in existing quantum devices.
\end{abstract}

\maketitle

\tableofcontents

\section{Introduction}

Phases of matter with intrinsic topological order are characterized by a pattern of long-range quantum entanglement~\cite{Wenbook,Chen_Gu_Wen,Kitaev_2003}. 
This entanglement structure endows such states with a rich phenomenology.
Indeed, they exhibit exotic properties such as a topological ground state degeneracy, ``fractionalized'' bulk excitations with novel quantum statistics, and quantized response properties~\cite{Wen_90, Wilczek82, Wen89, Read89, Kivelson89}.
While such phases were first discovered in the context of the fractional quantum Hall effect~\cite{Tsui82,Laughlin83}, they were later theoretically generalized to quantum spin systems and connected with Anderson's resonating valence bond liquid~\cite{ANDERSON,Anderson87,Wen_90,ReadSachdev} to establish the notion of a quantum spin liquid (QSL).
Since then, QSLs have been the subject of decades of sustained interest within the context of condensed matter physics~\cite{SpinLiquidsReviewBalents,Knolle19,WenRMP,moessner2021topological,Broholmeaay0668}.

In addition to their intriguing material properties, gapped QSLs can also be utilized as a platform for fault-tolerant quantum computation \cite{Kitaev_2003,Kitaev06,Nayak_RMP}.
Namely, the novel quantum statistics of bulk excitations above a QSL can be used to implement logical gates upon it.
Since the statistics of these excitations are topologically protected and information is stored non-locally in the state, such states are intrinsically fault-tolerant at the hardware level.
Accordingly, the prospect of building a ``topological quantum computer'' has stimulated large-scale investigations of such phases in the context of quantum information science.

As a consequence, there have been persistent efforts toward realizing QSLs in solid-state materials \cite{SpinLiquidsReviewBalents,Knolle19,WenRMP,moessner2021topological,Broholmeaay0668}.
The key challenge here is that the requirements for realizing topological order in equilibrium are very restrictive. 
In a broad set of circumstances, there are two essential ingredients.
The first is that, at low energies, the system is described by an \textit{emergent gauge theory}---its low-energy states satisfy a local energetic constraint typically due to either geometric or interaction frustration.
Such constraints define the notion of a local Gauss law and lead to an extensive number of energetically low-lying Gauss law-satisfying states.
The second ingredient is the existence of strong quantum resonances connecting these low-lying states, which stabilize a thermodynamically extensive quantum superposition of them.
Such a superposition ensures that the excitations of the emergent gauge theory are deconfined, leading to the celebration notion of \textit{anyons} \cite{Leinaas77,Wilczek82}.
While local constraints are routinely found in frustrated magnetic systems (see e.g. Refs.~\onlinecite{liebmann1986statistical,Bramwell_2001,Castelnovo2008}), strong quantum resonances between states satisfying these constraints often require many local rearrangements of the state.
Since naturally occurring Hamiltonians typically only contain few-body terms, such resonances must be generated perturbatively.
As a consequence, these resonances are typically very weak leading to a small energy gap above a putative topologically ordered phase, thereby restricting such phases to small portions of phase diagrams.

Recently, however, there have been a number of pioneering experiments that see signatures of QSLs in programmable quantum simulators~\cite{Semeghini21, Satzinger21}.
Notably, building on a prior theory proposal \cite{Verresen21}, a recent experiment \cite{Semeghini21} on Rydberg atom tweezer arrays \cite{Browaeys_2020} found QSL-like signatures by placing atoms on the bonds of a kagome lattice (i.e. the atoms live on the so-called ``ruby lattice'') which interact via the Rydberg blockade \cite{Lukin01,Jaksch00,Gaetan09,Urban09}.
Intriguingly, the experiment was able to find QSL signatures in a regime of parameter space where a careful numerical study predicted QSL order would not be present in the ground state.
The supplementary material of the aforementioned experimental paper \cite{Semeghini21} along with follow-up numerical and variational studies \cite{Giudici22dyn,Cheng_2021} provided strong evidence that the origin of the QSL signatures could be traced back to the dynamical state preparation protocol used to explore the ground state phase diagram of the experiment.
In particular, using small system size numerics, these two studies found that the process of preparing the quantum simulator in a trivial phase and then dynamically tuning the Hamiltonian to a parameter regime predicted to be in the confined phase of the system's emergent gauge theory led to QSL signatures consistent with those observed in the experiment.
Despite strong numerical evidence, the Rydberg atom experiment and the subsequent numerical study leave open an intriguing theoretical question regarding the precise mechanism underlying the non-equilibrium preparation of a QSL-like state.
This open question inspires the present work.

The overarching goal of this work is to identify whether and when unitary quantum dynamics can approximately prepare exotic states of matter, even when these are not the ground state.
In answering this question, we pinpoint the precise dynamical regime where a parameter sweep can prepare spin liquids of restricted sizes---which we christen \emph{quantum spin lakes (or puddles)}. These findings are of considerable interest for at least three complementary reasons.

First, our results provide insight into how unitary quantum dynamics can give rise to surprisingly \emph{structured} entangled states, even in non-equilibrium regimes where one might not have expected them.
Notably, we study a regime of dynamics that is not contained within the two typically studied paradigms: we are neither close to equilibrium where adiabatic approximations and universal scaling theories directly hold \cite{AdiabaticTheorem_DeRoeck, Zurek_1985, Zurek_Review, ZUREK_1996, Kibble_1976, KIBBLE_1980, Chandran_2013, Chandran_2012}, nor are we so far out of equilibrium that our final state lacks any of the characteristics of the emergent low-energy physics \cite{Abanin_Review, Else_Review, Vedika_review, srednicki1994chaos, fisher2022random}.
Indeed, we combine elements from these two approaches by studying systems with two emergent degrees of freedom and working in a dynamical regime that is \textcolor{edit}{adiabatic} relative to one and sudden relative to the other.

Second, in evincing the mechanism underlying the dynamical preparation of QSL-like states, we gain an understanding of how the preparation procedure scales as a function of system size.
This is both an important theoretical question to answer but also practically addresses the applicability of the mechanism in future quantum simulation experiments with potentially larger numbers of qubits.
In particular, we identify features of the Hamiltonian and lattice geometry that inevitably restrict the size of the resulting spin lake.
Consequently, we conclude that preparing a thermodynamically large spin liquid still requires the presence of a spin liquid ground state.

Finally, our results make it possible to prepare a wide-range of topological states in analog noisy-intermediate scale quantum (NISQ) devices~\cite{Preskill2018NISQ}, where probing quantum dynamics is more natural than cooling to a many-body ground state~\cite{QuantumSimulators}.
Notably, this goes well beyond the case of a $\mathbb{Z}_2$ spin lake which had been sighted in the Rydberg tweezer array context \cite{Semeghini21}.
In particular, we illustrate how non-equilibrium dynamics can even prepare certain states which do not appear as stable ground states in generic two-dimensional systems. For instance, we describe the preparation of a $U(1)$ spin liquid in a dimer model on a bipartite lattice, which appears in equilibrium as a fine-tuned Rokshar-Kivelson point \cite{SpinLiquidsReviewBalents,RK,RKhoneycomb, ReadSachdev1989PRL,READSACHDEV_1989_Nuc, ReadSachdev_PRB,SachdevED,PolyakovBook}.

Motivation in hand, in the next section we will provide an overview of the ideas underlying the dynamical preparation of the QSL-like state and will provide an outline for the rest of this work in Section~\ref{subsec-Outline}.

\section{Intuitive Overview and Key Ideas} \label{sec-Key}

In this section, we will provide an intuitive picture of the mechanism that underlies the dynamical preparation of the QSL-like state, which will be made more precise and supported numerically in subsequent sections.
In particular, in Section~\ref{subsec-SLinEq} we will start by recounting the basic physics of spin liquids in equilibrium, using the toric code as a paradigmatic example.
Subsequently, in Section~\ref{subsec-Slakedynamics} we will provide an provide a physical picture for the mechanism underlying the dynamical preparation of a QSL-like state.
We will conclude in Section~\ref{subsec-Outline} with an outline of the remainder of this paper.

\subsection{Spin Liquids in Equilibrium} \label{subsec-SLinEq}

\begin{figure}[t]
    \centering
    \includegraphics[width = 247pt]{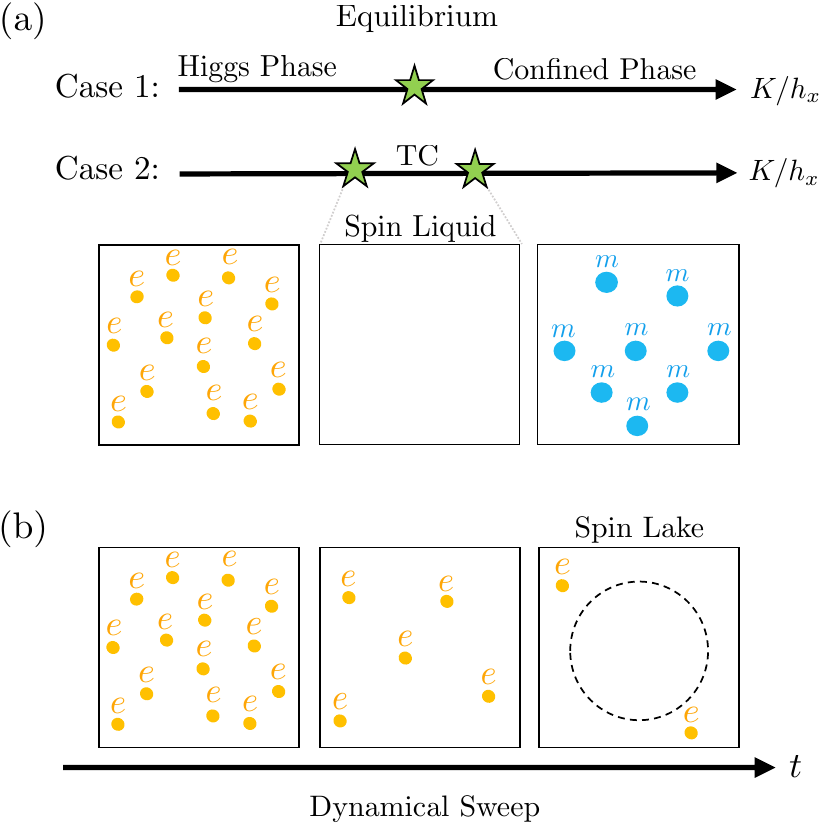}
    \caption{\textbf{Equilibrium Quantum Spin Liquids and Out-of-Equilibrium Quantum Spin Lakes.} (a) In equilibrium, QSLs are characterized by a lack of condensed $e$ and $m$-anyons (middle panel).
    When $K$ (defined in Eq.~\eqref{eq-KitaevTC}) is small, violations of the Gauss law ($e$-anyons) condense leading to a Higgs phase (left panel).
    Alternatively when $K$ is too large, perturbatively generated resonances are small relative to confining fields leading to the condensation of $m$-anyons (right panel). 
    The confined phase and Higgs phase are known to be adiabatically connected to one another but can occasionally be separated via a first-order phase transition. 
    (b) During a dynamical sweep, it is possible to remain in equilibrium relative to $e$-anyons but out-of-equilibrium relative to $m$-anyons.
    As a result, $e$-anyons are mostly equilibrated out during the sweep while $m$-anyons fail to be nucleated in due to experiencing a `sudden' approximation.
    This leads to a state that is nearly defectless over a large length scale, which we brand a quantum spin lake.
    }
    \label{fig:intuitivepicture}
\end{figure}

We start by recounting the physics of spin liquids in equilibrium.
Readers familiar with spin liquids may choose to skip this subsection and move to Section~\ref{subsec-Slakedynamics}.
As outlined in the introduction, the requirements for a broad class of spin liquids in equilibrium are twofold: (1) the presence of an energetic constraint on spin configurations that appear at low-energies and (2) the presence of terms in the Hamiltonian that connect such states. %
More precisely, the first condition gives us an effective constrained Hilbert space---typically no longer having a tensor product structure---where the local constraint can be interpreted as the Gauss law of an emergent gauge theory. The second condition introduces quantum fluctuations within this constrained space; if these fluctuations are large enough this can give rise to a ``deconfined'' or topological phase in the ground state \cite{Kitaev03}.

These two ingredients are manifest in Kitaev's famous toric code model \cite{Kitaev_2003}:
\begin{equation} \label{eq-KitaevTC}
    H_{\text{TC}} = -K\sum_v \begin{tikzpicture}[scale = 0.5, baseline = {([yshift=-.5ex]current bounding box.center)}]
    \draw[gray] (1.5,0) -- (-1.5, 0);
    \draw[gray] (0,1.5) -- (0,-1.5);
    \node at (0.75, 0) {\normalsize $Z$};
    \node at (-0.75, 0) {\normalsize $Z$};
    \node at (0, 0.75) {\normalsize $Z$};
    \node at (0, -0.75) {\normalsize $Z$};
\end{tikzpicture} - J \sum_p
\begin{tikzpicture}[scale = 0.5, baseline={([yshift=-.5ex]current bounding box.center)}]
\draw[gray] (-1, -1) -- (-1, 1) -- (1, 1) -- (1, -1) -- cycle;
\node at (0.0, -1) {\normalsize $X$};
\node at (0.0, 1) {\normalsize $X$};
\node at (-1, 0.0) {\normalsize $X$};
\node at (1, 0.0) {\normalsize $X$};
\end{tikzpicture}
\end{equation}
where qubits are placed at the links of the square lattice and the sum over $v$ and over $p$ indicate a sum over vertices and plaquettes of the square lattice respectively.
To be explicit, the first term in the Hamiltonian, enforces a constraint on low-energy spin configurations that:
\begin{equation} \label{eq-TCGaussLaw}
    G_v = \begin{tikzpicture}[scale = 0.5, baseline = {([yshift=-.5ex]current bounding box.center)}]
    \draw[gray] (1.5,0) -- (-1.5, 0);
    \draw[gray] (0,1.5) -- (0,-1.5);
    \node at (0.75, 0) {\normalsize $Z$};
    \node at (-0.75, 0) {\normalsize $Z$};
    \node at (0, 0.75) {\normalsize $Z$};
    \node at (0, -0.75) {\normalsize $Z$};
\end{tikzpicture} = +1
\end{equation}
This means that low-energy spin configurations satisfy the property that the number of down spins surrounding each vertex must be even.
If we treat our spins as $\mathbb{Z}_2$-valued electric fields $E$ 
$\left\{\ket{ \begin{tikzpicture}[scale = 0.5, baseline = {([yshift=-.5ex]current bounding box.center)}]
\draw[gray] (0, 0) -- (1, 0);
\node at (0.5, 0) {\normalsize $\downarrow$};
\end{tikzpicture}}= \ket{\frac{}{}\begin{tikzpicture}[scale = 0.5, baseline = {([yshift=-.5ex]current bounding box.center)}]
\draw[red,line width=0.5mm] (0.0, 0) -- (1, 0);
\end{tikzpicture}} \quad 
\ket{ \begin{tikzpicture}[scale = 0.5, baseline = {([yshift=-.5ex]current bounding box.center)}]
\draw[gray] (0, 0) -- (1, 0);
\node at (0.5, 0) {\normalsize $\uparrow$};
\end{tikzpicture}} = \ket{\frac{}{} \begin{tikzpicture}[scale = 0.5, baseline = {([yshift=-.5ex]current bounding box.center)}]
\draw[gray] (0, 0) -- (1, 0);
\end{tikzpicture}}\right\}$, then Eq.~\eqref{eq-TCGaussLaw} defines a local Gauss law constraint $(\nabla \cdot E) = 0 \text{ mod } 2$  and the low-energy manifold of states defines an emergent $\mathbb{Z}_2$ gauge theory consisting of all closed loops of electric fields.
The second term of the Hamiltonian commutes with the first and resonates between states that satisfy the Gauss law.
Consequently, the ground state of this Hamiltonian is an equal weight and equal phase superposition of all closed electric field loops:
\begin{equation}\label{eq-TCWF}
    \ket{\psi_\textrm{TC}} = \ket{\frac{}{} \quad } + \ket{\frac{}{} \hspace{-0.5mm} \begin{tikzpicture}[scale = 0.5, baseline = {([yshift=-.5ex]current bounding box.center)}]
    \draw[red,line width=0.5mm] (-0.5, -0.5) .. controls (-0.4, -0.7) and (-0.2, -0.9) .. (0.0, -0.5) -- (0.0, -0.5) .. controls (0.1, -0.425) and (0.5, -0.2) .. (0.25, 0.1) -- (0.25, 0.1) .. controls (0, 0.25) and (-0.3, 0.4) .. (-0.5, 0.1) -- (-0.5, 0.1) .. controls (-0.7, -0.2) and (-0.7, -0.4) .. (-0.5, -0.5);
    \draw[red,line width=0.5mm]
    (0.3, 0.5) ellipse (0.3 and 0.15) -- cycle;
    \end{tikzpicture}
    } + \ket{\hspace{-2mm}\frac{}{}  \begin{tikzpicture}[scale = 0.5, baseline = {([yshift=-.5ex]current bounding box.center)}]
    \draw[red,line width=0.5mm] (-0.5, -0.5) .. controls (-1, -0.2) and (-0.1, 0.2) .. (-0.5, 0.5) -- (-0.5, 0.5) .. controls (-0.9, 1) and (0.4, 1) .. (0.5, 0.5) -- (0.5, 0.5) .. controls (0.6, 0) and (0.9, -0.3) .. (0.5, -0.5) -- (0.5, -0.5) .. controls (0.1, -0.7) and (-0.1, -0.3) .. (-0.5, -0.5) -- cycle;
    \end{tikzpicture}\hspace{-1mm}}  +  \cdots
\end{equation}
The excitations above this TC state are so-called anyons. In fact, violations of the first (second) term in Eq.~(1) are called $e$-anyons ($m$-anyons), which are created at the end of string operators composed of products of Pauli-$X$ ($Z$) operators \cite{Kitaev_2003}.

As such, $X$ or $Z$ fields locally create anyon pairs, such that introducing strong fields gives a way of driving a transition out of the topological phase, which can be interpreted as ``condensing'' either of these anyons into the ground state.
This is captured by the minimal model \cite{FradkinShenker,tupitsyn_topological_2010, Nahum_Z2_2021}:
\begin{equation} \label{eq-KitaevTC-in-F}
    H_{\text{TC} + \text{f}} = H_{\text{TC}} - h_x \sum_\ell X_{\ell} - h_z \sum_\ell Z_{\ell}
\end{equation}
where $h_x$ drives $e$-condensation (where loops are broken into open strings) and $h_z$ $m$-condensation (where loops are no longer in a massive superposition).
Since $e$-anyons are the electric charges of this emergent $\mathbb Z_2$ gauge theory, one can also refer to the $e$-condensate as the ``Higgs phase''.
Due to the non-trivial braiding between $e$ and $m$, condensing the latter implies that the former is no longer deconfined, such that the $m$-condensate is also called the ``confined phase'' \cite{FradkinShenker}.
Although it is known that these two condensates form a single trivial phase \cite{FradkinShenker}, there can be an unnecessary (first-order) transition between them \cite{tupitsyn_topological_2010}. See Fig.~\ref{fig:intuitivepicture}(a) for two generic scenarios occurring in the parameter regimes that we will be exploring in this work; a detailed analysis of the phase diagram is found in Sec.~\ref{sec-DTC}).

\subsection{Spin Lakes from Quantum Dynamics}\label{subsec-Slakedynamics}

We now turn to understanding how a QSL-like state can be produced via dynamics even when the ground state of the system does not resemble a QSL.
More precisely, we envision the case where the ground state is still in a constrained Hilbert space, imposed by an energetic Gauss law, but we might not be in the deconfined phase (e.g., the $m$-condensate discussed above).

To be concrete, we will consider the model of Eq.~\eqref{eq-KitaevTC-in-F} without any plaquette resonances ($J = 0$), though the discussion below is quite general.
When $J = 0$, the ground state of the aforementioned model will fail to be a QSL aside from a small portion of the phase diagram (See Sec.~\ref{sec-DTC}).
Nevertheless, we will show in this subsection that short-time quantum dynamics (such as those available in analog NISQ devices) can create a state which has QSL-like signatures over large but inevitably finite patches of the system.

An intuitive picture for the dynamical preparation of a QSL-like state can be understood as follows and is depicted in Fig.~\ref{fig:intuitivepicture}(b).
Envision starting with a finite size system and initializing it in the ground state of the $e$-condensate, i.e., Higgs phase [$h_x \gg K, h_z$; left-most panel of Fig.~\ref{fig:intuitivepicture}(b)].
We then ramp up the value of $K$ (energetically enforcing the Gauss law) slow enough to be \emph{\textcolor{edit}{adiabatic}} with respect to the $e$-anyons such that the density of the $e$-anyons will go to zero at the end of the sweep.
At the same time, we will show that is possible to guarantee that this parameter sweep is much faster than the $m$-anyon energy scale, allowing for a \emph{sudden approximation} where $m$-anyon dynamics is frozen.
In conclusion, we equilibrate out the fast $e$-anyons present in the initial state, and prevent the nucleation of the relatively slow $m$-anyons.
As such, at the end of the sweep, the state prepared will be characterized by a lack of condensation of any anyons and the final state will be the deconfined phase of the emergent gauge theory.
We can use this effective picture to develop a prediction for the final state of the system following a sweep:
\begin{equation} \label{eq-sweeping-proj}
    \ket{\psi(T)} \; \propto \; \mathcal{P}_{G} \ket{\psi(0)}
\end{equation}
where $\ket{\psi(0)}$ is the state at the beginning of the sweep, $\ket{\psi(T)}$ is the state at the end of the sweep, $\mathcal{P}_{G}$ is the operator that projects out violations of the Gauss Law ($= \prod_v (1 + G_v)/2$ for the toric code example, see Eq.~\eqref{eq-TCGaussLaw}).

To illustrate above in the simplest possible setting, consider the initial state $\ket{\psi(0)} = \ket{+}^{\otimes N}$ which is the ground state of at $h_z=K=0$. 
Observe that by expanding this product state in the diagonal ($Z$) basis and using the above visual representation, it is the sum of all \emph{closed and open} string states:
\begin{equation}\label{eq-plus_state_strings}
    \ket{\psi(0)} = \ket{\frac{}{} \quad } + \ket{\frac{}{} \hspace{-0.5mm} \begin{tikzpicture}[scale = 0.5, baseline = {([yshift=-.5ex]current bounding box.center)}]
    \draw[red,line width=0.5mm] (-0.5, -0.5) .. controls (-0.4, -0.7) and (-0.2, -0.9) .. (0.0, -0.5) -- (0.0, -0.5) .. controls (0.1, -0.425) and (0.5, -0.2) .. (0.25, 0.1) -- (0.25, 0.1);
    \draw[red,line width=0.5mm] (-0.1, 0.4) ellipse (0.3 and 0.15);
    \end{tikzpicture}
    } + \ket{\hspace{-2mm}\frac{}{}  \begin{tikzpicture}[scale = 0.5, baseline = {([yshift=-.5ex]current bounding box.center)}]
    \draw[red,line width=0.5mm] (-0.5, -0.5) .. controls (-1, -0.2) and (-0.1, 
    0.2) .. (-0.5, 0.5) -- (-0.5, 0.5) .. controls (-0.9, 1) and (0.4, 1) .. (0.5, 0.5) -- (0.5, 0.5) .. controls (0.6, 0) and (0.9, -0.3) .. (0.5, -0.5) -- (0.5, -0.5) .. controls (0.1, -0.7) and (-0.1, -0.3) .. (-0.5, -0.5) -- cycle;
    \end{tikzpicture}\hspace{-1mm}} + \ket{\hspace{-2mm}\frac{}{}  \begin{tikzpicture}[scale = 0.5, baseline = {([yshift=-.5ex]current bounding box.center)}]
    \draw[red,line width=0.5mm] (-0.5, -0.5) .. controls (-1, -0.2) and (-0.1, 
    0.2) .. (-0.5, 0.5) -- (-0.5, 0.5) .. controls (-0.9, 1) and (0.4, 1) .. (0.5, 0.5) -- (0.5, 0.5);
    \end{tikzpicture}\hspace{-1mm}}  +  \cdots
\end{equation}
Hence, if we can project out all states containing open strings, we obtain the topological state in Eq.~\eqref{eq-TCWF}, i.e., $\ket{\psi_\textrm{TC}} \propto \mathcal P_G \ket{\psi(0)}$.
We claim that this projection can (approximately) be achieved by the aforementioned non-equilibrium parameter sweep, where we attempt to be adiabatic with respect to $e$-anyons (which gradually enforces $G_v=1$) and sudden with respect to $m$-anyons (i.e., keeping the coefficients in Eq.~\eqref{eq-TCWF} approximately constant).
In fact, in the fine-tuned case of $h_z=0$, quantum numbers prevent any $m$-anyon dynamics, but we will explore the more interesting and generic\footnote{While this example might suggest that our mechanism requires a nearby flux(plaquette)-conserving model, this is not the case; our ruby lattice example in Sec.~\ref{sec-Experiment} will illustrate this.} case of $h_z\neq 0$.

However, in the thermodynamic limit, we will argue that it will not be possible\footnote{Here, we consider the generic case, i.e., $h_z\neq 0$, such that the plaquette resonance is not a conserved quantity.} to sweep the Hamiltonian at a rate that is both slow relative to the energy scale of the initially condensed defects (such as the $e$-anyons above) and fast relative to energy scale within the constrained space (such as the $m$-anyons above). In those cases, the final state will not be a perfect QSL.
Nevertheless, we will argue that correlations in the final state will be similar to those found in a QSL in any large patch of the system [right-most panel of Fig.~\ref{fig:intuitivepicture}(b)].
As a consequence, it will be appropriate to brand the final state as either a \textit{quantum spin puddle} or \textit{quantum spin lake} (depending on one's philosophical bend).
Since the authors are glass-half-full, we will henceforth refer to such states somewhat optimistically as quantum spin lakes which, though not thermodynamic QSLs, are states that enable studying QSL physics in finite-size quantum simulation experiments available in the NISQ era.
More generally, the effective picture presented above provides a route to applying projection operators on quantum states by using non-equilibrium unitary dynamics!

\subsection{Outline of the Paper}\label{subsec-Outline}

The remainder of this work will be focusd on fleshing out the above intuitive idea, providing numerical confirmation, identifying its limitations, building a bridge to existing experimental data, and finally providing generalizations.
First, Section~\ref{sec-Qutrit} makes the above picture more precise in the simplest possible context: a single qutrit model that has the essential ingredients of the setup above.
Subsequently, in Section~\ref{sec-DTC} we will provide numerical support for this picture by performing large-scale matrix product state numerics on Eq.~\eqref{eq-KitaevTC-in-F} without explicit plaquette resonances ($J = 0$).
Equipped with the numerical evidence backing the intuitive picture presented above, we then turn to considering the validity of this picture for thermodynamically large systems in Section~\ref{sec-Summary} where we will make precise the notion of a quantum spin lake.
We use this notion to make comments on the relevance of these ideas for explaining the recent Rydberg atom experiment (Section~\ref{sec-Experiment}).
Driving the above intuition to its logical conclusion suggests that dynamically preparing QSL-like states works best in models with vanishing $m$-anyon dynamics, which we exemplify by simulating a model on a tree lattice in Sec.~\ref{sec-Tree}. 
Remarkably, we find that such tree numerics can even be used as a tool to accurately describe experimental data within the timescales used to prepare the quantum spin lake.
Although the bulk of the paper focuses on the preparation of $\mathbb{Z}_2$ spin lakes, we conclude by highlighting the generality of the mechanism by demonstrating the preparation of a $U(1)$ spin lake (Section~\ref{sec-U1}).

\section{Single Qutrit Toy Model for Dynamical QSL Preparation}\label{sec-Qutrit}

\begin{figure*}
    \centering
    \includegraphics[width = 485pt]{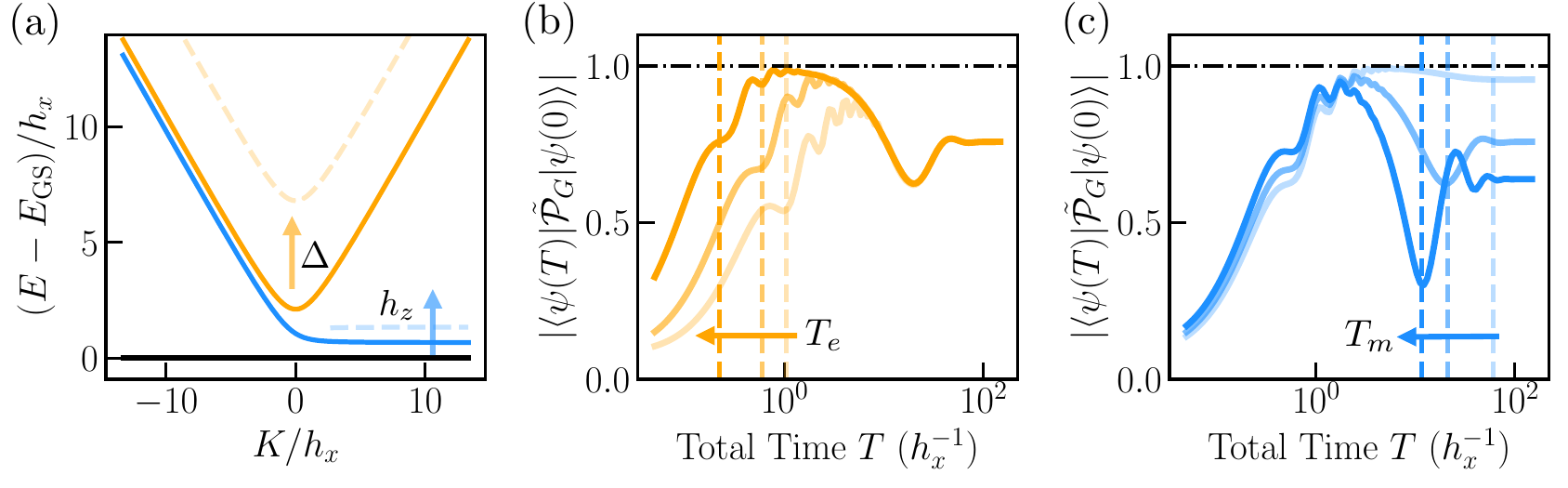}
    \caption{\textbf{Level Structure and Dynamics of Single Qutrit Model.} Our dynamical protocol for implementing the projection operator by combining adiabatic and sudden approximations can already be illustrated in a three-state toy model. (a) Representative level structure of single qutrit model as a function of $K/h_x$ (shown for parameters $h_z = 1/3$ and $\Delta = 0$). Here, the ground state is depicted in black, the first excited state in blue, and the second excited state in orange. Note that $\Delta$ controls the gap between the ground state and the orange level and $h_z$ controls the gap between the ground state and the blue level when $K > 0$.
    In panels (b, c), we start with an initial ground state $\ket{\psi(0)}$ at $K/h_x = -20$ (Eq.~\eqref{eq-qutritpsi0}) and dynamically sweep to $K/h_z = 20$ over a total time $T$: we plot the overlap between the final state $\ket{\psi(T)}$ and the initial state with constraint violations projected out $\mathcal{P}_G\ket{\psi(0)}$ (Eq.~\eqref{eq-qutritpsiproj}).
    We find an intermediate regime $[T_e, T_m]$ where this overlap is maximized, which qualitatively corresponds to sweep rates which are \emph{below} the energy scale set by the orange curve (in (a)) but \emph{above} the splitting $\sim h_z$ of the lowest two states for $K>0$; we are thus approximately adiabatic (sudden) with respect to the former (latter) branch.
    In (b), by plotting this projected overlap as a function of $T$ for $\Delta = 0, 7, 30$ (corresponding to the lightest to darkest orange lines) and fixed $h_z = 1/3$, we find that we can move $T_e$ to shorter times.
    In (c), by plotting this overlap as a function of $T$ for $\Delta = 7$ and $h_z = 1/15, 1/3, 2/3$ (lightest to darkest lines), we find that we can tune $T_m$ to shorter times as well.}
    \label{fig-SingleQutrit}
\end{figure*}

In Section~\ref{subsec-Slakedynamics}, we discussed an effective picture for how a QSL-like state is created during a dynamical sweep.
This picture suggested a natural but striking prediction for the final state of the dynamics (Eq.~\eqref{eq-sweeping-proj}): the final state is the initial state of the sweep but with Gauss law violations projected out.
Here, we show how this picture emerges and confirm this prediction in a truly minimal setting: a single qutrit model that mimics the setup of the last section.
In particular, consider the following Hamiltonian for a single qutrit:
\begin{equation} \label{eq-qutritHam}
    H_{\text{qutrit}} = -K \mathcal{Z}^2 - h_x \mathcal{X} - h_z\mathcal{Z}
\end{equation}
where $\mathcal{X}, \mathcal{Z}$ are spin-one Pauli matrices:
\begin{equation}
    \mathcal{Z} = \begin{pmatrix} 1 & 0 & 0 \\ 0 & 0 & 0 \\ 0 & 0 & -1 \end{pmatrix},\quad \mathcal{X} = \frac{1}{\sqrt{2}}\begin{pmatrix} 0 & 1 & 0 \\ 1 & 0 & 1 \\ 0 & 1 & 0 \end{pmatrix}.
\end{equation}
Such a Hamiltonian is a nice $(0 + 1)$D analogue of the Hamiltonian of Eq.~\eqref{eq-KitaevTC-in-F} with $J = 0$.
In particular, the feature we would like to focus on is that for large positive $K$, we have a ``constrained low-energy Hilbert space'' where $\mathcal Z^2 = 1$, i.e., $\{ \ket{1}, \ket{-1}\}$, where we define this basis such that $\mathcal{Z} \ket{\alpha} = \alpha \ket{\alpha}$, $\alpha = -1, 0, 1$.

\subsection{Projection and Superpositions via Dynamics}

We will start at large negative $K$, where the ground state is nearly classical:
\begin{equation}
    \ket{\psi(0)} \approx \ket{0} + \varepsilon \left( \ket{1} + \ket{-1} \right) \label{eq-qutritpsi0}
\end{equation}
provided that $|h_z|\ll |h_x| \ll |K|$, such that $\varepsilon \sim \frac{|h_x|}{|K|} \ll 1$.
Our claim is that one can use a non-equilibrium sweep towards large positive $K$ to effectively project our initial state into the constrained space defined by $\mathcal{Z}^2 = +1$, i.e., we obtain
\begin{equation}
\mathcal{P}_G\ket{\psi(0)} \propto \ket{1} + \ket{-1} . \label{eq-qutritpsiproj} 
\end{equation}
Note that this superposition of constrained states is in stark contrast to what would be the \emph{ground state} in this parameter regime: for any $h_z > 0$ and positive $K \gg h_x$, the ground state is approximately $\ket{1}$.

To justify this claim, it is useful to examine the spectrum of this three-level model as a function of $K/h_x$ at fixed but small $h_z$ which is shown in Fig.~\ref{fig-SingleQutrit}(a).
For large values of $K$, we see two low-energy states (the black and blue lines) corresponding to the ``constrained'' space spanned by $\ket{1}$ and $\ket{-1}$. Well-separated above this, we see $\ket{0}$ (orange line).
As such, starting with the state in Eq.~\eqref{eq-qutritpsi0} and sweeping from negative to positive $K$, we should throughout remain adiabatic with respect to third (orange) curve; as a result the final wavefunction will be in the constrained subspace. 
If at the same time we remain faster than the splitting of the blue and black lines in this constrained space, we can use the sudden approximation indicating that the portion of the initial wavefunction \eqref{eq-qutritpsi0} that was within this space does not time-evolve, achieving the projection in Eq.~\eqref{eq-qutritpsiproj}.

The above discussion highlighted the two necessary ingredients for dynamics to produce the desired projected state: the sweep rate should be slow relative to the orange curve and fast relative to the blue curve (for $K \geq 0$).
The validity and applicability of these conditions is in principle set by the following two parameters. 
First, $h_z$ determines the splitting between the two constrained states [as indicated in Fig.~\ref{fig-SingleQutrit}(a)].
Second, by pushing up the third level by an amount $\Delta$\footnote{We can do this by making our Hamiltonian time-dependent, adding a term $\Delta \cdot \mathcal{P}_3(t)$ which pushes the highest level up in energy at each instance by $\Delta$.}, we can tune the gap at the ``transition'' into the constrained space.
Hence, we expect the projection in Eq.~\eqref{eq-qutritpsiproj} to become a better approximation for the non-equilibrium time-evolution when $h_z$ is small and $\delta$ is large.
We now test and confirm these expectations quantitatively.

\subsection{Numerical Confirmation and Timescales}

We numerically confirm the expectations above by exactly simulating the dynamics of the qutrit.
In particular, we initialize the qutrit in its ground state at large negative $K = -20$ with $h_x = 1$ and $h_z$ fixed at three representative values, and then linearly increase $K$ to $K = 20$ over a total time $T$. 
We subsequently plot the overlap the normalized\footnote{Throughout this work, we will use $\tilde {\mathcal P}$ to denote a projector followed by normalization.} projected state defined in Eq.~\eqref{eq-qutritpsiproj} with the final state of the sweep $\ket{\psi(T)}$ as a function of $T$ [See Figs.~\ref{fig-SingleQutrit}(b, c)].

We find that, for any fixed value of $\Delta$ and $h_z$ [any of the curves in Fig.~\ref{fig-SingleQutrit}(b, c)], that the overlap with the projected state displays three distinct regimes as a function of total time $T$, demarcated by two time scales which we will call $T_{e}$ and $T_{m}$ ($T_e < T_m$).
Here, $1/T_e$ is the rate below which our ground state energy level is adiabatic with respect to the orange level throughout the evolution and $1/T_m$ is the rate above which the ground state level is sudden with respect to the blue level around $K = 0$. Within $T_e < T < T_m$, we find that the projected state \eqref{eq-qutritpsiproj} is a good approximation to the result of the non-equilibrium sweep.

As stated in the previous subsection, by increasing the value of $\Delta$ and fixing the value of $h_z$ in Fig.~\ref{fig-SingleQutrit}(b), we find that the time scale $T_m$ remains fixed and $T_e$ shifts to smaller times because it is possible to be adiabatic relative to the orange level while sweeping faster when $\delta$ is large.
Additionally, as $\Delta$ is increased, the approximation that the final state tends to the projected state becomes more exact as predicted.
If instead we fix the value of $\Delta$ and increase the value of $h_z$ [Fig.~\ref{fig-SingleQutrit}(c)], we find that $T_e$ remains fixed and $T_m$ shifts to smaller times.
This is because one needs to sweep faster in order to be sudden relative to the splitting between the ground state and blue level which confirms our expectations.

\subsection{Analogy with Toric Code}

We can reframe the results for the single qutrit model in a language that is closer to the one used to discuss the toric code.
A full dictionary between the two is enumerated in Table~\ref{table:single-qutrit}.
Notably, the constraint $\mathcal{Z}^2 = +1$ which holds when $K$ is large and positive can be reinterpreted as a ``Gauss law'' similar to the Gauss law of the toric code [Eq.~\eqref{eq-TCGaussLaw}].
Then, the orange level for $K > 0$ can be thought of as the ``$e$-anyon'' as it represents a violation of the Gauss law $\mathcal{Z}^2 = + 1$.
As such, when $K$ is large and negative, we can interpret the ground state as though this $e$-anyon has ``condensed'' (gained an expectation value in the ground state) corresponding to the Higgs phase of the toric code.
Similarly, the splitting between the ground state and the blue level can be thought of as the energy scale associated with the ``$m$-anyon'' as it respects the Gauss law.
At any finite $h_z$, the ground state can be interpreted as being the analogue of the toric code's ``confined phase.''

In this language, we can reinterpret the results of the dynamical sweep.
Namely, we prepare the projected state [which is equivalent to the deconfined phase due to being a superposition of constrained states (See Table.~\ref{table:single-qutrit})], when we remain in adiabatic relative to the energy level connected to the qutrit's $e$-anyon and sudden relative to the energy scale associated with the qutrit's $m$-anyon.
In being adiabatic relative to the $e$-anyon, it is equilibrated out as we exit the Higgs phase.
Moreover, in being sudden relative to the $m$-anyon, it fails to be nucleated in as we enter the confined phase.
Having tested and verified our intuition in this toy model, we study the analogous effect in a truly many-body system, namely the toric code model.

\begin{table}
\begin{center}
\begin{tabular}{||c c||}
 \hline
 Toric Code & Single Qutrit\\ [0.5ex] 
 \hline\hline
 Gauss Law & $G = \mathcal{Z}^2 = +1$ \\
 \hline
  $h_x X_{\ell}, h_z Z_{\ell}$ & $h_x \mathcal{X}, h_z \mathcal{Z}$ \\
 \hline
 Higgs & $\ket{0}$  \\ 
 \hline
 QSL & $\ket{\Omega} = \frac{1}{\sqrt{2}} (\ket{+1} + \ket{-1})$  \\
 \hline 
 Confined & $\ket{1}, \ket{-1}$  \\
 \hline 
 $e$-anyon & $e^{\dagger} \ket{\Omega} = \mathcal{X} \ket{\Omega}$  \\
 \hline
 $m$-anyon & $m^{\dagger} \ket{\Omega} = \mathcal{Z} \ket{\Omega}$ \\
 \hline
\end{tabular}
\caption{\textbf{Conceptual Dictionary Between Toric Code and Single Qutrit Model.}\label{table:single-qutrit} }
\end{center}
\end{table}

\section{Deformed Toric Code Model} \label{sec-DTC}

We now investigate how the effective picture of Section~\ref{sec-Key} and the conjecture of Eq.~\eqref{eq-sweeping-proj} appear in a many-body context.
In particular, let us consider the model of Eq.~\eqref{eq-KitaevTC-in-F} without explicit plaquette resonances ($J = 0$):
\begin{equation} \label{eq-TCf_2}
    H_{\text{TC + f}} = -K\sum_v \begin{tikzpicture}[scale = 0.5, baseline = {([yshift=-.5ex]current bounding box.center)}]
    \draw[gray] (1.5,0) -- (-1.5, 0);
    \draw[gray] (0,1.5) -- (0,-1.5);
    \node at (0.75, 0) {\normalsize $Z$};
    \node at (-0.75, 0) {\normalsize $Z$};
    \node at (0, 0.75) {\normalsize $Z$};
    \node at (0, -0.75) {\normalsize $Z$};
\end{tikzpicture} - h_x \sum_{\ell} X_{\ell} - h_z \sum_{\ell} Z_{\ell}
\end{equation}
which we will refer to as the \emph{deformed toric code} for brevity.
Before exploring the dynamical preparation of QSL-like states in this model, in Subsection~\ref{subsec-TCGS}, we first consider its ground state physics, where we will find a thin sliver of topological order.
Subsequently, in Subsection~\ref{subsec-TCSpinLakes}, we test the prediction of Eq.~\eqref{eq-sweeping-proj} by calculating the local overlap of the time-evolved state and the projected state. 
This indeed suggests a spin liquid-like state which is vastly more extended in the phase diagram compared to the ground state physics.
Here we focus on detecting these spin liquid-like properties in finite regions; we postpone the discussion of scaling and the thermodynamic limit to Section~\ref{sec-Summary}.

\subsection{Ground State Phase Diagram} \label{subsec-TCGS}

\begin{figure}
    \centering
    \includegraphics[width = 247pt]{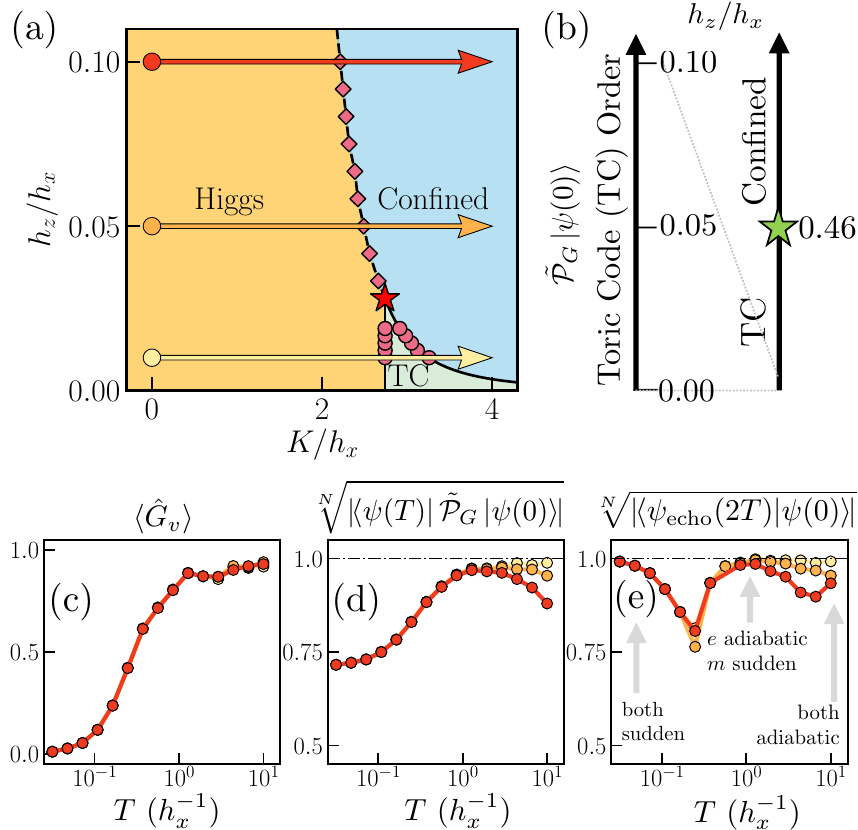}
    \caption{\textbf{Creation of a Quantum Spin Lake in a Deformed Toric Code Model.} We consider a version of the toric code model in a field \emph{without} plaquette resonances: Eq.~\eqref{eq-TCf_2}. (a) Ground state phase diagram of as a function of $h_z/h_x$ and $K/h_x$. There is only a tiny sliver of Toric Code (TC) topological order.
    (b) In contrast, projecting a product state (i.e., ground state at $K=0$ for different $h_z/h_x$) into the Gauss-law preserving states $G_v=1$ yields a robust spin liquid phase (see Sec.~\ref{subsec-TCproof} for an analytic derivation).
    In panels (c, d, e), we study the dynamical sweeps along the three colored arrows in panel (a) corresponding to $h_z=0.01,0.05,0.1$; we numerically confirm that non-equilibrium dynamics can effectively implement the aforementioned projection.
    (c) By increasing the total time of the sweep, $e$-anyons are pushed out of the final state as detected by the expectation value of the Gauss Law $\langle G_v \rangle \approx 1$ [defined in Eq.~\eqref{eq-TCGaussLaw}]. This (quasi-)adiabatic approximation for $e$-anyons holds for all three sweeps as evidenced by the overlapping curves.
    (d) Having established that we dynamically sweep into the constrained subspace, we test whether the resulting state $\ket{\psi(T)}$ has a large overlap with the corresponding projected state (the latter is in a topological phase as shown in (b)). 
    We observe a window $T_e < T <T_m$ where the overlap density between the final state of the sweep and the projected state is large; here we are approximately adiabatic (sudden) w.r.t. $e$-($m$-)anyons.
    (e) We make explicit the presence of three dynamical regimes by plotting the return probability for an echo experiment that sweeps back and forth through the transition.
    }
    \label{fig-dTC}
\end{figure}

Using the density matrix renormalization group (DMRG) \cite{White92, White93, Stoudenmire12, Hauschild18} on an infinite cylinder of circumference $L_y = 4$ (the qualitative properties of interest do not sensitively depend on this choice; see Appendix~\ref{app-Ly5}), we find the ground state phase diagram of the model of Eq.~\eqref{eq-TCf_2} in Fig.~\ref{fig-dTC}~(a), finding the three distinct phases which we schematically discussed in Section \ref{sec-Key}. (See Appendix~\ref{app-Ly5} for $L_y=5$ and further numerical details.)

The first phase we observe is the $e$-condensed (or Higgs) phase which occurs when $K, h_z \ll h_x$, its fixed-point limit being the product state in the $X$-basis, $\ket{+}^{\otimes N}$.
The second two phases---the toric code (TC) phase and the confined phase---can be understood as follows.
When $K \gg h_x, h_z$, the ground state manifold will be nearly degenerate, consisting of all states that satisfy the Gauss law $G_v=1$ [see Eq.~\eqref{eq-TCGaussLaw}].
The effect of $h_z$ and $h_x$ can then be treated perturbatively.
Namely, using degenerate perturbation theory in $h_z$ and $h_x$, the effective Hamiltonian governing these states will contain the plaquette term of Eq.~\eqref{eq-KitaevTC} with $J_{\text{eff}} \sim \mathcal{O}(h_x^4/K^3)$.
As a consequence, when $h_z \ll J_{\text{eff}}$, the model will be in the $\mathbb{Z}_2$ QSL phase of the toric code model of Eq.~\eqref{eq-KitaevTC}.
However, when $h_z \gg J_{\text{eff}}$, $m$-anyons in the system will condense corresponding to the confined phase, whose fixed-point limit is $\ket{\uparrow}^{\otimes N}$ as $h_z \to +\infty$.

As a final remark, we note that throughout the phase diagram of Fig.~\ref{fig-dTC}, the energy scale associated with $m$-anyon excitations is set by $h_z$ and potentially a plaquette term generated at fourth order in perturbation theory, both of which are small relative to $h_x$ and $K$ which set the $e$-anyon dynamics.
These small energy scales naturally signal that the dynamics of $m$-anyons will be slow relative to the $e$-anyons.

\subsection{Quantum Dynamics and Spin Lakes} \label{subsec-TCSpinLakes}

Given the equilibrium phase diagram, let us now contrast it with the state prepared via a dynamical sweep simulated using MPO methods \cite{Zaletel15}.
Due to the separation in energy scales between the $e$- and $m$-anyons, we anticipate that we will be able to prepare a QSL-like state [or quantum spin lakes (see Section~\ref{sec-Summary} for more details)], extending well beyond the spin liquids in the ground state phase diagram.
In particular, the discussion in Section~\ref{sec-Key} suggests that sweep rates which are slow with respect to $e$ and fast with respect to $m$ should approximately project out Gauss law violating states from the initial state [see Eq.~\eqref{eq-sweeping-proj}].
If we take initial states at $K=0$ (where the ground state is a product state), then projecting these into $G_v=1$ gives the phase diagram in Fig.~\ref{fig-dTC}(b) (which will be derived in the next subsection).
Crucially, we see that the topological (i.e., deconfined) phase extends over a broad range of parameter space, up to $h_z / h_x \lessapprox 0.46$.
This is in contrast to the tiny sliver of toric code phase found in the ground state [Fig.~\ref{fig-dTC}(a)].
We now numerically test the prediction that appropriate sweeping rates can approximately prepare this projected wavefunction through dynamics.

We initialize the system in the product state ground state at $K = 0$ and small values of $h_z$ [in the `Higgs phase' of Fig.~\ref{fig-dTC}(a)].
Subsequently, we ramp $K$ linearly at a rate $1/T$ and investigate the nature of the final state.
By simulating Eq.~\eqref{eq-TCf_2} on an infinite cylinder using matrix product state techniques, we are able to investigate properties of the final state numerically as a function of the total time $T$.
First, we verify that as we increase the total time $T$ (thereby decreasing the sweeping rate), there is a time-scale $T_e$ above which our dynamics are nearly in equilibrium relative to $e$-anyons.
In particular, above $T_e$, we expect that the density of $e$-anyons in the final state will be nearly zero similar to the ground state for $K$ large and greater than zero (though, for any finite $T$, the ground state will have a non-zero density of $e$-anyon defects; see Sec.~\ref{sec-Summary} for a discussion of finite sizes and scaling).
To verify this, in Fig.~\ref{fig-dTC}(c), we plot the expectation value of the Gauss law operator [Eq.~\eqref{eq-TCGaussLaw}] $\langle G_v \rangle$ as a function of the total time of the sweep and three values of $h_z$.
We find that above a characteristic value of $T_e \sim 0.5 h_x^{-1}$, the value of $\langle G_v \rangle$ rapidly increases and saturates to a near maximal value (consistent with the equilibrium value) independent of the value of $h_z$.
As $h_z$ controls the energetics of the $m$-anyon, this is to be expected.

Next, we confirm that beyond $T_e$, we enter a regime where the dynamics is simultaneously fast relative to $m$-anyons and slow relative to $e$-anyons.
Here, we expect that the final state will have a high overlap density with the normalized projected state $\tilde{\mathcal{P}}_G \ket{\psi(0)}$, which is a spin liquid for the parameters chosen (See Fig.~\ref{fig-dTC}(b) for phase diagram of the projected state, proved in the next subsection).
By plotting the overlap density per site between $\ket{\psi(T)}$ and $\tilde{\mathcal{P}}_G \ket{\psi(0)}$ in Fig.~\ref{fig-dTC}(d) (where the tilde simply denotes that we have normalized the state), we find that there is a window $[T_e, T_m]$ where indeed this occurs, in agreement with the prediction of Eq.~\eqref{eq-sweeping-proj}.
Furthermore, we find that as we increase $h_z$, the coupling responsible for nucleating $m$-anyons, $T_m$ decreases and hence the window shrinks.
This is consistent with our expectations that, as we increase $h_z$, the time-scale in which $m$-anyons are nucleated decreases and hence our dynamics can be slow relative to both $e$ and $m$ (i.e. quasi-\textcolor{edit}{adiabatic}) at faster rates (shorter total times).
We confirm in Appendix~\ref{app-GSrecovery} that indeed, beyond $T_m$, the system recovers the ground state.
This provides strong numerical evidence for our effective picture wherein $e$-anyons are in equilibrium and $m$-anyons are frozen for intermediate time sweeps. 

We can independently verify the existence of the two time-scales $T_e$ and $T_m$ as well as the intermediate regime via the follow numerical ``echo'' experiment.
Namely, we consider sweeping $K$ linearly from $K = 0$ to a maximal value of $K$ for a total time $T$ and subsequently sweeping $K$ linearly backwards back to zero for the same amount of time.
Generically, the state that one will recover will not be initial state of the sweep.
Nevertheless, we expect that when the dynamics is purely \textcolor{edit}{adiabatic} or purely sudden, the initial state will be recovered.
Moreover, since our proposed mechanism involves the dynamics relative to $e$ being quasi-\textcolor{edit}{adiabatic} and the dynamics relative to $m$ being quasi-sudden, a non-trivial prediction of our effective picture is that, in the intermediate regime, the initial state will also be recovered.
We numerically simulate this experiment and plot the overlap per site of the final state $\ket{\psi(2T)}$ with the initial state in Fig.~\ref{fig-dTC}(e).
Let us first observe that when we are fully out-of-equilibrium relative to $e$ and $m$ ($T \ll T_e$) or nearly in-equilibrium relative to both ($T \gg T_m$), we indeed find that this overlap is near maximal.
More interestingly, when we are deep within the regime $T_e \ll T \ll T_m$, we also see a very large revival of the initial state, consistent with certain degrees of freedom being quasi-\textcolor{edit}{adiabatic} and others being frozen.

\begin{figure}
    \centering
    \includegraphics[width = 247pt]{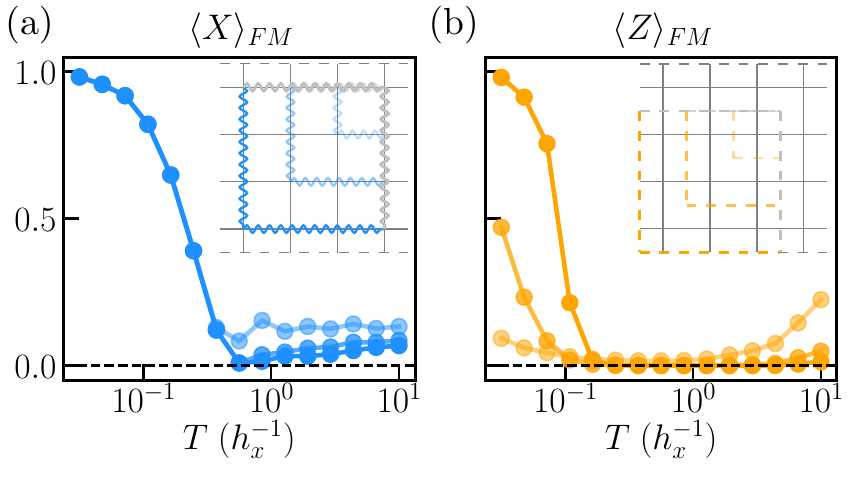}
    \caption{\textbf{Fredenhagen-Marcu Order Parameter for Final State of Dynamics.} We consider the same dynamical sweeps in the deformed toric code model as introduced in Fig.~\ref{fig-dTC}. (a) Here, we report the maximum value of the $\langle X \rangle_{\text{FM}}$ order parameter (for three different string lengths) obtained during a dynamical sweep that takes place over a total time $T$. The particular sweep we take linearly ramps the value of $K/h_x$ from $0$ to $4$ at $h_z/h_x = 0.1$ (this corresponds to the red arrow in Fig.~\ref{fig-dTC}). We find that during the window $[T_e, T_m]$, its value flows downwards with increasing length. (b) In this panel, we plot the value of $\langle Z \rangle_{\text{FM}}$ versus total time $T$ at the same time in the sweep where $\langle X \rangle_{\text{FM}}$ was maximized. Here, for total times  $T \in [T_e, T_m]$, the value of $\langle Z \rangle_{FM}$ is nearly zero and is confirmed to flow downwards. The decreasing flow of these two order parameters during the window $[T_e, T_m]$ cements the fact that in this window, the system exhibits QSL-like properties, consistent with the large overlap between the time-evolved state and the topologically ordered projected wavefunction in Fig.~\ref{fig-dTC}(d). For numerics performed, we used a bond- dimension of $\chi = 256$ and trotter step size of $dt = 0.0025$ (with convergence in both parameters verified). Moreover, see Appendix~\ref{app-Phase-and-CP-extraction}
    for FM order parameters in equilibrium. } 
    \label{fig-FMdTC}
\end{figure}

Until now, we have only verified that states produced through dynamics in the intermediate regime $[T_e, T_m]$ look like QSL's by utilizing the overlap density with the projected state (which is provably a QSL; see next subsection).
We conclude this subsection by independently verifying this through the use of the so-called Fredenhagen-Marcu (FM) order parameter~\cite{Fredenhagen83,Fredenhagen86, Marcu86,Gregor11,Chandran_2013,Verresen21}.
To define the FM order parameter, we first introduce the following string operators:
\begin{equation}
\begin{tikzpicture}[scale = 0.5, baseline={([yshift=-.5ex]current bounding box.center)}]
\foreach \i in {0,...,1} {
    \draw[gray] (\i, -0.5) -- (\i, 1.5);
    \draw[gray] (-0.5, \i) -- (1.5, \i);
}
\draw[orange(ryb), dashed, line width = 0.3 mm] (-0.5, 0.5) -- (1.5, 0.5);
\end{tikzpicture}
\ =\ 
\begin{tikzpicture}[scale = 0.5, baseline={([yshift=-.5ex]current bounding box.center)}]
\foreach \i in {0,...,1} {
    \draw[gray] (\i, -0.5) -- (\i, 1.5);
    \draw[gray] (-0.5, \i) -- (1.5, \i);
}
\node at (0, 0.5) {\normalsize $Z$};
\node at (1, 0.5) {\normalsize $Z$};
\end{tikzpicture}
\quad \quad 
\begin{tikzpicture}[scale = 0.5, baseline={([yshift=-.5ex]current bounding box.center)}]
\foreach \i in {0,...,1} {
    \draw[gray] (\i, -0.5) -- (\i, 1.5);
    \draw[gray] (-0.5, \i) -- (1.5, \i);
}
\draw[dodgerblue, decorate, decoration={snake, segment length=1.2mm, amplitude =.4mm}, line width = .25mm, opacity = 1]  (-0.5, 0) -- (1.5, 0);
\end{tikzpicture}
\ =\ 
\begin{tikzpicture}[scale = 0.5, baseline={([yshift=-.5ex]current bounding box.center)}]
\foreach \i in {0,...,1} {
    \draw[gray] (\i, -0.5) -- (\i, 1.5);
    \draw[gray] (-0.5, \i) -- (1.5, \i);
}
\node at (-0.5, 0) {\normalsize $X$};
\node at (0.5, 0) {\normalsize $X$};
\node at (1.5, 0) {\normalsize $X$};
\end{tikzpicture}
\end{equation}
the first (second) of which is called the 't Hooft (Wilson) line operator and creates $m$($e$)-anyons at its endpoints.
String operators in hand, the FM order parameter is defined as:
\begin{align} \label{eq-FMOP}
\langle Z \rangle_{\text{FM}} = \frac{\left \langle\begin{tikzpicture}[scale = 0.5, baseline={([yshift=-.5ex]current bounding box.center)}]
\draw[orange(ryb), dashed, line width = 0.4 mm] (2,0) arc(0:-180:1);
\node at (0, 0) {\normalsize $m$};
\node at (2, 0) {\normalsize $m$};
\end{tikzpicture} \right \rangle}{\sqrt{\left \langle\begin{tikzpicture}[scale = 0.5, baseline={([yshift=-.5ex]current bounding box.center)}]
\draw[orange(ryb), dashed, line width = 0.4 mm] (0,0) circle (1);
\end{tikzpicture} \right \rangle}\ } \quad \ 
\langle X \rangle_{\text{FM}} = \frac{\left \langle\begin{tikzpicture}[scale = 0.5, baseline={([yshift=-.5ex]current bounding box.center)}]
\draw[dodgerblue, decorate, decoration={snake, segment length=1.2mm, amplitude =.4mm}, line width = .25mm, opacity = 1] (2,0) arc(0:-180:1);
\node at (0, 0) {\normalsize $e$};
\node at (2, 0) {\normalsize $e$};
\end{tikzpicture} \right \rangle}{\sqrt{\left \langle\begin{tikzpicture}[scale = 0.5, baseline={([yshift=-.5ex]current bounding box.center)}]
\draw[dodgerblue, decorate, decoration={snake, segment length=1.2mm, amplitude =.4mm}, line width = .25mm, opacity = 1] (0,0) circle (1);
\end{tikzpicture} \right \rangle}\ }.
\end{align}
where the length of the string in the denominator has twice the length of the numerator and we have drawn schematically the $e$ and $m$ anyon excitations at the endpoints of the string.
Broadly speaking, the FM order parameter detects the lack of condensation of anyons in a topologically ordered phase.
In particular, the numerator is similar to a two-point function for either the $e$ or $m$-anyon, with the endpoints connected by a string.
Since any string operator will generically have some line tension causing it to decay exponentially regardless of whether the anyons are condensed or not, the denominator is chosen to cancel the contribution of this line tension.
The expectation is then that in a topologically ordered phase, the value of both $\langle \mathcal{Z} \rangle_{\text{FM}}$ and $\langle X \rangle_{\text{FM}}$ will go to zero with increased string length.
Meanwhile, in either the Higgs or confined phase, it will tend to a non-zero value.

In Fig.~\ref{fig-FMdTC}, we use the FM order parameter to diagnose the presence of QSL-like order in the final state of the sweep performed at $h_z = 0.1$.
In particular, in Fig.~\ref{fig-FMdTC}(a), we show the value of the minimum value of $\langle X \rangle_{\text{FM}}$ (for three different string lengths) obtained during a sweep of total time $T$\footnote{The minimum value is reported because the FM order parameter oscillates towards the end of the sweep. The time trace for this oscillation is shown in Appendix~\ref{app-FM}.}.
Similarly, we show the value of $\langle Z \rangle_{\text{FM}}$ obtained at the same time that $\langle X \rangle_{\text{FM}}$ was minimized.
In doing so, we find that, indeed, the intermediate regime $[T_e, T_m]$ is characterized by both order parameters decaying to zero with increased string length.
This confirms that the intermediate regime $[T_e, T_m]$ displays QSL-like signatures.

\subsection{When is the Projected State a Spin Liquid?}\label{subsec-TCproof}

So far, we have presented strong numerical evidence that, for a window of sweep rates, the final state of a dynamical sweep will be the initial state with Gauss law violations projected out.
Nevertheless, we have yet to discuss when such a projected state will be a quantum spin liquid.
This can be answered analytically, but first we provide an intuitive explanation.

Note that the deep in the Higgs phase, if $h_z = 0$, then the ground state has no $m$-anyons because the expectation value of the plaquette term will be $+1$ everywhere.
As such, when we project out Gauss law violations, the resulting state now has no $e$-anyons while remaining free of $m$-anyons, and is consequently a quantum spin liquid.
More precisely, the ground state deep in the Higgs phase when $h_z = 0$ is the state where all the qubits are in the $\ket{+}$ state.
When this state is expanded in the $Z$ basis, it looks like the equal weight super position of all open and closed string states [in the electric field representation of our spins (See Sec.~\ref{subsec-SLinEq})].
As such, when we project out all Gauss law violations (equiv. states with open strings), the resulting state is the sum of all closed string configurations which is precisely the toric code ground state of Eq.~\eqref{eq-TCWF}.
If we instead start with a product state for a nonzero $h_z$, there will be a small number of virtual $m$-anyon fluctuations in the initial state (or equivalently, the strings in the wavefunction will have a small line tension).
Hence, as we increase $h_x$ beyond some threshold value, we will eventually fail to create a QSL under projection.

To make the above arguments mathematically precise, note that the initial state of the dynamical sweep, when $K = 0$, is a product state of the form:
\begin{equation}
\ket{\psi(0)} = \bigotimes_{\ell} \left[ \cos(\theta) \ket{+} + \sin(\theta) \ket{-} \right]_{\ell} = \frac{e^{\frac{\beta}{2} \sum_{\ell} Z_{\ell}} }{\mathcal{Z}}\ket{+}^{\otimes N}
\end{equation}
where $\theta = \frac{1}{2}\arctan\left( \frac{h_z}{h_x}\right)$ and the second equality is an exact reparameterization in terms of $\tanh(\beta/2) = \tan(\theta)$.
We want to know when:
\begin{equation}\label{eq-proj_state_pre_KW}
\mathcal{P}_{G}\ket{\psi(0)}
\propto
e^{\frac{\beta}{2} \sum_{\ell} Z_{\ell}} \ket{\psi_{TC}}
\end{equation}
is a quantum spin liquid.
In the above equation, we used the fact that $\mathcal{P}_{G}$ commutes with $e^{\frac{\beta}{2} Z_{\ell}}$ and $\mathcal{P}_{G} \ket{+}^{\otimes N} \propto \ket{\psi_{TC}}$ where $\ket{\psi_{TC}}$ is the toric code wavefunction defined in Eq.~\eqref{eq-TCWF}.
To do so, we first map the state in Eq.~\eqref{eq-proj_state_pre_KW} to its dual under the Kramers-Wannier map defined in this case as: 
\begin{equation}
    \begin{tikzpicture}[scale = 0.5, baseline={([yshift=-.5ex]current bounding box.center)}]
    \draw[gray] (0, -1) -- (0, 1);
    \draw[gray] (-1.5, 1) -- (1.5, 1);
    \draw[gray] (-1.5, -1) -- (1.5, -1);
    \node at (0,0){\normalsize $Z$};
    \end{tikzpicture} \Longleftrightarrow \begin{tikzpicture}[scale = 0.5, baseline={([yshift=-.5ex]current bounding box.center)}]
    \draw[gray] (0, -1) -- (0, 1);
    \draw[gray] (-1.5, 1) -- (1.5, 1);
    \draw[gray] (-1.5, -1) -- (1.5, -1);
    \node at (-1,0){\normalsize $Z$};
    \node at (1,0){\normalsize $Z$};
    \end{tikzpicture}
\end{equation}
and
\begin{equation}
    \begin{tikzpicture}[scale = 0.5, baseline={([yshift=-.5ex]current bounding box.center)}]
\draw[gray] (-1, -1) -- (-1, 1) -- (1, 1) -- (1, -1) -- cycle;
\node at (0.0, -1) {\normalsize $X$};
\node at (0.0, 1) {\normalsize $X$};
\node at (-1, 0.0) {\normalsize $X$};
\node at (1, 0.0) {\normalsize $X$};
\end{tikzpicture} \Longleftrightarrow \begin{tikzpicture}[scale = 0.5, baseline={([yshift=-.5ex]current bounding box.center)}]
\draw[gray] (-1, -1) -- (-1, 1) -- (1, 1) -- (1, -1) -- cycle;
\node at (0.0, 0.0) {\normalsize $X$};
\end{tikzpicture} 
\end{equation}
which maps our model defined on the links of the square lattice to a model defined on the plaquettes of the square lattice.
Note that under this duality, $G_v$ is restricted to be $+1$ and hence, the toric code wavefunction, which is stabilized by the $G_v$ operator and the plaquette resonance operator, is mapped to the the $+1$ eigenstate of the $X_p$ operator.
Hence, 
\begin{equation}
    e^{\frac{\beta}{2} \sum_{\ell} Z_{\ell}} \ket{\psi_{TC}} \Longleftrightarrow e^{\frac{\beta}{2}\sum_{\langle p, p' \rangle } Z_p Z_{p'}} \ket{+}^{\otimes N}.
\end{equation}

We thus obtain a state whose diagonal correlations are set by the classical Ising model at inverse temperature $\beta$.
It is known that the disordered phase of the classical Ising model maps to the QSL phase of the toric code model under Kramers-Wannier.
Thus, the projected state is a quantum spin liquid when the initial state is such that $\beta > \beta_c$ where $\beta_c =\frac{\ln(1+\sqrt{2})}{2}$ is the transition temperature of the classical Ising model \cite{kramersandwannier}.
Translating this to our parameters, we obtain that there is a transition out of the topological phase at:
\begin{align}
\frac{h_z}{h_x} &= \tan \left( 2 \arctan \left(  \tanh \left( \frac{\ln(1+\sqrt{2})}{4} \right) \right) \right) \\
&= \sqrt{ \frac{\sqrt{2}-1}{2} } \approx 0.4551,
\end{align}
as shown in Fig.~\ref{fig-dTC}(b).

We should remark that before our dynamical protocol reaches such values of $h_z/h_x$, our system will fail to satisfy our dynamical requirements of being slow with respect to dyanmics of $e$-anyons and fast with respect the dynamics of $m$-anyons.
As a consequence, for all values of $h_x$ wherein our mechanism applies, we see a QSL-like state.
In the next section, we remark on what happens as the system fails to satisfy the aforementioned dynamical requirements. In particular, we will argue that as we increase our system size, the requirements are eventually bound to fail, leading to a finite-size `quantum spin lake'.

\section{Scaling and Limitations of the Mechanism}
\label{sec-Summary}

In the previous section, we saw QSL-like properties emerge in the non-equilibrium dynamics of the toric code model of Eq.~\eqref{eq-TCf_2}, even when the ground state was not topologically ordered.
This could be understood in a nice effective picture.
By dynamically sweeping the value of $K$ in the Hamiltonian at a rate that was slow relative to $e$-anyons, we gradually pushed them out of the initial state.
If this rate was simultaneously fast relative to the dynamics of $m$-anyons, they were effectively not created during the sweep.
This led to a toric-code-like state as evidenced by the overlap density with the projected state and FM order parameters.

In this section, we will make this picture more precise by clarifying both what is meant by slow and fast, and investigating the fate of this mechanism as we scale the system to the thermodynamic limit.
We will argue that, as we increase the system size, it will not be possible to globally remain in equilibrium relative to $e$-anyons and out-of-equilibrium relative to $m$-anyons due to (1) the presence of a phase transition as we exit the Higgs phase and (2) a finite $m$-anyon energy scale. 
Nevertheless, the local correlations of the system will present QSL-like signatures defining the quantum spin lake.

In what follows, we will first consider the case where our dynamical sweep crosses the second order transition (Subsection~\ref{subsec-secondorder}) and then consider what happens when we cross the first order transition (Subsection~\ref{subsec-firstorder}). 
In both cases, we argue that a finite-size spin lake is created and explain the dependence on sweeping rate and total time.
We conclude by numerically testing and confirming the scaling of this mechanism (Subsection~\ref{subsec-limitationsnumerics}).

\subsection{Crossing the Second Order Phase Transitions} \label{subsec-secondorder}

In this subsection, we discuss the dynamics of both $e$ and $m$-anyons during a dynamical sweep that crosses a second order phase transition [e.g. the sweep shown with the yellow arrow on Fig.~\ref{fig-dTC}(a)].

To get a better understanding of what will happen across the transition, it will be useful to recall lessons learned from studying the single qutrit model.
In that model, our system started off in the qutrit's ground state and adiabatically followed it until we approached the parameter regime around $K/h_x = 0$.
In particular, here, for a sufficiently fast rate, our system fell out of equilibrium relative to both the orange level and the blue level (See Fig.~\ref{fig-SingleQutrit}(a)). 
By pushing up the orange level through $\Delta$, we found that we were able to avoid this issue and always remain in equilibrium relative to the orange level.
Nevertheless, after entering the regime close to $K/h_x = 0$, we were always out-of-equilibrium relative to the blue level.
It was for this reason that, after $K/h_x = 0$, our wavefunction remained orthogonal to the orange level (thereby being within the constrained subspace), but its dynamics were slow within the constrained subspace.
Under an assumption that these dynamics were perfectly slow, the final wavefunction would just be the wavefunction at the instance when it fell out of equilibrium relative to the blue level, but with constraint violations projected out.
In the rest of this subsection, we will argue that a similar picture arises in the many-body context by leveraging universal properties of the transition out of the Higgs phase.
    
In the many-body context, far before the transition, a similar story plays out: while the system is deep in the Higgs phase, it has a large gap and the dynamics are adiabatic, largely tracking the many-body ground state \cite{AdiabaticTheorem_DeRoeck}.
However, as we approach the critical point of the transition, this will no longer be the case.
In particular, in the vicinity of the transition, there will be an emergent notion of $e-$ and $m$-anyons.
The former will uncondense across the transition and as such will have a gap at the critical point scaling as $\sim 1/L$~\cite{cardy1996scaling}.
On the other hand, the $m$-anyon excitation at the critical point will not generically be gapless.
Nevertheless, we know that deep on the other side of the transition, the gap and bandwidth of single $m$-anyon excitations are small because they are set by small microscopic energy scales and perturbatively generated resonances.
As such, generically we can assume that the bandwidth of $m$-anyons remains small close to the transition and remains small as we move far past the transition.
This makes precise what is meant by the phrase ``fast with respect to $m$-anyons'': the rate at which one crosses the transition is faster than the time-scale associated with the bandwidth of the \textit{emergent} $m$-anyon excitation in the critical regime around the transition, which is presumed to not drastically change as we move past the transition (and hence, can be estimated through microscopics).
The small gap to both $e$ and $m$-anyons across the transition, implies that prior to the transition the system will fall out of equilibrium and the long-distance dynamics of $e$ and $m$-anyons will be slow near the transition.
Indeed, in the parlance of the Kibble-Zurek mechanism, this corresponds to the so-called Kibble-Zurek ``freeze-out'' regime \cite{Kibble_1976, KIBBLE_1980, Zurek_1985, ZUREK_1996, Zurek_Review, Chandran_2013, Chandran_2012}.
    
After exiting the Kibble-Zurek regime, the gap to $e$-anyon excitations will rapidly increase and the ``frozen'' system will go back into equilibrium relative to the $e$-anyon (similar to how the dynamics in the single qutrit went back into equilibrium with the orange level).
This approach to equilibrium is believed to occur through a process called coarsening wherein constraint-satisfying regimes will grow in size until the final state hosts a dilute density of $e$-anyons, $n_e$.
Kibble-Zurek makes a sharp prediction (tested in Subsection~\ref{subsec-limitationsnumerics}) that this density will be determined by the rate that one crosses the transition; given a fixed rate, the density of $e$-anyon defects is predicted to remain roughly constant provided that one has spent sufficiently long after the transition to have ``coarsened.''
This density further defines a length scale $L_e = 1/\sqrt{n_e}$ in which our system will ``look'' like it obeys the Gauss law.
This length scale will be the size over which our system will exhibit QSL-like signatures and defines the size of our spin lake.
Moreover, this discussion clarifies that the phrase ``slow relative to $e$-anyons'' implies that one travels at a sufficiently slow rate such that $L_e$ is of appreciable size.
In Subsection~\ref{subsec-limitationsnumerics}, we will numerically test the prediction of Kibble-Zurek that the density of $e$-anyons in the final state is determined by the rate.

Apart from the dynamics of $e$-anyons, after the Kibble-Zurek ``freeze-out regime'', our dynamics will continue to be sudden relative to dynamics of single $m$-anyon excitations.
Similar to the single qutrit case, under the approximation that these dynamics were perfectly sudden, the only dynamics of the system across the transition would be to equilibrate out $e$-anyons.
Assuming that the wavefunction when the system fell out of equilibrium is similar (up to a short-depth unitary) to the initial wavefunction of the sweep, this motivates the ansatz that the final wavefunction is the initial wavefunction with Gauss law violations projected out.
Of course, since the energy scale associated with $m$-anyons is finite, there is a time-scale above which we will start to nucleate $m$-anyon defects above our state.
As such, we can predict that the density of $m$-anyon defects will increase as we sweep for longer times around and past the transition.
This is in contrast to the case of the $e$-anyons where the total time swept is irrelevant as long as the rate is kept constant.
This prediction will also be tested in Subsection~\ref{subsec-limitationsnumerics}.

Given the discussion above, a few remarks are in order.
First and most importantly, the above considerations on the density of $e$ and $m$-anyons imply that it will be impossible to produce a full thermodynamic QSL via our dynamical mechanism--- the divergence of the $e$-anyon timescale and the finiteness of the $m$-anyon timescale implies that it is impossible to be slow with respect to the former and fast with respect to the latter. 
Nevertheless, depending on the local energy scale of the $m$-anyons, as mentioned earlier, it will be possible to respect the time-scale conditions and create a QSL-like state over a length scale of size $L_e$, which precisely defines the quantum spin lake.
Second, we remark that the discussion in the previous paragraphs is quite general; we expect that dynamical sweeps into constrained subspaces with multiple emergent excitations, some of which are fast and others of which are slow, can be used to prepare exotic finite-size orders.
In the examples that we discuss in this paper, these excitations have a nice microscopic description which enables us to make predictions as to the final state of the sweep (i.e. via the projection formula of Eq.~\ref{eq-sweeping-proj}).
Nevertheless, in general, this need not be the case; dynamical sweeps could project out emergent degrees of freedom.

\subsection{Crossing the First Order Phase Transition} \label{subsec-firstorder} 

We now turn to the case where we cross a first order transition during our dynamical sweep.
In the the model of Eq.~\eqref{eq-TCf_2}, this first order transition is between the Higgs phase and the confined phase and occurs when a level crossing occurs between the two phases. 
While such a level crossing leads to sharp and discontinuous change in the nature of the ground state, such a level crossing does not impact the dynamics in the vicinity of the transition.
This is because the Higgs ground state and confined ground state are macroscopically distinct from one another and hence the ground state transition cannot be detected by local dynamics.
Deep enough beyond a first-order transition, we expect to become sensitive to false vacuum decay~\cite{Vidal_False_Vacuum, Coleman_1977, devoto2022false}, but this process is mediated by $m$-anyon dynamics which we have assume to be slow relative to total time of our dynamical sweep.
Hence, our local dynamics will effectively encounter instead the above second order transition and the considerations of the previous section will follow.
We now test this prediction.

\subsection{Numerical Confirmation} \label{subsec-limitationsnumerics}

\begin{figure}
    \centering
    \includegraphics[width = 247 pt]{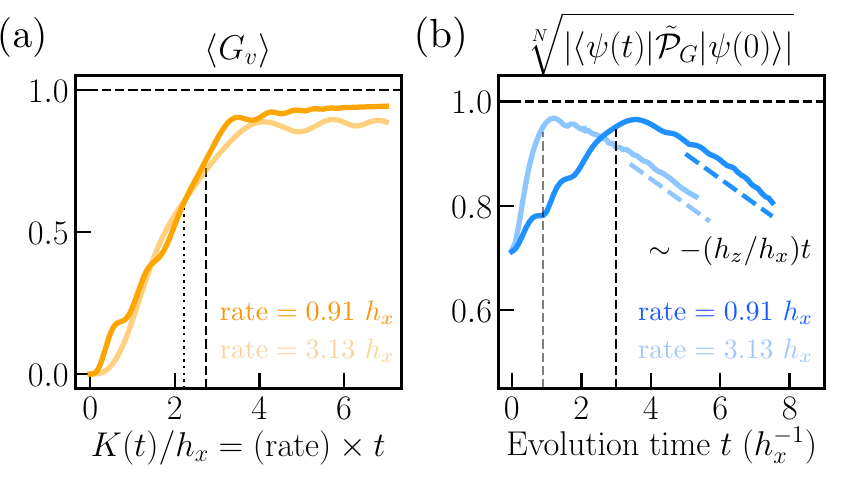}
    \caption{ \textbf{Numerical Confirmation of Rate and Evolution Time Scaling of $e$- and $m$-Anyon Generation.} Based on Kibble-Zurek type arguments, we expect that the density of $e$-anyons in the final state is controlled by the rate with which one traverses the transition.
    In panel (a), we check this expectation numerically by simulating a dynamical sweep which ramps $K$ linearly from $0$ to $7$ at two rates and fixed $h_z/h_x = 0.1$ (red arrow in Fig.~\ref{fig-dTC}(a)). 
    We find that after crossing the transition, the expectation value of $\langle G_v \rangle$,  which measures the Gauss law, is largely insensitive to the final time provided that one allows for the system to sufficiently coarsen after crossing the transition but depends sensitively on the rate, as predicted.
    We remark that, at $h_z = 0.1$, the dynamics strictly crosses a first order phase transition (shown with the dotted line) but is insensitive to its presence (see discussion in main text).
    However, the value of $\langle G_v \rangle$ starts to change dramatically shortly after a putative or `hidden' second order transition (dashed line).
    In panel (b), we check the dynamics of $m$-anyons by plotting the overlap density of our instantaneous wavefunction $\ket{\psi(t)}$ with the projected state as a function of evolution time $t$.
    We find that following the putative transition (shown in the dashed lines) and coarsening, when $e$-anyons are back in equilibrium, $m$-anyons decrease the projected overlap density linearly with evolution time as expected.
    In our numerics, we utilized a bond dimension of $\chi = 256$ and a trotter step size of $dt = 0.0025$. \label{fig:scaling}}
\end{figure}

We now seek to numerically verify the predictions of the last two subsections.
To summarize, our predictions are three-fold.
First, we predicted that across the local dynamics of our system is insensitive to the presence of the first-order transition and instead effectively sees the presence of a second-order transition.
Second, we predicted that the density of $e$-anyons (as detected by $\langle G_v\rangle$) at the end of the sweep is set by the rate $(dK/dt)/h_x$ that one crosses the effective second-order transition as opposed to the total time of the sweep.
Finally, we predicted that the density of $m$-anyons produced during the sweep is determined by the total time spent after the transition as opposed to the rate that we sweep across the transition.

To test these, we start by plotting the expectation value of the Gauss law operator as we cross the transition as a function of location $K(t)/h_x$ along the sweep for sweep done at two different rates [See Fig.~\ref{fig:scaling}(a)].
We first find that the expectation value of the Gauss law operator shows no signature as we cross the point where the ground state undergoes the first-order transition indicating that indeed our system is insensitive to its presence.

Moreover, we find that after crossing the putative location of the effective second order phase transition, the expectation value of the Gauss law operator approximately saturates to a constant value.
This constant value appears set by the rate with the value of $\langle G_v \rangle$ increasing with decreasing (slower) rate.
This is consistent with the predictions of Kibble-Zurek.
We remark that there is a slight increase in the value of $\langle G_v \rangle$ as the value of $K(t)$ increases.
This is due to the fact that as we increase $K$, the emergent Gauss law of the low-energy constrained subspace becomes closer to the bare Gauss law: $G_v$.
This effect similarly occurs within the ground state.

Lastly, we want to confirm whether the total $m$-anyon density is set by the total time that the system evolves passed the transition (as opposed to the rate).
To do so, we plot the value of the overlap of the instantaneous wavefunction $\ket{\psi(t)}$ with the projected wavefunction $\mathcal{N} \cdot \mathcal{P}_{G} \ket{\psi(0)}$ as a function of the evolution time $t$.
We do so for two different rates that we cross the transition.
Since we have confirmed that the rate determines the $e$-anyon density, decrease in the projected overlap with time signals the nucleation of $m$-anyons.
Our prediction would signal that the slope with which the projected overlap decreases should be independent of rate.
Remarkably, we find that this is indeed the case in Fig.~\ref{fig:scaling}(b)!
This is strong evidence in support of our predictions.

\section{Experimental Relevance: Rydberg Atom Ruby Dimer Liquid} \label{sec-Experiment}

\begin{figure}
    \centering
    \includegraphics[width = 247 pt]{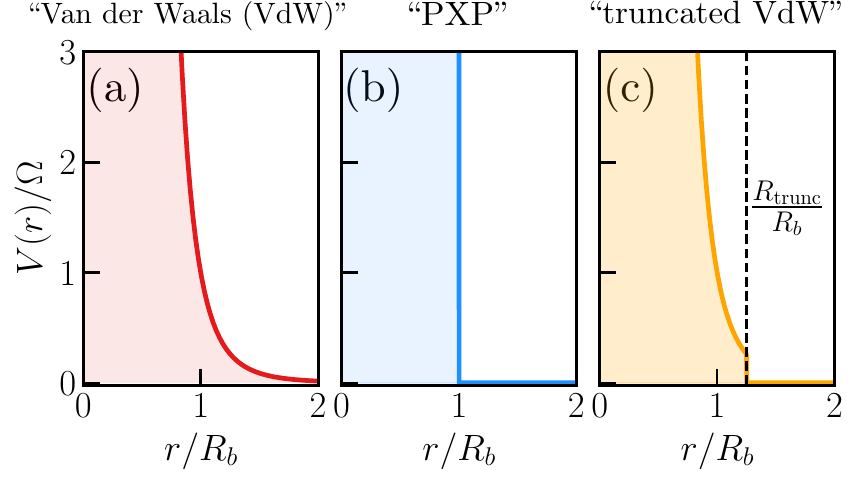}
    \caption{\textbf{Interaction Potentials Utilized to Study Rydberg Atom Systems.} (a) Typically, the potential between Rydberg atoms is treated as a standard van der Waals potential which falls off with distance $r$ as $\sim (R_b/r)^6$ where $R_b$ is a characteristic radius called the blockade radius \cite{Lukin01, Jaksch00}. (b) Since the van der Waals interaction potential is much larger within the blockade radius than outside it, it is typically approximated to be nearly infinite in the blockade radius and zero outside the blockade radius. In this limit, the effective Hamiltonian governing the dynamics of Rydberg atoms is the so-called PXP model of Eq.~\eqref{eq-actualHPXP}. (c) To incorporate the effects of the long-range tails either numerically or analytically, to study a truncated version of the full van der Waals potential. Here, the full $(R_b/r)^6$ of the van der Waals model is taken into account for $r \leq R_{\text{trunc}}$ but the interaction potential is sharply truncated for  $r > R_{\text{trunc}}$. }
    \label{fig-RydbergPotential}
\end{figure}

Thus far we have carefully studied the mechanism for creating a quantum spin lake, first in the qutrit toy model (Section~\ref{sec-Qutrit}) and subsequently in a genuine many-body toric code model (Section~\ref{sec-DTC}), which we then also used to study and display how it does (not) scale (Section~\ref{sec-Summary}). In this section, we present a concrete application of our theory. In particular, we consider how it applies to the Rydberg atom quantum simulator experiment of Ref.~\onlinecite{Semeghini21} based on the proposal by Ref.~\onlinecite{Verresen21}.
Therein, Rubidium-87 atoms are placed at the links of the kagome lattice (equivalently the sites of the ruby lattice):
\vspace{-2mm}
\begin{equation*}
    \begin{tikzpicture}[scale = 1, baseline={([yshift=-.5ex]current bounding box.center)}]
    \foreach \i in {0,...,1}{
        \foreach \j in {0,...,1}{
            \draw[gray] ({\i + \j * 1/2}, {\j * 0.866025404}) -- ({\i + \j * 1/2 + 1/2}, {\j * 0.866025404}) -- ({\i + \j * 1/2 + 1/2*1/2}, {\j * 0.866025404 + 1/2*0.866025404}) -- cycle;
            \draw[gray] ({\i + \j * 1/2}, {\j * 0.866025404}) -- ({\i + \j * 1/2 - 1/2}, {\j * 0.866025404}) -- ({\i + \j * 1/2 - 1/2*1/2}, {\j * 0.866025404 - 1/2*0.866025404}) -- cycle;
            \filldraw  ({\i + \j * 1/2 + 1/4}, {\j * 0.866025404}) circle (1 pt);
            \filldraw  ({\i + \j * 1/2 - 1/4}, {\j * 0.866025404}) circle (1 pt);
            \filldraw  ({\i + \j * 1/2 + 1/8 }, {\j * 0.866025404 + 1/4 * 0.866025404}) circle (1 pt);
            \filldraw  ({\i + \j * 1/2 - 1/8 }, {\j * 0.866025404 - 1/4 * 0.866025404}) circle (1 pt);
            \filldraw  ({\i + \j * 1/2 + 1/8 + 1/4 }, {\j * 0.866025404 + 1/4 * 0.866025404}) circle (1 pt);
            \filldraw  ({\i + \j * 1/2 - 1/8 -1/4}, {\j * 0.866025404 - 1/4 * 0.866025404}) circle (1 pt);
        }
    }
    \draw[red, fill=red, fill opacity=0.1, dashed, thick](1/4,0) circle (0.51);
    \filldraw[red] (1/4, 0) circle (1 pt);
    \draw [-stealth, thick] (1/4,0) -- (1/4 - 0.44, 0.25);
    \draw [stealth-stealth, line width = 0.1 mm] (5/4 + 0.1 + 0.04,-0.1 + 0.01 ) -- (5/4 + 0.25 + 0.025, 0.25 - 0.1 + 0.025 );
    \node at ({-1/4 -1/6}, 0.4) {\normalsize $R_b$};
    \node at (5/4 + 0.25 + 0.2, 0.25 - 0.15 ) {\normalsize $a$};
    \end{tikzpicture}
\end{equation*}
Each atom encodes a qubit (or hardcore boson) using a hyperfine atomic ground state $\ket{\downarrow}$ and Rydberg state $\ket{\uparrow}$ of the atom.
These atoms then interact via the following Hamiltonian~\cite{Fendley04}:
\begin{equation} \label{eq-HPXP}
    H = \frac{\Omega}{2} \sum_{i} X_i - \delta \sum_i n_i + \frac{1}{2}\sum_{i,j} V_{i, j} n_{i} n_{j} 
\end{equation}
where $i, j$ runs over all qubits on the lattice and $n_i = \frac{1}{2}(1 + Z_i)$.
In the experiment, $\Omega$ and $\delta$ corresponds to the Rabi frequency and frequency detuning of the laser that addresses the ground-Rydberg transition and $V_{i, j}$ is the van der Waals interaction potential of two atoms in their Rydberg states which $\approx \Omega \times \left( \frac{R_b}{r_i-r_j} \right)^6$ [See Fig.~\ref{fig-RydbergPotential}(a)], where $R_b$ is called the blockade radius (shown above to be at least $2a$) and $r_i$ is the position of atom $i$.

\subsection{Rydberg Model in Equilibrium} \label{subsec-RydEq}

Before discussing the dynamics in the experiment, we briefly review the equilibrium physics of the Rydberg model, following Ref.~\onlinecite{Verresen21}.
One of its key features is that, since the energy of exciting two atoms within the blockade radius is very large, at low-energies, states satisfy the ``blockade constraint'' wherein two nearby atoms cannot be simultaneous excited.
If we represent the states of our Rydberg qubits with dimers, $\ket{ \begin{tikzpicture}[scale = 0.5, baseline = {([yshift=-.5ex]current bounding box.center)}]
\draw[gray] (0, 0) -- (1, 0);
\node at (0.5, 0) {\normalsize $\uparrow$};
\end{tikzpicture}}= \ket{\frac{}{}\begin{tikzpicture}[scale = 0.5, baseline = {([yshift=-.5ex]current bounding box.center)}]
\draw[gray] (0, 0) -- (1, 0);
\draw[red, fill = red] (0.5,0) ellipse (0.45 and 0.1);
\end{tikzpicture}}$ 
and $\ket{ \begin{tikzpicture}[scale = 0.5, baseline = {([yshift=-.5ex]current bounding box.center)}]
\draw[gray] (0, 0) -- (1, 0);
\node at (0.5, 0) {\normalsize $\downarrow$};
\end{tikzpicture}} = \ket{\frac{}{} \begin{tikzpicture}[scale = 0.5, baseline = {([yshift=-.5ex]current bounding box.center)}]
\draw[gray] (0, 0) -- (1, 0);
\end{tikzpicture}}$, then the blockade constraint, defined such that $R_b$ contains the six nearest neigbors (i.e., $2a < R_b < \sqrt{7} a$), implies that two dimers cannot share the same vertex.

If the detuning $\delta$ in the Eq.~\eqref{eq-HPXP} is large, low-energy states will have as many atoms in their Rydberg states as possible while still obeying the blockade constraint.
As such, in such a regime, the system will behave like a dimer model \cite{Verresen21} characterized by the Gauss law:
\begin{equation}\label{eq-RydbergGauss}
G_v = \begin{tikzpicture}[scale = 1.2, baseline={([yshift=-.5ex]current bounding box.center)}]
\draw[gray] (0,0) -- ({-0.5 * 1/2}, {0.5 * 0.866025404}) -- ({0.5 * 1/2}, {0.5 * 0.866025404}) -- cycle;
\draw[gray] (0,0) -- ({-0.5 * 1/2}, {-0.5 * 0.866025404}) -- ({0.5 * 1/2}, {-0.5 * 0.866025404}) -- cycle;
\draw[orange(ryb), dashed, line width = 0.4 mm] (0,0) circle (7 pt);
\end{tikzpicture} = -1 \quad \quad
\begin{tikzpicture}[scale = 1.2, baseline={([yshift=-.5ex]current bounding box.center)}]
\draw[gray] (0,0) -- ({-1.5*(0.5) * 1/2}, {-(1.5*0.5) * 0.866025404}) -- ({(1.5*0.5) * 1/2}, {-(1.5*0.5) * 0.866025404}) -- cycle;
\draw[orange(ryb), dashed, line width = 0.4 mm] (-1.5*0.3, 1.5*-0.25) -- (1.5*0.3, 1.5*-0.25);
\end{tikzpicture}\ \ =\ \  
\begin{tikzpicture}[scale = 1.2, baseline={([yshift=-.5ex]current bounding box.center)}]
\draw[gray] (0,0) -- ({-1.5*(0.5) * 1/2}, {-(1.5*0.5) * 0.866025404}) -- ({(1.5*0.5) * 1/2}, {-(1.5*0.5) * 0.866025404}) -- cycle;
\node at ({-1.5*(0.5) * 1/4}, {-(1.5*0.5) * 0.866025404/2}) {\normalsize $Z$};
\node at ({-1.5*(0.5) * 1/4 + 0.75/2}, {-(1.5*0.5) * 0.866025404/2}) {\normalsize $Z$};
\end{tikzpicture}
\end{equation}
where we have defined the 't Hooft loop operator shown in orange and $v$ refers to a particular vertex of the kagome lattice.
The presence of this local Gauss law implies that, at low-energies, the Rydberg model is an emergent $\mathbb{Z}_2$ gauge theory.
The deconfined phase of this gauge theory can be characterized by its fixed-point wavefunction \cite{RK,Moessner_2001,Misguich02}:
\begin{equation} \label{eq-RydRVB}
    \ket{\text{RVB}} = \ket{   \includegraphics[scale = 0.3, valign = c]{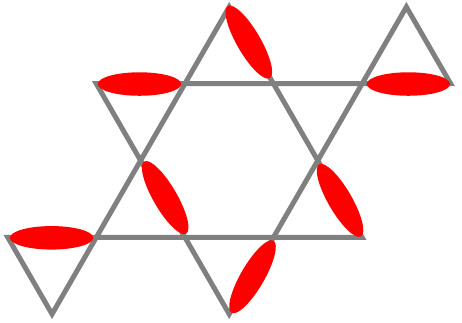} } + \ket{   \includegraphics[scale = 0.3, valign = c]{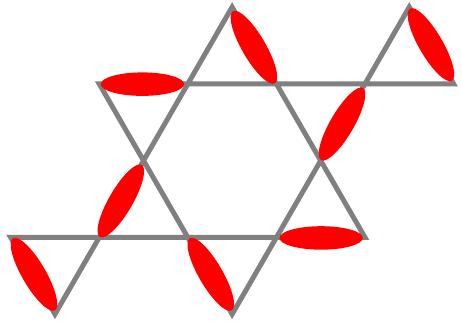} } + \ket{   \includegraphics[scale = 0.3, valign = c]{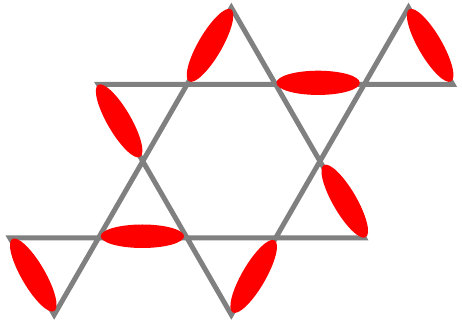} } + \cdots  
\end{equation}
which is the equal-weight, equal-phase superposition of all full-packing dimer configurations (dimer configurations that have no untouched vertices) and is the dimer analogue of Anderson's resonating valence bond (RVB) state of singlets \cite{ANDERSON}. Such a superposition of dimer states represents a $\mathbb Z_2$ spin liquid owing to the kagome being a nonbipartite lattice \cite{Read91,Sachdev92,Sachdev_Triangle,Moessner01,Misguich02} (for approaches to emergent dimer models on other lattices see Refs.~\onlinecite{Glaetzle_2014,Samajdar_2021,SamajdarJoshi22,Yan22}).
The above state is the unique state that is stabilized by the 't Hooft loop of Eq.~\eqref{eq-RydbergGauss} and the following Wilson loop operator \cite{unifying}:
\begin{equation} \label{eq-RydbergWilson}
W_{\varhexagon} = \begin{tikzpicture}[scale = 1, baseline={([yshift=-.5ex]current bounding box.center)}]
\draw[gray] (0,0) -- ({-0.5 * 1/2}, {0.5 * 0.866025404}) -- ({0}, {0.866025404}) -- ({0.5}, {0.866025404}) -- ({ 0.5 +0.5 * 1/2}, {0.5 * 0.866025404}) -- (0.5, 0) -- cycle;
\draw[dodgerblue,decorate, decoration={snake, segment length=1.5mm}, line width = .35mm] (0,0) -- ({-0.5 * 1/2}, {0.5 * 0.866025404});
\draw[dodgerblue,decorate, decoration={snake, segment length=1.5mm}, line width = .35mm] ({-0.5 * 1/2}, {0.5 * 0.866025404}) -- ({0}, {0.866025404});
\draw[dodgerblue,decorate, decoration={snake, segment length=1.5mm}, line width = .35mm] ({0}, {0.866025404}) --  ({0.5}, {0.866025404});
\draw[dodgerblue,decorate, decoration={snake, segment length=1.5mm}, line width = .35mm] ({0.5}, {0.866025404}) -- ({ 0.5 +0.5 * 1/2}, {0.5 * 0.866025404});
\draw[dodgerblue,decorate, decoration={snake, segment length=1.5mm}, line width = .35mm] ({ 0.5 +0.5 * 1/2}, {0.5 * 0.866025404}) -- (0.5, 0);
\draw[dodgerblue,decorate, decoration={snake, segment length=1.5mm}, line width = .35mm] (0.5, 0) -- (0,0);
\end{tikzpicture} = +1
\quad \quad
\begin{tikzpicture}[scale = 1.2, baseline={([yshift=-.5ex]current bounding box.center)}]
\draw[gray] (0,0) -- ({-1.5*(0.5) * 1/2}, {-(1.5*0.5) * 0.866025404}) -- ({(1.5*0.5) * 1/2}, {-(1.5*0.5) * 0.866025404}) -- cycle;
\draw[dodgerblue,decorate, decoration={snake, segment length=1.5mm}, line width = .35mm] ({-1.5*(0.5) * 1/2}, {-(1.5*0.5) * 0.866025404}) -- ({(1.5*0.5) * 1/2}, {-(1.5*0.5) * 0.866025404});
\end{tikzpicture}  = \begin{cases}
\begin{tikzpicture}[scale = 1.2, baseline={([yshift=-.5ex]current bounding box.center)}]
\draw[gray] (0,0) -- ({-.7*(0.5) * 1/2}, {-(.7*0.5) * 0.866025404}) -- ({(.7*0.5) * 1/2}, {-(.7*0.5) * 0.866025404}) -- cycle;
\draw[red, fill = red, rotate around={60:({-.35*(0.5) * 1/2}, {-(.35*0.5) * 0.866025404})}] ({-.35*(0.5) * 1/2}, {-(.35*0.5) * 0.866025404}) ellipse (0.175 and 0.03);
\end{tikzpicture} \leftrightarrow
\begin{tikzpicture}[scale = 1.2, baseline={([yshift=-.5ex]current bounding box.center)}]
\draw[gray] (0,0) -- ({-.7*(0.5) * 1/2}, {-(.7*0.5) * 0.866025404}) -- ({(.7*0.5) * 1/2}, {-(.7*0.5) * 0.866025404}) -- cycle;
\draw[red, fill = red, rotate around={-60:({(.35*0.5) * 1/2}, {-(.35*0.5) * 0.866025404})}]  ({(.35*0.5) * 1/2}, {-(.35*0.5) * 0.866025404})  ellipse (0.175 and 0.03);
\end{tikzpicture}
\\
\begin{tikzpicture}[scale = 1.2, baseline={([yshift=-.5ex]current bounding box.center)}]
\draw[gray] (0,0) -- ({-.7*(0.5) * 1/2}, {-(.7*0.5) * 0.866025404}) -- ({(.7*0.5) * 1/2}, {-(.7*0.5) * 0.866025404}) -- cycle;
\end{tikzpicture}
\leftrightarrow
\begin{tikzpicture}[scale = 1.2, baseline={([yshift=-.5ex]current bounding box.center)}]
\draw[gray] (0,0) -- ({-.7*(0.5) * 1/2}, {-(.7*0.5) * 0.866025404}) -- ({(.7*0.5) * 1/2}, {-(.7*0.5) * 0.866025404}) -- cycle;
\draw[red, fill = red] ({0}, {-(.7*0.5) * 0.866025404}) ellipse (0.175 and 0.03); 
\end{tikzpicture}
\end{cases}
\end{equation}
where $\varhexagon$ refers to a particular hexagon on the kagome lattice.
Similar to the case of the toric code, we can define $e$-anyons above this state to be violations of the Gauss law of Eq.~\eqref{eq-RydbergGauss} and $m$-anyons to be violations of the equal phase condition of Eq.~\eqref{eq-RydRVB} (equivalently violations of the Wilson loop stabilizer Eq.~\eqref{eq-RydbergWilson}).

\subsubsection{Phase Diagram in the Absence of Long-Range Tails} \label{subsec-PXP_PD}

Since the effect of the blockade is largest effect of the interactions $V$, as a first approximation the interaction can be to replace $V(r \leq R_b) = + \infty$ and $V(r > R_b) = 0$, neglecting the effect of long-range $1/R^6$ tails [See Fig.~\ref{fig-RydbergPotential}(b)].
In this limit, the Rydberg model is traditionally called a ``PXP model'' \cite{Fendley04,Bernien17,Turner18a,Turner18b,Choi19,Lin_2019,Shiraishi19,Mark20,Moudgalya20,Iadecola20,Surace20,Lin20,Michailidis20}.
This is because the effective Hamiltonian in this limit is simply the Pauli-$X$ operator projected into the blockade constraint satisfying subspace along with the detuning term:
\begin{equation} \label{eq-actualHPXP}
    H_{\text{PXP}} = \frac{\Omega}{2} \sum_i \mathcal{P}_{\text{blockade}} X_i \mathcal{P}_{\text{blockade}} - \delta \sum_i n_i
\end{equation}
where $\mathcal{P}_{\text{blockade}} = \prod_{i, j: |r_i - r_j| \leq R_b} \left( 1 - n_i n_j \right)$ removes configurations that violate the blockade constraint.

The ground state phase diagram of this PXP model on the ruby lattice was found in Ref.~\onlinecite{Verresen21} to be:
\begin{equation*}
\hspace{7 mm}\begin{tikzpicture}[scale = 1, baseline={([yshift=-.5ex]current bounding box.center)}]
\draw[-stealth, line width=0.4mm] (0, 0) -- (5, 0);
\node at (5.5,0) {\normalsize $\delta/\Omega$};
\draw[line width=0.4mm] (1.2, 0) -- (1.2, 0.2);
\node at (0.6,0.3) {\small Higgs};
\draw[line width=0.4mm] (3.3, 0) -- (3.3, 0.2);
\node at (4.2,0.3) {\small Confined};
\node at (2.3,0.3) {\small $\mathbb{Z}_2$ QSL};
\node at (1.2, -0.2) {\small $\approx 1.4$};
\node at (3.3, -0.2) {\small $\approx 2.1$};
\end{tikzpicture}
\end{equation*}
which contains three phases.
In particular, when $\delta/\Omega$ is small or negative, $e$-anyon excitations condense, yielding the Higgs phase which is adiabatically connected to the state with no dimers (equivalently no excited Rydberg atoms).
Moreover, in an intermediate regime of $\delta/\Omega$ we get the deconfined phase which shares the properties of the RVB state of Eq.~\eqref{eq-RydRVB}.
Finally, when $\delta/\Omega$ is large, we can treat the effect of $\Omega$, which generates violations of the Gauss law (analogous to $h_x$ term in Eq.~\eqref{eq-KitaevTC-in-F}), perturbatively.
This will generate a resonance between dimer states given by \cite{Verresen21}:
\begin{equation}\label{eq-Rydberg-Resonances}
H_{\text{eff}} = -  \frac{3\Omega^6}{32\delta^5} \sum_{\varhexagon}
\ket{\begin{tikzpicture}[scale = 0.7, baseline={([yshift=-.5ex]current bounding box.center)}]
\draw[gray] (0,0) -- ({-0.5 * 1/2}, {0.5 * 0.866025404}) -- ({0}, {0.866025404}) -- ({0.5}, {0.866025404}) -- ({ 0.5 +0.5 * 1/2}, {0.5 * 0.866025404}) -- (0.5, 0) -- cycle;
\draw[red, fill = red] (0.25,0) ellipse (0.25 and 0.05);
\draw[red, fill = red, rotate around={60:({-0.25 * 1/2}, {0.75 * 0.866025404})}] ({-0.25 * 1/2}, {0.75 * 0.866025404}) ellipse (0.25 and 0.05);
\draw[red, fill = red, rotate around={-60:({1.25 * 1/2}, {0.75 * 0.866025404})}] ({1.25 * 1/2}, {0.75 * 0.866025404}) ellipse (0.25 and 0.05);
\end{tikzpicture}}
\bra{\begin{tikzpicture}[scale = 0.7, baseline={([yshift=-.5ex]current bounding box.center)}]
\draw[gray] (0,0) -- ({-0.5 * 1/2}, {0.5 * 0.866025404}) -- ({0}, {0.866025404}) -- ({0.5}, {0.866025404}) -- ({ 0.5 +0.5 * 1/2}, {0.5 * 0.866025404}) -- (0.5, 0) -- cycle;
\draw[red, fill = red] (0.25,{0.866025404}) ellipse (0.25 and 0.05);
\draw[red, fill = red, rotate around={-60:({-0.25 * 1/2}, {0.25 * 0.866025404})}] ({-0.25 * 1/2}, {0.25 * 0.866025404}) ellipse (0.25 and 0.05);
\draw[red, fill = red, rotate around={60:({1.25 * 1/2}, {0.25 * 0.866025404})}] ({1.25 * 1/2}, {0.25 * 0.866025404}) ellipse (0.25 and 0.05);
\end{tikzpicture}}
\end{equation}
This resonance alone (as well as an eigth order perturbatively generated term) yields a confined ``valence bond solid'' phase \cite{Nikolic03}.
Such a phase is a condensate of $m$-anyons and the ground state wavefunction corresponds to a localized superposition of a subset of full-packing dimer configurations.

\subsubsection{Effect of Long-Range Tails} \label{subsubsec-EffectofLRtails}

Having reviewed the basic physics of the Rydberg PXP model in the absence of long-range tails, we now analyze the effect of the long-range tails beyond the blockade radius.
Since the long-range density-density interactions such as $n_i n_j$ commute with the Gauss law of Eq.~\eqref{eq-RydbergGauss} but fail to commute with the Wilson loop of Eq.~\eqref{eq-RydbergWilson}, they contribute to the energy-scale associated with the creation of $m$-anyons.
Concretely, in the absence of long-range tails and local resonances, the Rydberg blockade treats all dimer configurations on equal footing, inducing no splittings between such states; our goal is to estimate how much these tails lead to energy density splittings between dimer configurations.
One might expect that the leading order contribution of these tails is due to the first interaction outside of the blockade radius, namely at $r = \sqrt{7} a$.
However, we demonstrate a new result that instead this contribution is due to the \emph{second} interaction outside of the blockade and occurs with a much smaller coefficient than one would expect; roughly speaking, this decreases the apparent effect by at least an order of magnitude.

To estimate effect of the long-range tails, we consider the effect of the six leading contributions of the Rydberg interaction outside of the blockade radius: 
\begin{equation} \label{eq-RydbergDistances}
    \begin{tikzpicture}[scale = 1, baseline={([yshift=-.5ex]current bounding box.center)}]
    \foreach \i in {0,...,1}{
        \foreach \j in {0,...,1}{
            \draw[gray] ({\i + \j * 1/2}, {\j * 0.866025404}) -- ({\i + \j * 1/2 + 1/2}, {\j * 0.866025404}) -- ({\i + \j * 1/2 + 1/2*1/2}, {\j * 0.866025404 + 1/2*0.866025404}) -- cycle;
            \draw[gray] ({\i + \j * 1/2}, {\j * 0.866025404}) -- ({\i + \j * 1/2 - 1/2}, {\j * 0.866025404}) -- ({\i + \j * 1/2 - 1/2*1/2}, {\j * 0.866025404 - 1/2*0.866025404}) -- cycle;
            \filldraw  ({\i + \j * 1/2 + 1/4}, {\j * 0.866025404}) circle (1 pt);
            \filldraw  ({\i + \j * 1/2 - 1/4}, {\j * 0.866025404}) circle (1 pt);
            \filldraw  ({\i + \j * 1/2 + 1/8 }, {\j * 0.866025404 + 1/4 * 0.866025404}) circle (1 pt);
            \filldraw  ({\i + \j * 1/2 - 1/8 }, {\j * 0.866025404 - 1/4 * 0.866025404}) circle (1 pt);
            \filldraw  ({\i + \j * 1/2 + 1/8 + 1/4 }, {\j * 0.866025404 + 1/4 * 0.866025404}) circle (1 pt);
            \filldraw  ({\i + \j * 1/2 - 1/8 -1/4}, {\j * 0.866025404 - 1/4 * 0.866025404}) circle (1 pt);
        }
    }
    \draw [-stealth, thick, dodgerblue] (1/4,0) -- (1/2- 1/8 -1/4, 1* 0.866025404 - 1/4 * 0.866025404);
    \draw [-stealth, thick, red, dashed] (1/4,0) -- (1/2 - 1/4, 1* 0.866025404);
    \draw [-stealth, thick, red] (1/4 + 1/2,0) -- (1/2 + 1/2 - 1/4, 1* 0.866025404);
    \draw [-stealth, thick, forest] (1/4 + 1/2,0) -- (1/2 + 1/2 - 1/4 + 1/8, 1* 0.866025404 + 0.866025404/4);
    \draw [-stealth, thick, orange(ryb)] (1/4 + 1/8,0.866025404/4) -- (1/2 + 1/4, 1* 0.866025404);
    \draw [-stealth, thick, blue] (1/4 + 1/8, 0.866025404/4) -- (1/2 + 1/8, 5/4* 0.866025404 );
    \draw [-stealth, thick, magenta] (1/4 + 1/8 + 1, 0.866025404/4) -- (1/2 + 3/8 + 1, 5/4* 0.866025404 );
    \end{tikzpicture}
    \quad
    \begin{tikzpicture}[scale = 0.4, baseline={([yshift=-.5ex]current bounding box.center)}]
    \node at (0, 2.5) {\footnotesize $\textcolor{dodgerblue}{R_1 = \sqrt{7} a}$};
    \node at (0, 1.5) {\footnotesize $\textcolor{orange(ryb)}{R_2 = 3 a}$};
    \node at (0, 0.5) {\footnotesize $\textcolor{red}{R_3 = 2\sqrt{3} a}$};
    \node at (0, -0.5) {\footnotesize $\textcolor{blue}{R_4 = \sqrt{13} a}$};
    \node at (0, -1.5) {\footnotesize $\textcolor{magenta}{R_5 = 4 a}$};
    \node at (0, -2.5) {\footnotesize $\textcolor{forest}{R_6 = \sqrt{19} a}$};
    \end{tikzpicture}
    \quad
    \begin{tikzpicture}[scale = 1, baseline={([yshift=-.5ex]current bounding box.center)}]
    \foreach \i in {0,...,1}{
        \foreach \j in {0,...,1}{
            \draw[gray] ({\i + \j * 1/2}, {\j * 0.866025404}) -- ({\i + \j * 1/2 + 1/2}, {\j * 0.866025404}) -- ({\i + \j * 1/2 + 1/2*1/2}, {\j * 0.866025404 + 1/2*0.866025404}) -- cycle;
            \draw[gray] ({\i + \j * 1/2}, {\j * 0.866025404}) -- ({\i + \j * 1/2 - 1/2}, {\j * 0.866025404}) -- ({\i + \j * 1/2 - 1/2*1/2}, {\j * 0.866025404 - 1/2*0.866025404}) -- cycle;
        }
    }
    \node at (1/4,0) {\tiny $1$};
    \node at  (1/4 + 1/8, 0.866025404/4) {\tiny $2$};
    \node at  (1/4 - 1/8, 0.866025404/4) {\tiny $3$};
    \node at (-3/8 + 1/2, 0.866025404*3/4 ) {\tiny $4$};
    \node at (-1/8 + 1/2, 0.866025404*3/4 ) {\tiny $5$};
    \node at (-1/4 + 1/2, 0.866025404 ) {\tiny $6$};
    \node at (1/4 + 1/2, 0.866025404 ) {\tiny $7$};
    \node at (1/8 + 1/2, 5/4*0.866025404 ) {\tiny $9$};
    \node at (3/8 + 1/2, 5/4*0.866025404 ) {\tiny $8$};
    \node at (5/8 - 1/16, -1/4*0.866025404 ) {\tiny $10$};
    \node at (7/8 + 1/16, -1/4*0.866025404 ) {\tiny $11$};
    \node at (3/4, 0) {\tiny $12$};
    \node at (1 + 1/4,0) {\tiny $13$};
    \node at  (1 + 1/16 + 1/4 + 1/8, 0.866025404/4) {\tiny $14$};
    \node at  (1 - 1/16 + 1/4 - 1/8, 0.866025404/4) {\tiny $15$};
    \node at (1 - 1/16 -3/8 + 1/2, 0.866025404*3/4 ) {\tiny $16$};
    \node at (1 + 1/16 -1/8 + 1/2, 0.866025404*3/4 ) {\tiny $17$};
    \node at (1 -1/4 + 1/2, 0.866025404 ) {\tiny $18$};
    \end{tikzpicture}
\end{equation}
where the colors refer to different distance couplings \footnote{We remark that the distances $R_n$ are the square roots of the so-called ``Loeschian numbers.''} and we will utilize the numeric labeling of the sites on the right in the following discussion.
For convenience, we will denote the distance-$R_n$ coupling as $V_n = \Omega (R_b/R_n)^6 \sum_{|r_i - r_j| = R_n} n_i n_j$.

Naively, the leading effect of the long-range tails will be due to the distance $R_1$ coupling $V_1$.
For practical experimental values of $R_b$ and $\Omega$, this coupling can be quite large (See Subsubsection~\ref{subsubsec-NumericalEstimates} for more details) which would suggest that the $m$-anyons would proliferate and strongly confine the QSL.
However, one can prove that such a coupling must be constant across all full-packing dimer configurations (i.e. $V_1 \ket{\psi} = c\ket{\psi}$ if $\ket{\psi}$ is a full-packing dimer configuration, where $c$ does not depend on $\ket{\psi}$).
To see that this is the case, we remark that any full-packing dimer configuration must obey the Gauss law and as such $\sum_{\alpha \in v} n_{\alpha} = 1$ (where $\alpha$ label sites neighboring vertex $v$).
Using this fact, the distance-$R_1$ coupling, $V_1$, acts on full-packing dimer configurations $\ket{\psi}$ as:
\begin{align}
n_1(n_4 + n_5) \ket{\psi} &= n_1 (1 - n_2 - n_3) \ket{\psi} \nonumber\\
&= [n_1 - n_1 (n_2 + n_3)]\ket{\psi} = n_1\ket{\psi}
\end{align}
where in the first line we used the Gauss law and in the second line we noted that $n_1 (n_2 + n_3) \ket{\psi} = 0$ by the blockade constraint.
As such, the effect of the distance $R_1$ coupling on $\ket{\psi}$ is $V_1 \ket{\psi} = \Omega (R_b/R_1)^6\sum_i n_i \ket{\psi}$ and hence it simply renormalizes the detuning of the model.
Since the number of dimers in any full-packing dimer configurations is the same, $V_1 \ket{\psi} = c \ket{\psi}$ as claimed and hence $V_1$ does not split dimer configurations (i.e. does not contribute to the $m$-anyon energy scale).

As a consequence of the above, the leading order effect of the long-range tail is due to the distance-$R_2$ coupling $V_2$.
Once again, although a naive estimate of the energy density splitting due to these terms is $\Omega (R_b/R_2)^6$, this turns out to not be the case.
To see why, first note that this coupling pairs atoms that are within a hexagon: 
\begin{equation} \label{eq-Hexagon1}
   \begin{tikzpicture}[scale = 1, baseline={([yshift=-.5ex]current bounding box.center)}]
    \foreach \i in {0,...,1}{
        \foreach \j in {0,...,1}{
            \draw[gray] ({\i + \j * 1/2}, {\j * 0.866025404}) -- ({\i + \j * 1/2 + 1/2}, {\j * 0.866025404}) -- ({\i + \j * 1/2 + 1/2*1/2}, {\j * 0.866025404 + 1/2*0.866025404}) -- cycle;
            \draw[gray] ({\i + \j * 1/2}, {\j * 0.866025404}) -- ({\i + \j * 1/2 - 1/2}, {\j * 0.866025404}) -- ({\i + \j * 1/2 - 1/2*1/2}, {\j * 0.866025404 - 1/2*0.866025404}) -- cycle;
        }
    }  
    \draw [stealth-stealth, thick, orange(ryb)] (1/4 + 1/8,0.866025404/4) -- (1/2 + 1/4, 1* 0.866025404);
    \draw [stealth-stealth, thick, orange(ryb)] (3/4 + 1/4 + 1/8,0.866025404/4) -- (1/2 + 1/4, 1* 0.866025404);
    \draw [stealth-stealth, thick, orange(ryb)] (1/4 + 1/8,0.866025404/4) -- (3/4 + 1/4 + 1/8,0.866025404/4);
    \draw [stealth-stealth, thick, orange(ryb)] (-1/8 + 1/2, 0.866025404*3/4 ) -- (3/4, 0);
    \draw [stealth-stealth, thick, orange(ryb)] (-1/8 + 1/2 + 3/4, 0.866025404*3/4 ) -- (3/4, 0);
    \draw [stealth-stealth, thick, orange(ryb)](-1/8 + 1/2 + 3/4, 0.866025404*3/4 )-- (-1/8 + 1/2, 0.866025404*3/4 );
    \node at  (1/4 + 1/8, 0.866025404/4) {\tiny $2$};
    \node at (1/4 + 1/2, 0.866025404 ) {\tiny $7$};
    \node at  (1 - 1/16 + 1/4 - 1/8, 0.866025404/4) {\tiny $15$};
    \node at (-1/8 + 1/2, 0.866025404*3/4 ) {\tiny $5$};
    \node at (3/4, 0) {\tiny $12$};
    \node at (1 - 1/16 -3/8 + 1/2, 0.866025404*3/4 ) {\tiny $16$};
    \end{tikzpicture}
\end{equation}
As such, $V_2$ can be written as the sum of terms localized to hexagons.
By exploiting the Gauss law once again, we can rewrite the action of one of these terms on a full-packing dimer configuration $\ket{\psi}$ as:
\begin{align}
    n_2 n_7 \ket{\psi} = \frac{1}{2} \left[\left(n_2 - n_2 n_5 - n_2 n_6 - n_2 n_9  \right) \right. \\
    \left.  + \left(n_7 - n_5 n_7 - n_4 n_7 - n_3 n_7  \right) \right]\ket{\psi}
\end{align}
Note that the first terms in both parenthesis will renormalize the detuning.
The second terms in both parenthesis will be zero when acting on any full-packing dimer configuration due to the blockade constraint.
The third terms look like a distance-$R_1$ density-density interaction and as such, when we sum over hexagons, these terms will just renormalize the detuning as per the discussion in the previous paragraph.
Therefore, the only remaining non-trivial term will be the fourth terms which look like distance $R_4$ density density interactions.
Put succinctly, we have found that $n_2 n_7 = \frac{1}{2} \left(-n_2 n_9 - n_3 n_7 + \cdots \right)$ where the ``$\cdots$'' indicates terms that renormalize the detuning.
This fact implies that the distance-$R_4$ couplings will partially cancel out the distance-$R_2$ couplings as: 
\begin{align}\label{eq-numericalestimate_prelim}
&\Omega R_b^6 \left[ \frac{n_2 n_7}{R_2^6} + \frac{ n_2 n_9}{R_4^6} +\frac{ n_3 n_7}{R_4^6} \right] \ket{\psi} \nonumber \\
&= \Omega R_b^6 \left[ \frac{1}{R_2^6} - \frac{2}{R_4^6} \right] n_2 n_7 \ket{\psi} = \alpha \Omega\left(\frac{R_b}{a}\right)^6 n_2 n_7 \ket{\psi}
\end{align}
where $\alpha
\approx 5 \cdot 10^{-4}$---which is reduced from the naive estimate by a factor of $3$.
As a remark, since there are two distance $R_4$-couplings per distance $R_2$-coupling, the above implies that the effect of $V_4$ has been completely canceled.

Finally, we can show that the distance $R_3$-couplings $V_3$ also serve to help cancel the effects of the $R_2$-coupling.
To see why, first note that there are two types of distance $R_3$-couplings.
The first is shown with a dashed red arrow in Eq.~\eqref{eq-RydbergDistances} and occurs between diametric ends of triangle pairs.
This coupling is identically zero on the space of full-packing dimer configurations.
The second type couples atoms at diametrically opposite ends of the hexagons:
\begin{equation}
   \begin{tikzpicture}[scale = 1, baseline={([yshift=-.5ex]current bounding box.center)}]
    \foreach \i in {0,...,1}{
        \foreach \j in {0,...,1}{
            \draw[gray] ({\i + \j * 1/2}, {\j * 0.866025404}) -- ({\i + \j * 1/2 + 1/2}, {\j * 0.866025404}) -- ({\i + \j * 1/2 + 1/2*1/2}, {\j * 0.866025404 + 1/2*0.866025404}) -- cycle;
            \draw[gray] ({\i + \j * 1/2}, {\j * 0.866025404}) -- ({\i + \j * 1/2 - 1/2}, {\j * 0.866025404}) -- ({\i + \j * 1/2 - 1/2*1/2}, {\j * 0.866025404 - 1/2*0.866025404}) -- cycle;
        }
    }  
    \draw[stealth-stealth, thick, red] (1/4 + 1/2, 0.866025404 ) --  (3/4, 0);
    \draw[stealth-stealth, thick, red] (-1/8 + 1/2, 0.866025404*3/4 ) --  (1 + 1/4 - 1/8, 0.866025404/4);
    \draw[stealth-stealth, thick, red] (1/4 + 1/8, 0.866025404/4) --  (1 -3/8 + 1/2, 0.866025404*3/4 );
    \node at  (1/4 + 1/8, 0.866025404/4) {\tiny $2$};
    \node at (1/4 + 1/2, 0.866025404 ) {\tiny $7$};
    \node at  (1 - 1/16 + 1/4 - 1/8, 0.866025404/4) {\tiny $15$};
    \node at (-1/8 + 1/2, 0.866025404*3/4 ) {\tiny $5$};
    \node at (3/4, 0) {\tiny $12$};
    \node at (1 - 1/16 -3/8 + 1/2, 0.866025404*3/4 ) {\tiny $16$};
    \end{tikzpicture}
\end{equation}
As such, $V_3$ can also be written as the sum of terms localized to the hexagons on the ruby lattice.
Using the Gauss law again, one of these terms can be rewritten as:
\begin{align} \label{eq-n2n16}
n_2 n_{16} \ket{\psi} &= \frac{1}{4} \left[ n_2 (2 - n_7 - n_8 - n_{18} - n_{17} - n_{15} - n_{14}) \right. \nonumber \\ 
&+ \left.(2 - n_5 - n_4 - n_3 - n_1 - n_{12} - n_{10}) n_{16}\right] \ket{\psi}
\end{align}
where the terms in parenthesis lead to renormalization of the detuning as well as a distance-$R_2$, distance-$R_5$, distance-$R_6$, distance-$R_6$, distance-$R_2$, and distance-$R_5$ coupling in that order.
First, note that for every distance-$R_3$ coupling, there are exactly four distance-$R_6$ couplings.
Consequently, the above will eliminate the distance-$R_6$ coupling, $V_6$, and we can discard the $n_i n_j$ terms in Eq.~\eqref{eq-n2n16} with $|r_i - r_j| = R_6$ due to near perfect cancellation with $V_6$: $1/4 (a/R_3)^6 - (a/R_6)^6 \approx -10^{-6}$.
Next, the $n_i n_j$ terms in Eq.~\eqref{eq-n2n16} with $|r_i - r_j| = R_2$ will help cancel the magnitude of the distance-$R_2$ coupling, $V_2$ further.
In particular, the coefficient $\alpha$ in Eq.~\eqref{eq-numericalestimate_prelim} will be lowered to $\alpha \to \beta =  \alpha - \frac{1}{4} \left(\frac{a}{R_3}\right)^6
\approx 3 \cdot 10^{-4} $.
Finally, terms such as $n_2 n_8$ can be destructively interfered with the distance-$R_5$ couplings, $V_5$, and will appear with magnitude $(a/R_5)^6 - 1/4 (a/R_3)^6  \approx   10^{-4} \approx \frac{1}{2} (a/R_5)^6$.
Note that unlike the case with $V_6$ and $V_4$, not all terms in $V_5$ are cancelled by using $V_3$. 
In particular, terms that coupling atoms within a ``line'' of the ruby lattice such as: 
\begin{equation}
    \begin{tikzpicture}[scale = 0.8, baseline={([yshift=-.5ex]current bounding box.center)}]
    \foreach \i in {0,...,0}{
        \foreach \j in {0,...,1}{
            \draw[gray] ({\i + \j * 1/2}, {\j * 0.866025404}) -- ({\i + \j * 1/2 + 1/2}, {\j * 0.866025404}) -- ({\i + \j * 1/2 + 1/2*1/2}, {\j * 0.866025404 + 1/2*0.866025404}) -- cycle;
            \draw[gray] ({\i + \j * 1/2}, {\j * 0.866025404}) -- ({\i + \j * 1/2 - 1/2}, {\j * 0.866025404}) -- ({\i + \j * 1/2 - 1/2*1/2}, {\j * 0.866025404 - 1/2*0.866025404}) -- cycle;
            \filldraw  ({\i + \j * 1/2 + 1/4}, {\j * 0.866025404}) circle (1 pt);
            \filldraw  ({\i + \j * 1/2 - 1/4}, {\j * 0.866025404}) circle (1 pt);
            \filldraw  ({\i + \j * 1/2 + 1/8 }, {\j * 0.866025404 + 1/4 * 0.866025404}) circle (1 pt);
            \filldraw  ({\i + \j * 1/2 - 1/8 }, {\j * 0.866025404 - 1/4 * 0.866025404}) circle (1 pt);
            \filldraw  ({\i + \j * 1/2 + 1/8 + 1/4 }, {\j * 0.866025404 + 1/4 * 0.866025404}) circle (1 pt);
            \filldraw  ({\i + \j * 1/2 - 1/8 -1/4}, {\j * 0.866025404 - 1/4 * 0.866025404}) circle (1 pt);
        }
    }
    \draw [-stealth, thick, magenta] (1/8, 0.866025404/4) -- (1/4 + 3/8, 5/4* 0.866025404 );
    \end{tikzpicture}
\end{equation}
are not canceled out.
Hence, the remaining Hamiltonian projected into the space of full-packing dimer configurations will be:
\begin{equation} \label{eq-projected_ham}
H_{LR} \approx \Omega \left( \frac{R_b}{a} \right)^{\hspace{-1mm} 6} \left[ \beta \hspace{-3mm} \sum_{|r_i - r_j| = R_2 } n_i n_j + \hspace{-3mm} \sum_{|r_i - r_j| = R_5} \frac{ \gamma_{i, j} n_i n_j}{(R_5/a)^6} \right]
\end{equation}
where $\gamma_{i, j}$ equals $1$ or $1/2$ depending on whether the term was cancelled out by $V_3$ or not.

We will estimate the $m$-anyon energy scale by summing an estimate for the maximum possible energy density (per qubit) from the first and second terms independently.
Since the $m$-anyon energy scale corresponds to energy density splittings between dimer configurations and both terms in Eq.~\eqref{eq-projected_ham} are positive semi-definite, this will provide an upper bound on these splittings.

Note that the first term couples qubits in the manner illustrated in Eq.~\eqref{eq-Hexagon1}.
First and foremost, the eigenstate with maximum eigenvalue under this $\sum_{|r_i - r_j| = R_2} n_i n_j$ is a valence bond solid configuration on the kagome lattice analyzed in Ref.~\onlinecite{Nikolic03}.
Such a configuration has a twelve hexagon unit cell with the value of $\sum_{|r_i - r_j| = R_2} n_i n_j$ on the unit cell being six (corresponding to two ``perfect'' hexagons).
As such, the maximum energy density per hexagon of the first term will be $\Omega \left( \frac{R_b}{a} \right)^6 \frac{\beta}{2}$ which per qubit is $\Omega \left( \frac{R_b}{a} \right)^6 \frac{\beta}{12}$.

For the second term, it is harder to precisely determine the maximum energy density per qubit.
To gain an estimate for the scale of this term, we note that the distance-$R_5$ coupling pairs qubits that are far relative to a blockade radius $2a <R_b < \sqrt{7} a$, one might expect that the maximum eigenvalue of  $n_i n_j$  (with $|r_i - r_j| = R_5$) will on average be close to the uncorrelated value for a dimer configuration of $\langle n_i \rangle \langle n_j \rangle  = 1/16$.
Hence, since there are three distance-$R_5$ couplings per qubit with two occuring with strength $\Omega \left( \frac{R_b}{a} \right)^6 \times \frac{1}{2(R_5/a)^6}$ and one feeling $\Omega \left( \frac{R_b}{a} \right)^6 \cdot \frac{1}{(R_5/a)^6}$, an estimate for the rough scale of the maximum energy density per qubit will be $\Omega \left( \frac{R_b}{a} \right)^6 \frac{2}{ (R_5/a)^6} \times \frac{1}{16}$.
(In fact, for the distance considered in the previous paragraph, which is considerably shorter, the analysis gave an effective $1/12$ which is already close to the uncorrelated value of $1/16$).
Thus, our estimate for the $m$-anyon energy scale is:
\begin{align} \label{eq-numericalestimate}
E_m \lesssim \Omega \left( \frac{R_b}{a} \right)^6 \left[ \frac{\beta}{12} + \frac{a^6}{8 R_5^6} \right]
\end{align}
where the term in parenthesis is approximately $5.6 \times 10^{-5}$ which is nearly $25$ times smaller than the naive expectation.
We note that the above gives a soft upper bound on the energy scale of splittings induced by the Rydberg interaction (See Section~\ref{subsubsec-NumericalEstimates} for discussion of this energy scale compared to experimental time scales).

\subsubsection{Phase Diagram with Long-Range Tails}

Since the long-range tails of the Rydberg interaction can generate $m$-anyon fluctuations, they could potentially confine the QSL phase of the model in the absence of these tails.
As a consequence, Refs.~\onlinecite{Verresen21,Semeghini21} studied the ground state phase diagram of the Rydberg model with the presence of the long-range tails in addition to a study of the PXP model phase diagram. In particular, we consider the truncated VdW model in Fig.~\ref{fig-RydbergPotential}, where we keep the Van der Waals interactions within a distance $r<R_\textrm{trunc}$.

Let us first consider the particular instance of ruby lattice defined by the qubits on the bonds of the kagome lattice. In this case, Ref.~\onlinecite{Verresen21} found that upon including the effects of the long-range Rydberg interaction around $R_\textrm{trunc} \approx 6a$, the QSL eventualy disappears and the phase diagram is:
\begin{equation*}
\hspace{7 mm}\begin{tikzpicture}[scale = 1, baseline={([yshift=-.5ex]current bounding box.center)}]
\draw[-stealth, line width=0.4mm] (0, 0) -- (5, 0);
\node at (5.5,0) {\normalsize $\delta/\Omega$};
\draw[line width=0.4mm] (2.5, 0) -- (2.5, 0.2);
\node at (1.25,0.3) {\small Higgs};
\node at (3.75,0.3) {\small Confined};
\node at (2.5, -0.2) {\small $\approx 3.5$};
\end{tikzpicture}
\end{equation*}
where the Higgs and confined phase are separated by a first-order phase transition.
As a consequence, for the lattice geometry simulated in the Rydberg atom experiment \cite{Semeghini21}, the system did not have a QSL phase in its ground state phase diagram.

We note that Ref.~\onlinecite{Verresen21} showed that the ground state spin liquid can persist by considering an elongated ruby lattice, where the triangles are placed further apart. In particular, while for the bonds of the kagome lattice the aspect ratio of the rectangles of the ruby lattice is $\rho = \sqrt{3}$, increasing this to $\rho = 3$ stabilizes a spin liquid in the ground state, even as one arbitrarily increases $R_\textrm{trunc}$.

\subsection{Dynamical Preparation of Quantum Spin Lake with Rydberg Atoms}

Equilibrium physics in hand, we can see that the physics is precisely in the regime where one would expect to dynamically prepare the quantum spin lake.
In particular, equivalent to the toric code case, the dynamics of the $e$-particle will be fast as it is set by the strong Rydberg interaction and Rabi oscillation scale, and the dynamics of the $m$ particle will be slow as it is set by terms generated at high orders in perturbation theory and small energy scales occuring due to the long-range tails of the Rydberg interaction.

In this subsection, we address why we would expect that dynamics that are slow relative to $e$-anyons and fast relative to $m$-anyons would produce a quantum spin lake in the Rydberg system.
Subsequently, we will make numerical estimates for the $m$-anyon scale in the experiment and demonstrate that the time and energy scales used in the Rydberg atom experiment place us in the regime for producing a quantum spin lake.

\subsubsection{Quantum Spin Lakes from Rydberg Atoms}

We aim to show that the ground state of the Higgs phase in the Rydberg model yields a quantum spin liquid when we project out Gauss law violations.
By translation invariance and the low-entangled nature of the Higgs phase, a mean-field ansatz for the initial state of the sweep can be expressed as:
\begin{equation}\label{eq-RydbergMFansatz}
    \ket{\psi(0)} \propto \mathcal{P}_{\text{blockade}} \bigotimes_{i} \left( \ket{\downarrow} + \varepsilon \ket{\uparrow} \right)
\end{equation}
where $\mathcal{P}_{\text{blockade}}$ is a projector onto blockade satisfying states (defined below Eq.~\eqref{eq-actualHPXP}). 
Then, by Eq.~\eqref{eq-sweeping-proj}, the final state under the dynamics will be:
\begin{equation} \label{eq-RydbergProj}
    \ket{\psi(T)} \propto \mathcal{P}_{G} \ket{\psi(0)} = \ket{\text{RVB}}
\end{equation}
where $\mathcal{P}_{G} = (1 - G_v)/2$, $G_v$ is defined in Eq~\eqref{eq-RydbergGauss}, and $\ket{\text{RVB}}$ is defined in Eq.~\eqref{eq-RydRVB}.
Crucially, the above follows from the fact that each dimer configuration has the same number of dimers and thus each enters with the same amplitude in Eq.~\eqref{eq-RydbergMFansatz}.
Consequently, the state prepared in dynamics will resemble a QSL.

\subsubsection{Numerical Estimates for Regime of the Experiment} \label{subsubsec-NumericalEstimates}

We conclude by numerically estimating what dynamical regime the Rydberg atom experiment of Ref.~\onlinecite{Semeghini21} was in.
Since the Rabi frequency and the Gauss law (enforced by the detuning and Rydberg blockade) are both large energy scales in the problem, the energy scales governing the equilibration of $e$-anyons is large as required. 
As such, here, we aim to estimate a figure of merit for the density of $m$-anyons produced during the sweep of the experiment.
In particular, we aim to compute $E_m T_{\text{exp}}$ where $E_m$ is the energy scale associated with the dynamics of $m$-anyons and $T_{\text{exp}}$ is the amount of time that the the experiment spends in the regime of parameter space with a constrained low-energy subspace.

To do so, we remark that in the Rydberg atom experiment, the Rabi frequency and blockade radius were reported to be $\Omega =  2\pi  \times 1.4\text{ MHz}$ and $R_b/a = 2.4$.
Ignoring the sixth order plaquette resonance term, we use Eq.~\eqref{eq-numericalestimate} to get an estimate for $E_m = 0.011 \cdot \Omega = 0.096 \text{ MHz}$ which is two orders of magnitude smaller than the characteristic energy scale for $e$-anyons!
Moreover, to get a rough estimate of $T_{\text{exp}}$ (which we roughly estimate to be around the amount of time in the experiment spent past $\Delta/\Omega \approx 3.5$) is $T_{\text{exp}} = 0.5\ \mu\text{s}$.
As such, the dimensionless figure of merit $E_m T_{\text{exp}} \approx 0.048 \ll 1$.
As a consequence, the density of $m$-anyons nucleated during the dynamical sweep in the Rydberg experiment of Ref.~\onlinecite{Semeghini21} is expected to be low, putting the experiment in the regime for preparation of the quantum spin lake.

We conclude by remarking that the large separation between the energy scales controlling the dynamics of $e$-anyons and $m$-anyons suggest that it should be possible to ignore the effects of $m$-anyons when numerically and analytically studying the experimental settings such as the Rydberg atom experiment.
This will be explored and confirmed in further detail in the following section.

\section{Resonating without Resonances: Spin Lakes on Trees} \label{sec-Tree}

The discussions of the previous two sections concluded with two findings.
First, in Section~\ref{sec-Summary}, we found that the preparation of a quantum spin lake was limited by the energy scale of $m$-anyon excitations, which is set by any confining fields in the problem and a perturbatively generated resonance term: the larger the energy scale of $m$-anyons, the smaller the spin lake one can prepare.
A natural conclusion of this is that, in the absence of confining fields, the presence of perturbatively generated resonance terms is what limits the preparation of a quantum spin lake on the ruby lattice!
This is a striking reversal of logic relative to the equilibrium case where a proper combination of resonances are precisely what stabilize the QSL.

Second, in Section~\ref{sec-Experiment}, we found that, in experimentally relevant settings, the aforementioned perturbative resonances and confining terms are quite small and hence are predicted to not influence the short-time dynamics accessible in experiments.
As a consequence, as alluded to in the previous section, it should be possible within this time-frame to study the dynamics  numerically and analytically by ignoring the effects of $m$-anyons.

\begin{figure}
    \centering
    \includegraphics[width = 247pt]{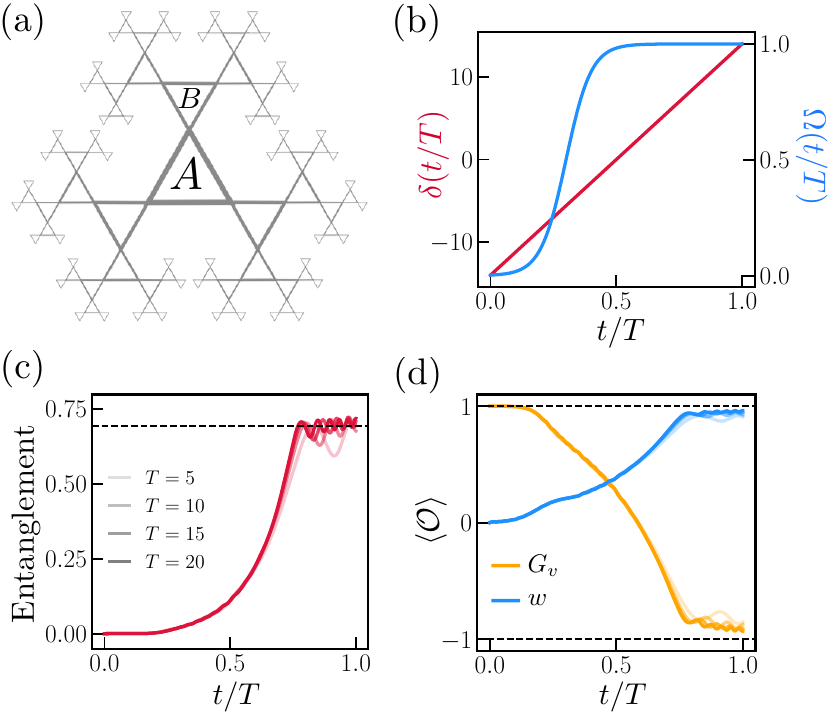}
    \caption{\textbf{$\mathbb{Z}_2$ Spin Lakes on the Husimi Cactus.}
    (a) The Husimi cactus lattice is a tree version of the kagome lattice. Here, we simulate the PXP model for Rydberg atoms living on the links of the Husimi cactus by using a two-tensor infinite tree tensor network ansatz with the first tensor encoding the state of the $A$ triangles and the second encoding the state of the $B$ triangles.  (b) In our simulations, we dynamically sweep the values of the detuning $\delta$ and Rabi drive $\Omega$ in such a way that our sweep adiabatically prepares the ground state for $\Omega = 1$ and $\delta$ large and negative. Subsequently, $\Omega$ remains constant and set to $1$ as $\delta$ sweeps into a Gauss law satisfying regime. At the end of our dynamical sweep, the resulting state displays QSL-like signatures. In particular, in panel (c), we show that the state has the same entanglement as the fixed point RVB state on the Husimi cactus and in panel (d) we show that the final state in approximately stabilized by the stabilizers of the RVB (defined in Eq.~\eqref{eq-RydbergGauss} and Eq.~\eqref{eq-treewilson}). For both the entanglement and the stabilizers, the value better approaches the fixed point value with increased total time of the sweep. In our numerics, we use a bond-dimension of $\chi_{\alpha} = 7$ ($\alpha = a, b, c$) and trotter step size of $dt = 0.005$.}. 
    \label{fig:Z2Tree}
\end{figure}

In this section, we culminate these two observations by studying the Rydberg model on a tree lattice version of the ruby lattice.
In particular, we envision putting qubits on the links of the so-called Husimi cactus lattice: a version of the kagome lattice with no hexagonal loops [See Figure~\ref{fig:Z2Tree}(a) and Subsection~\ref{subsec-Z2TreeModel} for more detail].
The motivation to do so is due to a unique feature of this tree lattice.
Namely, resonances generated through the $\Omega$ term of the Rydberg model do not occur at any finite order in perturbation theory---the Rydberg model on this lattice has no resonances!
By using infinite tree tensor network methods (described in Subsection~\ref{subsec-Z2TTNMethod}), we numerically demonstrate the preparation of a quantum spin lake for the PXP model [defined by Eq.~\eqref{eq-HPXP} with $V_{ij}$ taken to be that of Fig.~\ref{fig-RydbergPotential}(b)] of the Husimi cactus in Subsection~\ref{subsec-Z2TreePXPNumerics}, thereby confirming the aforementioned reversal of logic in the most extreme setting.
The complete absence of $m$-anyon dynamics allows one to prepare an arbitrarily large quantum spin lake.

In additional to its conceptual value, we show that the Rydberg model on the tree can correctly approximate the experimental setup within time-scales wherein one does not resolve the $m$-anyon dynamics.
Indeed, in Subsection~\ref{subsec-Z2TreeVdWNumerics}, we show that tree tensor network simulations of the Rydberg model with the more experimentally faithful truncated VdW potential [given by Fig.~\ref{fig-RydbergPotential}(c) with $R_{\text{trunc}} = 2 \sqrt{3} a$] on the tree lattice are able to match the experimental data from the Rydberg experiment just as well as cylinder matrix product state simulations of true ruby lattice. 
Moreover, we find that such simulations are roughly two orders of magnitude faster than the cylinder matrix product state simulations that are traditionally used to study dynamics of such systems.
As such, this identifies tree tensor network methods as an ideal numerical tool for studying the dynamical preparation of QSL-like order in analog NISQ devices.

\subsection{Rydberg Models on the Husimi Cactus} \label{subsec-Z2TreeModel}

As stated earlier, we want to study a version of the Rydberg model with qubits on the links of the Husimi cactus lattice, a tree version of the kagome lattice [See Fig.~\ref{fig:Z2Tree}(a)].
While the global structure of the Husimi cactus differs from the kagome lattice, the local structure and connectivity of the lattice is identical.
As such, we can consider a version of the Rydberg model on links of the Husimi cactus.
In particular, the PXP model Eq.~\eqref{eq-HPXP} can be directly carried over to this tree geometry, where we understand the blockade interactions to project out any two neighboring bonds from both being occupied with a dimer.

Later in this section we will also consider a slightly modulated version: while we will not include longer-range $\sim 1/r^6$ interactions (which admittedly requires care to define on a tree geometry), we will make the interactions within the blockade radius spatially dependent, choosing the strengths we had on the planar lattice for the experimental choice of blockade radius $R_b = 2.4a$. In particular, while the shortest intra-triangle interactions are still infinitely strong (i.e., there is never more than one dimer per triangle), we set the second nearest neighbor to be $\Omega \left( \frac{2.4}{\sqrt{3}} \right)^6 \approx 7.08 \Omega$ and the third to be $\Omega \left( \frac{2.4}{2} \right)^6 \approx 2.99 \Omega$.

\subsection{Tree Tensor Network Numerical Method}\label{subsec-Z2TTNMethod}

To analyze either Rydberg model, our approach will be to numerically simulate the dynamics using an infinite tree tensor network approach \cite{Vidal_Tree, Vidal_Tree_2, Murg_Tree}.
In particular, we will make the following translationally invariant ansatz for the wavefunction defined on the lattice of Fig.~\ref{fig:Z2Tree}(a):
\begin{equation} \label{eq-TTN}
    \includegraphics[width = 70pt, valign = c]{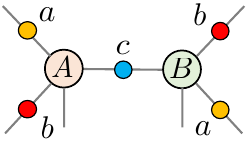}
\end{equation}
where $A$ and $B$ are $4 \times \chi_a \times \chi_b \times \chi_c$ tensors that encode the state of the two triangles forming the unit cell of the lattice and the $\{\chi_{\alpha}\}$ are called the bond dimensions and refers to the ranks of the non-dangling legs of the tensors.
The  physical legs of the $A$ and $B$ tensors are rank-4 corresponding to the following four states:
\begin{equation} \label{eq-Atriangle}
d = 
\quad 
\begin{tikzpicture}[scale = 2, baseline={([yshift=-.5ex]current bounding box.center)}]
\draw[gray] (0,0) -- ({-.7*(0.5) * 1/2}, {-(.7*0.5) * 0.866025404}) -- ({(.7*0.5) * 1/2}, {-(.7*0.5) * 0.866025404}) -- cycle;
\node at (0, {-(.7*0.25) * 0.866025404 -0.03}) {\small $0$};
\end{tikzpicture}
\quad
\begin{tikzpicture}[scale = 2, baseline={([yshift=-.5ex]current bounding box.center)}]
\draw[gray] (0,0) -- ({-.7*(0.5) * 1/2}, {-(.7*0.5) * 0.866025404}) -- ({(.7*0.5) * 1/2}, {-(.7*0.5) * 0.866025404}) -- cycle;
\draw[red, fill = red] ({0}, {-(.7*0.5) * 0.866025404}) ellipse (0.175 and 0.03);
\node at (0, {-(.7*0.25) * 0.866025404 -0.03}) {\small $1$};
\end{tikzpicture}
\quad
\begin{tikzpicture}[scale = 2, baseline={([yshift=-.5ex]current bounding box.center)}]
\draw[gray] (0,0) -- ({-.7*(0.5) * 1/2}, {-(.7*0.5) * 0.866025404}) -- ({(.7*0.5) * 1/2}, {-(.7*0.5) * 0.866025404}) -- cycle;
\draw[red, fill = red, rotate around={60:({-.35*(0.5) * 1/2}, {-(.35*0.5) * 0.866025404})}] ({-.35*(0.5) * 1/2}, {-(.35*0.5) * 0.866025404}) ellipse (0.175 and 0.03);
\node at (0, {-(.7*0.25) * 0.866025404 -0.03}) {\small $2$};
\end{tikzpicture} \quad
\begin{tikzpicture}[scale = 2, baseline={([yshift=-.5ex]current bounding box.center)}]
\draw[gray] (0,0) -- ({-.7*(0.5) * 1/2}, {-(.7*0.5) * 0.866025404}) -- ({(.7*0.5) * 1/2}, {-(.7*0.5) * 0.866025404}) -- cycle;
\draw[red, fill = red, rotate around={-60:({(.35*0.5) * 1/2}, {-(.35*0.5) * 0.866025404})}]  ({(.35*0.5) * 1/2}, {-(.35*0.5) * 0.866025404})  ellipse (0.175 and 0.03);
\node at (0, {-(.7*0.25) * 0.866025404 -0.03}) {\small $3$};
\end{tikzpicture}
\end{equation}
for $A$ and:
\begin{equation}\label{eq-Btriangle}
d =\quad \begin{tikzpicture}[scale = 2, baseline={([yshift=-.5ex]current bounding box.center)}]
\draw[gray] (0,0) -- ({-0.7*(0.5) * 1/2}, {0.7*0.5 * 0.866025404}) -- ({0.7*0.5 * 1/2}, {0.7*0.5 * 0.866025404}) -- cycle;
\node at (0, {(.7*0.25) * 0.866025404 +0.03}) {\small $0$};
\end{tikzpicture}
\quad
\begin{tikzpicture}[scale = 2, baseline={([yshift=-.5ex]current bounding box.center)}]
\draw[gray] (0,0) -- ({-0.7*(0.5) * 1/2}, {0.7*0.5 * 0.866025404}) -- ({0.7*0.5 * 1/2}, {0.7*0.5 * 0.866025404}) -- cycle;
\node at (0, {(.7*0.25) * 0.866025404 +0.03}) {\small $1$};
\draw[red, fill = red, rotate around={-60:({-0.35*(0.5) * 1/2}, {0.35*0.5 * 0.866025404})}]  ({-0.35*(0.5) * 1/2}, {0.35*0.5 * 0.866025404})  ellipse (0.175 and 0.03);
\end{tikzpicture}
\quad
\begin{tikzpicture}[scale = 2, baseline={([yshift=-.5ex]current bounding box.center)}]
\draw[gray] (0,0) -- ({-0.7*(0.5) * 1/2}, {0.7*0.5 * 0.866025404}) -- ({0.7*0.5 * 1/2}, {0.7*0.5 * 0.866025404}) -- cycle;
\node at (0, {(.7*0.25) * 0.866025404 +0.03}) {\small $2$};
\draw[red, fill = red, rotate around={60:({0.35*(0.5) * 1/2}, {0.35*0.5 * 0.866025404})}]  ({0.35*(0.5) * 1/2}, {0.35*0.5 * 0.866025404})  ellipse (0.175 and 0.03);
\end{tikzpicture}
\quad
\begin{tikzpicture}[scale = 2, baseline={([yshift=-.5ex]current bounding box.center)}]
\draw[gray] (0,0) -- ({-0.7*(0.5) * 1/2}, {0.7*0.5 * 0.866025404}) -- ({0.7*0.5 * 1/2}, {0.7*0.5 * 0.866025404}) -- cycle;
\node at (0, {(.7*0.25) * 0.866025404 +0.03}) {\small $3$};
\draw[red, fill = red]  ({0}, {0.7*0.5 * 0.866025404})  ellipse (0.175 and 0.03);
\end{tikzpicture}
\end{equation}
for $B$.
By encoding the local Hilbert space of the triangles of the lattice in the manner above, we explicitly enforce the blockade constraint inside the triangles, which amounts to assuming that the Rydberg interaction is effectively infinite for qubits within the same triangle.
This explicit enforcement is exact for the PXP model and is a good approximation for the truncated van der Waals model with the $1/R^6$ tails where for $R_b = 2.4 a$, the interaction within the triangles is $\approx 191 \Omega$ (two orders of magnitude larger than every other coupling in the system).
As such, our ansatz enables studying both models.

Our ansatz contains three additional tensors $a, b,$ and $c$ that are diagonal matrices that live on the bonds between the $A$ and $B$ tensors.
These tensors encode the Schmidt values of the tree tensor network state under bipartitioning, similar to the diagonal tensors in the mixed canonical form of the matrix product state \cite{Li_TTN, Hauschild18}.
Using such an ansatz, we can efficiently simulate trotterized dynamics on this system \cite{Li_TTN}.

\subsection{Large Spin Lakes in the PXP Model on a Tree}\label{subsec-Z2TreePXPNumerics}

\begin{figure}
    \centering
    \includegraphics[width = 247 pt]{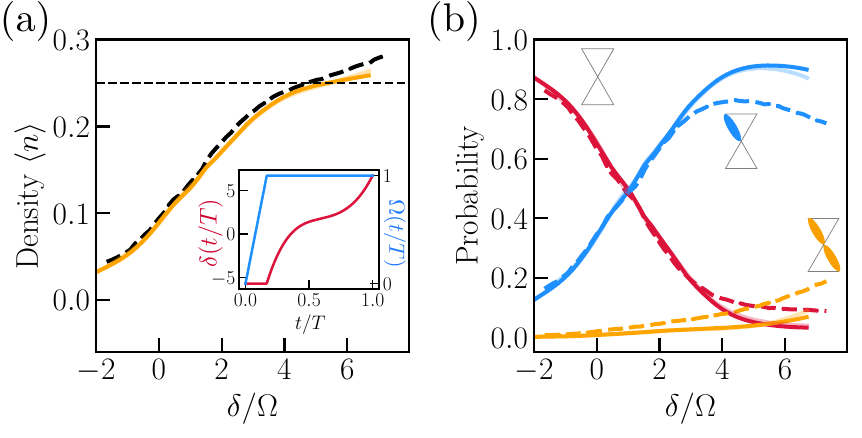}
    \caption{ \textbf{Comparing Tree Simulations, Matrix Product State Cylinder Simulations, and Experimental Data for Rydberg Atom System.} We simulate the dynamical sweep performed in the Rydberg atom experiment of Ref.~\onlinecite{Semeghini21} by simulating a tree lattice instead of the usual ruby lattice. The dynamical sweep is pulled directly from the experimental paper and is show in the inset of panel (a). In panel (a), we compare the time trace of the density $\langle n(t) \rangle$ between tree tensor network simulations (dark orange), cylinder matrix product state simulations (light orange), and experiment (black, dashed). We find excellent agreement between the three, with the two numerical methods being virtually indistinguishable. In panel (b), we compare the probability of observing monomers (empty vertices), dimers, and double dimers in the wavefunction at $\delta(t)/\Omega(t)$. We once again find near perfect agreement between conventional matrix product state simulations and tree lattice simulations. Both simulations qualitatively reproduce the results of the experiment for all times and quantitatively reproduce the results of the experiments at early times. At late times, the experiment quantitatively finds a higher probability of observing monomers and double dimers.
    In our tree lattice numerics, we use a bond dimension of $\chi_{\alpha} = 7$ ($\alpha = a, b, c$) and a trotter step size of $dt = 0.01$.}
    \label{fig:experimental_trees}
\end{figure}

Numerical method in hand, we now simulate the dynamics of the PXP model defined on the links of the Husimi cactus.
Since the dynamics of $m$-anyons is infinitely slow in this model, our goal is to demonstrate the emergence of a quantum spin lake in this model that increases in fidelity as we decrease the sweep rate.

We start by initializing our state in the ground state of the Higgs phase ($\delta/\Omega = -14$).
To ensure we initialize the state properly, we start by setting $\Omega = 0$ and initializing the state with no dimers, the exact ground state.
Subsequently, we ramp $\delta$ and $\Omega$ in the fashion shown in Fig.~\ref{fig:Z2Tree}(b) to prepare the ground state of the Higgs phase adiabatically and then sweep to the Gauss law satisfying phase ($\delta/\Omega = 14$) \footnote{The exact nature of the ground state at $\delta/\Omega$ large and positive is not important to the discussion in this section. It is sufficient that at low energies, the system will satisfy Eq.~\eqref{eq-RydbergGauss}}.
We now use two approaches to diagnose the onset of the quantum spin lake.

First, we compute the entanglement across a bond of the tree tensor network (equivalently,  a vertex of the original Husimi cactus lattice), and compare to the expected value of the fixed point $\ket{\text{RVB}}$ state on the tree lattice which we find to be $\log(2)$ (See Appendix.~\ref{app-fprvb} for an exact tree tensor network for the RVB state from which the entanglement can be computed).
Indeed, by plotting the entanglement entropy as a function of time in Fig.~\ref{fig:Z2Tree}(c) (in units of the total time of the sweep which we vary), we find that the entanglement entropy saturates to $\log(2)$ after crossing the transition, with convergence improving as a function of total time.
This is consistent with the emergence of the quantum spin lake.

Our second approach will be to measure the stabilizers of the original ruby lattice Rydberg model.
While the Gauss law ('t Hooft) loop of Eq.~\eqref{eq-RydbergGauss} will remain unchanged, the resonance (Wilson) loop of Eq.~\eqref{eq-Rydberg-Resonances} becomes an infinitely long line on the Husimi lattice:
\begin{equation*}
    \begin{tikzpicture} [scale = 0.9, baseline={([yshift=-.5ex]current bounding box.center)}]
        \draw[gray] (0, 0) -- (1, 0) -- (1/2, 0.866025404) -- cycle;
        \draw[gray] (0, 0) -- ( -1/4, -0.866025404/2) -- (-1/2, 0) -- cycle;
        \draw[gray] (1, 0) -- ( 1 + 1/4, -0.866025404/2) -- (1 + 1/2, 0) -- cycle;
        \draw[gray] (1/2, 0.866025404) -- (1/2 - 1/4, 0.866025404 * 3/2) -- (1/2 + 1/4, 0.866025404 * 3/2) -- cycle;
        \draw[dodgerblue, decorate, decoration={snake, segment length=3.6mm}, line width = .35mm] ( -1/4 * 1.4, -0.866025404/2 * 1.4) -- ({(1/2 + 1/4)*1.2}, {(0.866025404 * 3/2) * 1.2});
        \draw[-stealth, dodgerblue, line width = 0.5mm] ({(1/2 + 1/4)*1.1}, {(0.866025404 * 3/2) * 1.1 - 0.5})  -- ({(1/2 + 1/4)*1.5}, {(0.866025404 * 3/2) * 1.5 - 0.5});
        \draw[-stealth, dodgerblue, line width = 0.5mm] ({(1/2 + 1/4)*(0.2)}, {(0.866025404 * 3/2) * (0.2) - 0.5})  -- ({(1/2 + 1/4)*(-0.2)}, {(0.866025404 * 3/2) * (-0.2) - 0.5});
    \end{tikzpicture}
\end{equation*}
Since even the smallest perturbation of the RVB from its fixed point states will generically endow the Wilson line with a line tension rendering its expectation value zero, we wish to compute the expectation value of the Wilson line per unit length.
This can be done using the tree tensor network ansatz by constructing the following mixed transfer matrix:
\begin{equation} \label{eq-treewilson}
 T = \includegraphics[width = 75pt, valign = c]{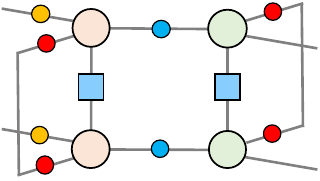}
\end{equation}
where the blue squares represent the action of the Wilson line (left of Eq.~\eqref{eq-RydbergWilson}).
Subsequently, we compute the square root of the largest eigenvalue of $T$, which encodes the expectation value of the Wilson line per unit length.

With these two stabilizers, we can plot the expectation value of the gauss loop $\langle G_v\rangle $ and the Wilson line per unit length $\langle w \rangle$ as a function of time across the sweep [Fig.~\ref{fig:Z2Tree}(d)] in units of the total time.
We find that both stabilizers approach and saturate their fixed point value with the deviation from the fixed point decreasing with increased sweep time, once again signaling the onset of the quantum spin lake!

Having numerically demonstrated the preparation of a quantum spin lake for the PXP model on the links of the Husimi cactus lattice---a model with no dimer resonances---we now raise the queston, where did the resonances in the final state come from. 
The answer is elucidated by the projection formula of Eq.~\eqref{eq-RydbergProj}.
In particular, during the dynamics, when the initially condensed $e$-anyons are equilibrated (projected) out, onsite quantum fluctuations of the initial state get elevated to many-body fluctuations in the final state.

\subsection{ Tree Simulations as Numerical Tools for Experiments} \label{subsec-Z2TreeVdWNumerics}

So far, we have examined a model of Rydberg atoms on the links of the Husimi cactus to highlight a conceptual point about the dynamical preparation of quantum spin lakes---preparation benefits from the perfectly slow $m$-anyon dynamics of Rydbergs on a tree.
While the dynamics of $m$-anyons on normal lattices are never perfectly slow, they are often significantly slower than the dynamics of $e$-anyons.
Indeed, in Section~\ref{subsubsec-NumericalEstimates}, we showed that the energy scale controlling the dynamics of the $m$-anyons is an order of magnitude smaller than the one for $e$-anyons for the Rydberg atom experiment in Ref.~\onlinecite{Semeghini21}.
As a consequence, one might postulate that the tree model studied in this section may be able to approximately capture the dynamics of the aforementioned experiment. 
Here, we show that, indeed, this is the case.
By numerically simulating the truncated van der Waals Rydberg model on the Husimi cactus lattice (defined in Section~\ref{subsec-Z2TreeModel}),  we show that the tree model is able to capture the results of the Rydberg atom experiment nearly as well as matrix product state simulations done for the regular ruby lattice.
Moreover, the tree tensor network numerics have roughly a two order of magnitude speed up compared to the matrix product state simulations performed.
As such, we propose that tree tensor network simulations of tree lattices can actually be used as a practical experimental aid to study the dynamical preparation of QSL-like order in NISQ devices.

We now numerically simulate the dynamical sweep performed in the experiment for the truncated van der Waals model on the tree lattice.
To accurately reproduce the dynamics of the experiment, we use the same sweep profiles of $\Omega(t)$ and $\delta(t)$ as used in the experiment (inset of Fig.~\ref{fig:experimental_trees}(a)) and the same value of the blockade radius $R_b = 2.4 a$ \cite{Semeghini21}.
We now compare the results from our tree tensor network simulations of Rydberg atoms on the Husimi cactus, the matrix product state simulations of Rydberg atoms on the ruby lattice \cite{zenodoQSL} (from the supplemental information of Ref.~\onlinecite{Semeghini21}), and the experimental data from the Rydberg atom experiment \cite{semeghinidata} (from the main text of Ref.~\onlinecite{Semeghini21}).
In particular, first, we plot the density $\langle n \rangle$ of Rydberg atoms as a function of $\delta/\Omega$ for each of these three methods in Fig.~\ref{fig:experimental_trees}(a).
We find excellent agreement between the three for all values of $\delta/\Omega$.
We note that our tree tensor network simulations very slightly deviate from the experimental value towards the end of the sweep but matches the matrix product state results throughout.

Beyond comparing the density, we additionally compare the probability of dimerless vertices, vertices touching a single dimer, and vertices touching two dimers between the experiment, tree tensor network simulations, and matrix product state simulations in Fig.~\ref{fig:experimental_trees}.
We find that near perfect agreement between the matrix product state simulations and the tree tensor network simulations.
Moreover, both types of simulations quantitatively reproduce the results of the experiment near the beginning of the sweep but only qualitatively capture the experiment towards the end.
Generally, the experiment shows a higher density of dimerless vertices and vertices touching two dimers.
The fact that the tree numerics and cylinder DMRG---which rely on entirely different approximations---give virtually identical results suggests that for these timescales and parameter values, we roughly obtain the true result for the 2D lattice. It would be interesting for future work to pinpoint the source of the experimental deviation, which nevertheless qualitatively agrees.

\section{Generalizations: U(1) Spin Lake} \label{sec-U1}

\begin{figure}
    \centering
    \begin{tikzpicture}
    \node at (0,0) {\includegraphics[width = 247pt]{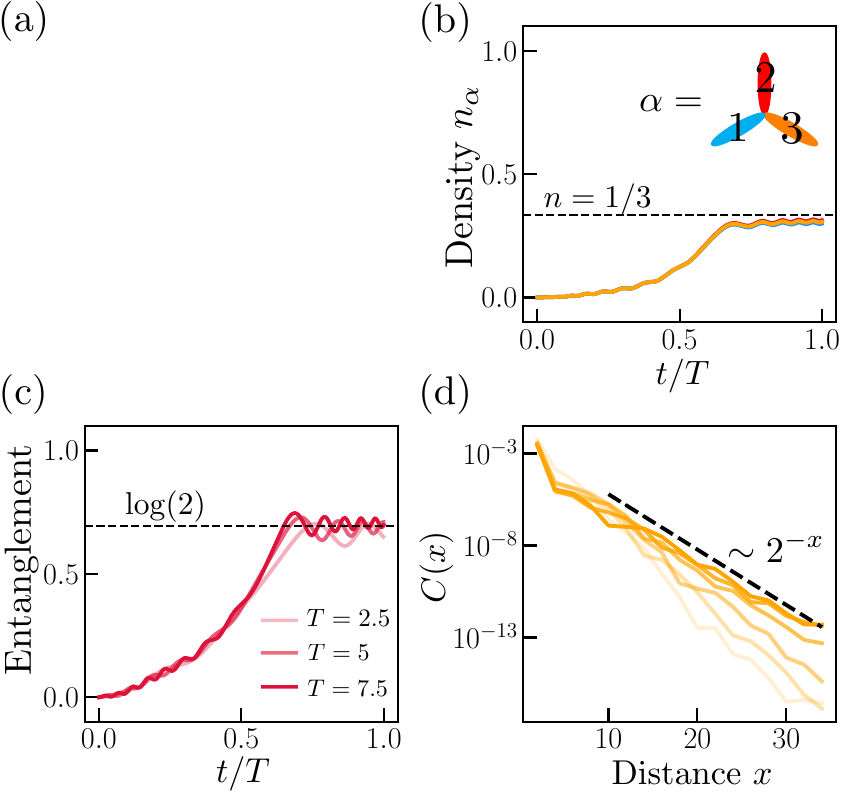}};
    \node at (-2.2,2.15) {\includegraphics[scale=0.3]{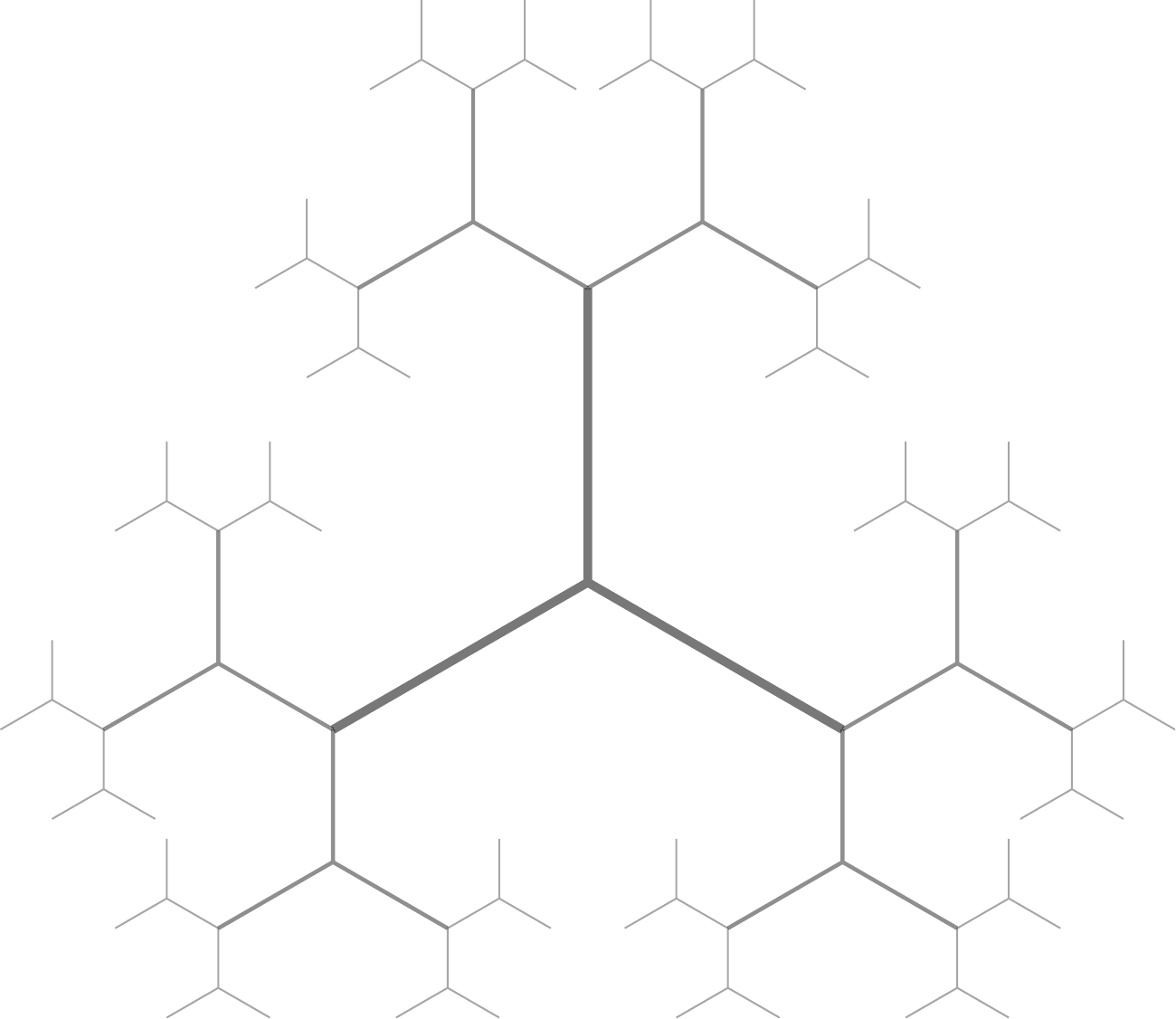}};
    \end{tikzpicture}
    \caption{\textbf{U(1) Spin Lakes on the Bethe Lattice.} (a) We simulate the PXP model on the links of the Bethe lattice---a tree version of the honeycomb lattice. Since the vertices of the lattice are bipartite, the RVB wavefunction on this lattice is a $U(1)$ QSL and is analogous to the fine-tuned Rokhsar-Kivelson point wavefunction. To simulate dynamics, we use the same dynamical sweep as Fig.~\ref{fig:Z2Tree}(b) but with $\delta$ going from $\delta = -10$ to $10$.  In panel (a), we show that the density $\langle n_{\alpha} \rangle$ is isotropic indicating the state is not a VBS that breaks the discrete rotation symmetries of the model. Since our tree tensor network ansatz explicitly preserves lattice translation symmetry, any candidate VBS state would be a cat state with $\log(2)$ bipartite entanglement, the same as the RVB. In panel (c), we show that the entanglement we observe is consistent with either the VBS cat state or the RVB. However, in panel (d), we plot density-density correlations as a function of distance [using a rolling average of $3$ sites for easier visualization (See Appendix~\ref{app-U1} for raw data)] for different total sweep times  (from light to dark, $T \in \{2.5, 3.5, 4.5, 4.75, 5, 7.5\}$) on the tree lattice.
    We find that they fall off as $2^{-x}$ which is the predicted fall off of gapless states on tree lattices \cite{Laumann09}.
    This is inconsistent with a VBS cat state and hence, the above provides strong numerical evidence for the emergence of a $U(1)$ spin lake on the Bethe lattice. For our numerics, we use a bond dimension of $\chi_{\alpha} = 10$ ($\alpha = a, b, c$) and $dt = 0.005$. } 
    \label{fig:U1QSL}
\end{figure}

Thus far, we have focused on the dynamical preparation of $\mathbb{Z}_2$ quantum spin lakes.
In this section, we generalize our results to a broader class of spin liquid states by demonstrating that non-equilibrium dynamics can also prepare a $U(1)$ quantum spin lake.
This is particularly surprising because as ground states, U(1) quantum spin liquids are unstable, being described by a compact U(1) gauge theory that is known to be typically confining \cite{PolyakovBook}.

To see the emergence of a $U(1)$ quantum spin lake, we will first introduce a model \textit{capable} of hosting an (unstable) $U(1)$ QSL.
For reasons that will reviewed in Section~\ref{subsec-U1equil}, a clear option will be to place Rydberg atoms on the bonds of the (bipartite) honeycomb lattice.
In the subsections that follow, we will consider an extremal version of the Rydberg model on the honeycomb lattice with infinitely large plaquettes.
This will once again exemplify the reversal of logic from the last section that the absence of dimer resonances \textit{helps} in the dynamical preparation of QSLs.
Moreover, as we evidenced in the previous section, such a tree geometry offers a good approximation of the sweep dynamics on the true physical planar lattice, suggesting that this indeed offers a realistic route to a $U(1)$ spin lake accessible with current Rydberg atom tweezer array platforms.

\subsection{$U(1)$ QSLs from Rydberg Atoms}\label{subsec-U1equil}

Let us consider the `PXP' Rydberg model of Section~\ref{sec-Experiment} (Eq.~\eqref{eq-HPXP} without long-range tails) placed on the links of the honeycomb lattice, or equivalently, the vertices of the kagome lattice:
\begin{equation*}
    \begin{tikzpicture}[scale = 1, baseline={([yshift=-.5ex]current bounding box.center)}]
    \foreach \i in {0,...,1}{
        \foreach \j in {0,...,1}{
            \draw[gray] ({\i + \j * 1/2}, {\j * 0.866025404}) -- ({\i + \j * 1/2 + 1/2}, {\j * 0.866025404}) -- ({\i + \j * 1/2 + 1/2*1/2}, {\j * 0.866025404 + 1/2*0.866025404}) -- cycle;
            \draw[gray] ({\i + \j * 1/2}, {\j * 0.866025404}) -- ({\i + \j * 1/2 - 1/2}, {\j * 0.866025404}) -- ({\i + \j * 1/2 - 1/2*1/2}, {\j * 0.866025404 - 1/2*0.866025404}) -- cycle;
            \filldraw  ({\i + \j * 1/2}, {\j * 0.866025404}) circle (1 pt);
            \filldraw  ({\i + \j * 1/2 - 1/2}, {\j * 0.866025404}) circle (1 pt);
            \filldraw  ({\i + \j * 1/2 + 1/4 }, {\j * 0.866025404 + 1/2 * 0.866025404}) circle (1 pt);
            \filldraw  ({\i + \j * 1/2 - 1/4 }, {\j * 0.866025404 - 1/2 * 0.866025404}) circle (1 pt);
            \filldraw  ({\i + \j * 1/2 + 1/2}, {\j * 0.866025404}) circle (1 pt);
        }
    }
    \draw[red, fill=red, fill opacity=0.1, dashed, thick](1/2,0) circle (0.6);
    \filldraw[red] (0.5, 0) circle (1 pt);
    \draw [-stealth, thick] (1/2,0) -- (1/2 - 0.44/0.5 * 0.6, 0.25/0.5 * 0.6);
    \node at ({-1/4 - 1/16}, 0.4) {\normalsize $R_b$};
    \draw [stealth-stealth, line width = 0.1 mm] (0,0.866025404 + 0.15) -- (1/2 ,0.866025404 + 0.15);
    \node at (0.25, 0.866025404 + 0.15 + 0.15) {\normalsize $a$};
    \end{tikzpicture}
\end{equation*}
If we choose the blockade radius $R_b$ such that the blockade radius encloses only a qubit's four nearest neighbors (as shown in the above schematic), i.e.:
\begin{equation} \label{eq-honeycombRb}
    1 < R_b /a < \sqrt{3}, 
\end{equation}
then we exclude two neighboring bonds from both being occupied. Hence, by appropriately tuning the chemical potential $\delta/\Omega$ (to achieve a density $\langle n_i \rangle \approx 1/3$) we will approximately realize a dimer model on the honeycomb lattice.

Unlike the case studied above for the ruby lattice (where we obtained a kagome dimer model), by putting atoms on the kagome lattice, we obtain an effective dimer model on the honeycomb lattice which is \emph{bipartite}. If we label the two sublattices $A$ and $B$, we can assign each dimer an orientation pointing from the $A$ sublattice to the $B$ sublattice:
\begin{equation}
    \begin{tikzpicture}[scale = 0.7, baseline={([yshift=-.5ex]current bounding box.center)}]
        \draw[gray] (0, 0) -- (0, 1) -- (0.86602540378, 1 + 1/2);
        \draw[gray] (0, 1) -- (-0.86602540378, 1 + 1/2);
        \draw[gray] (0.86602540378, -1/2) -- (0, 0) -- (-0.86602540378, -1/2);
        \node at (0,0) {\normalsize $A$};
        \node at (0,1) {\normalsize $B$};
        \draw[red, fill = red] (0,0.5) circle (0.075);
        \draw[black, fill = black] (0.86602540378/2, -1/4) circle (0.075);
        \draw[black, fill = black] (-0.86602540378/2, -1/4) circle (0.075);
        \draw[black, fill = black] (0.86602540378/2, 1+1/4) circle (0.075);
        \draw[black, fill = black] (-0.86602540378/2, 1+1/4) circle (0.075);
    \end{tikzpicture} \quad \longrightarrow \quad
    \begin{tikzpicture}[scale = 0.7, baseline={([yshift=-.5ex]current bounding box.center)}]
        \draw[gray] (0, 0) -- (0, 1) -- (0.86602540378, 1 + 1/2);
        \draw[gray] (0, 1) -- (-0.86602540378, 1 + 1/2);
        \draw[gray] (0.86602540378, -1/2) -- (0, 0) -- (-0.86602540378, -1/2);
        \draw[-stealth, red, line width = 0.75 mm] (0,0) -- (0,1);
    \end{tikzpicture}
\end{equation}

The ability to orient dimers on bipartite lattice has a striking consequence.
In particular, it implies that the $\mathbb{Z}_2$ Gauss law of the ruby lattice (Eq.~\eqref{eq-RydbergGauss}) gets promoted to:
\begin{equation}\label{eq-U1RydbergGauss}
G_v = \begin{cases} \ \begin{tikzpicture}[scale = 0.4, baseline={([yshift=-.5ex]current bounding box.center)}]
\draw[gray] (0.86602540378, -1/2)--(0, 0) -- (-0.86602540378, -1/2);
\draw[gray] (0, 0) -- (0, 1);
\draw[orange(ryb), dashed, line width = 0.5mm] (0, 0) circle (0.5);
\end{tikzpicture} \\ \\
\ \begin{tikzpicture}[scale = 0.4, baseline={([yshift=-.5ex]current bounding box.center)}]
\draw[gray] (0.86602540378, 1 + 1/2)--(0, 1) -- (-0.86602540378, 1+1/2);
\draw[gray] (0, 0) -- (0, 1);
\draw[orange(ryb), dashed, line width = 0.5mm] (0, 1) circle (0.5);
\end{tikzpicture}
\end{cases} = \begin{cases}
    +1 \text{ if } v \in A \text{ sublattice} \\
    -1 \text{ if } v \in B \text{ sublattice}
\end{cases} 
\end{equation}
where the small orange loop operator counts the number of outgoing arrows subtracted by the number of incoming arrows.
As a consequence, the number of outgoing arrows subtracted by the number of incoming arrows from any closed loop will equal the number of $A$ sublattices subtracted from the number of $B$ sublattices enclosed which can be any integer.
Consequently, the emergent gauge theory is a compact $U(1)$ gauge theory \cite{Baskaran88,ReadSachdev_PRB,ReadSachdev1989PRL,READSACHDEV_1989_Nuc,Fradkin90}.

We will again be interested in the resonating valence bond state given by the equal weight and equal phase superposition of all dimer configurations:
\begin{equation}
    \ket{\text{RVB}} = \ket{\includegraphics[width = 40 pt, valign = c]{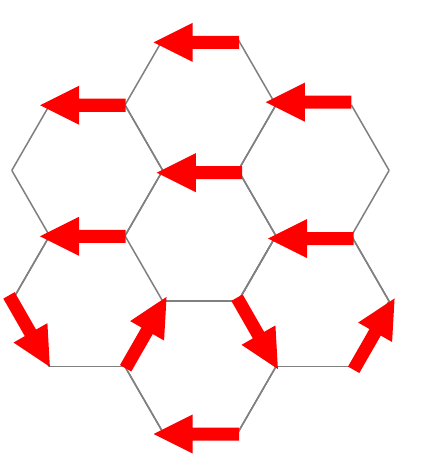}} + \ket{\includegraphics[width = 40 pt, valign = c]{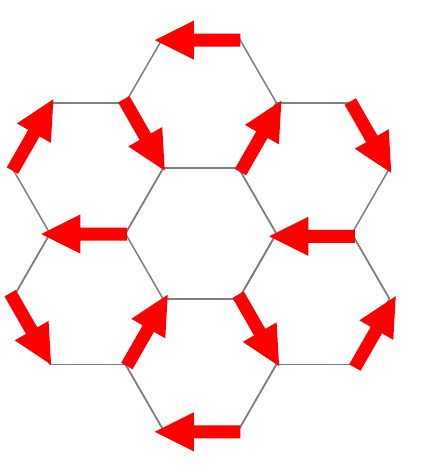}} + \ket{\includegraphics[width = 40 pt, valign = c]{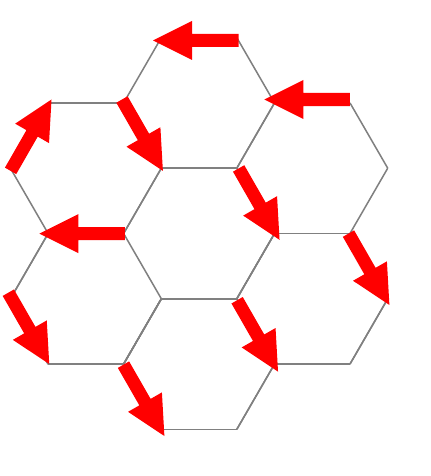}} + \cdots \label{eq-U1RVB}
\end{equation}
As before, this RVB state corresponds to the deconfined phase of this gauge theory, analogous to electromagnetism, and is a fixed point representative of the $U(1)$ QSL.

Unlike the $\mathbb Z_2$ QSL, this $U(1)$ QSL has algebraic correlations. In two spatial dimensions, it is known that these gapless excitations make this into a fine-tuned point, which is unstable to generic deformations \cite{ReadSachdev1989PRL,READSACHDEV_1989_Nuc,ReadSachdev_PRB,SachdevED}.
This is well-illustrated by the Rokshar-Kivelson (RK) model \cite{RK}, which both on the (bipartite) square \cite{RK} and honeycomb lattices \cite{RKhoneycomb} admit exactly solvable
RK 
points where the ground state is this RVB state, surrounded by nearby gapped phases.

Hence, in $2+1$D one does not expect to observe a pure $U(1)$ spin liquid without excessive fine-tuning.
In particular, although we are primarily interested in dynamical preparation, we take a moment to establish our expectations for the equilibrium phase diagram of the Rydberg model in the honeycomb dimer regime ($a<R_b<\sqrt{3} a$; $\delta > 0$).
In this regime, we would expect that the long-range nature of the Rydberg interaction would push dimers onto opposite sides of a hexagon, giving rise to the staggered or columnar ground state of the honeycomb RK model \cite{RKhoneycomb} (i.e., the regime $t<V$ of the RK model).
It is worth noting that a recent work~\cite{Samajdar_2021} studied the equilibrium properties of Rydberg atoms on the Kagome lattice.
While they mainly focused on larger blockade radii and approximate triangular, rather than honeycomb, dimer models, they did indeed report a solid phase for $R_b\sim 1.2 a$ that matches the staggered phase expected from the preceding arguments.
%
While such a ground state is far from the $U(1)$ QSL state of Eq.~\eqref{eq-U1RVB}, we will now see that, in contrast to the equilibrium physics of the Rydberg model, \emph{dynamics} can lead to a robust \emph{$U(1)$ quantum spin lake}

\subsection{From Honeycomb to Bethe Lattice and Numerical Implementation}

To explore the dynamical preparation of a $U(1)$ quantum spin lake in the above model, we perform the same approximation as in Sec.~\ref{sec-Tree} by taking the plaquette size to infinity. This makes the problem more easily numerically tractable and moreover defines a limit where the energy scale of the magnetic particles (or visons or fluxons) is effectively zero. In addition to its conceptual value, we discussed in Sec.~\ref{sec-Tree} how this drastic approximation can, in fact, offer a good simulation for a realistic experiment on a genuine 2D lattice for timescales where we are fast with respect to the magnetic excitations. Indeed, for a dimer model on the honeycomb lattice, the dimer resonances occur at the same order as for the kagome dimer model studied in Eq.~\eqref{eq-Rydberg-Resonances}, which thus occurs at sixth order in perturbation theory. Moreover, similar to our analysis in Section~\ref{subsubsec-EffectofLRtails} and Section~\ref{subsubsec-NumericalEstimates}, one can show that the $\sim 1/r^6$ tails of the Rydberg interactions only lead to small splittings in the dimer subspace.
For example, by repeating a version of the analysis performed in Section~\ref{subsubsec-EffectofLRtails}, we found that the induced splittings are on the order of $\sim 10^{-2} \times \Omega$ for $R_b = 1.2 a$.

More precisely, the tree version of the honeycomb lattice is called the Bethe lattice \cite{bethe1935statistical} and is shown in Fig.~\ref{fig:U1QSL}(a). In particular, this is for the coordination number $z=3$. It is known that the correlation length for physical states (i.e. cat states excluded) is restricted to be below $ \xi = 1/\log(z - 1)$ (owing to the $(z-1)$-fold branching rate of the tree) with gapless systems saturating this bound \cite{Laumann09, nagy2012simulating}. Hence, although the RVB dimer wavefunction on the honeycomb lattice \eqref{eq-U1RVB} has algebraic correlations (and thus an infinite correlation length), the deconfined phase of $U(1)$ gauge theory on the Bethe lattice will have $\xi = 1/\log 2$.

To numerically simulate the Rydberg model on the Bethe lattice, we first start by doubling each qubit degree of freedom on the Bethe lattice:
\begin{equation}
    \begin{tikzpicture}[scale = 0.7, baseline={([yshift=-.5ex]current bounding box.center)}]
        \draw[gray] (0, 0) -- (0, 1) -- (0.86602540378, 1 + 1/2);
        \draw[gray] (0, 1) -- (-0.86602540378, 1 + 1/2);
        \draw[gray] (0.86602540378, -1/2) -- (0, 0) -- (-0.86602540378, -1/2);
        \draw[black, fill = black] (0,0.75) circle (0.075);
        \draw[black, fill = black] (0,0.25) circle (0.075);
        \draw[stealth-stealth, black] (0.3, 0.25) -- (0.3, 0.75);
        \draw[black, fill = black] (0.86602540378 * 3/4, -1/2 * 3/4) circle (0.075);
        \draw[black, fill = black] (0.86602540378 * 1/4, -1/2 * 1/4) circle (0.075);
        \draw[black, fill = black] (-0.86602540378 * 3/4, -1/2 * 3/4) circle (0.075);
        \draw[black, fill = black] (-0.86602540378 * 1/4, -1/2 * 1/4) circle (0.075);
        \draw[black, fill = black] (0.86602540378 * 3/4, 1+1/2*3/4) circle (0.075);
        \draw[black, fill = black] (0.86602540378 * 1/4, 1+1/2*1/4) circle (0.075);
        \draw[black, fill = black] (-0.86602540378 * 3/4, 1+1/2*3/4) circle (0.075);
        \draw[black, fill = black] (-0.86602540378 * 1/4, 1+1/2*1/4) circle (0.075);
    \end{tikzpicture}
\end{equation}
Subsequently, we add to our model strong Ising ferromagnetic interactions between our doubled qubits so that at energies below that scale, the system behaves like the undoubled system.
With this doubling and the blockade constraint, we can utilize the tree tensor network ansatz of Eq.~\eqref{eq-TTN} to encode the wavefunction of our system.
The physical legs of the $A$ tensor will encode the following states:
\begin{equation}
    d = \quad \begin{tikzpicture}[scale = 0.5, baseline={([yshift=-.5ex]current bounding box.center)}]
    \draw[gray] (0.86602540378, -1/2)--(0, 0) -- (-0.86602540378, -1/2);
\draw[gray] (0, 0) -- (0, 1);
    \node at (0, 0) {\normalsize $0$};
    \end{tikzpicture}
    \quad \begin{tikzpicture}[scale = 0.5, baseline={([yshift=-.5ex]current bounding box.center)}]
    \draw[gray] (0.86602540378, -1/2)--(0, 0) -- (-0.86602540378, -1/2);
\draw[gray] (0, 0) -- (0, 1);
    \draw[-stealth, red, line width = 0.75 mm] (0, 0) -- (-0.86602540378, -1/2);
    \node at (0, 0) {\normalsize $1$};
    \end{tikzpicture}
    \quad \begin{tikzpicture}[scale = 0.5, baseline={([yshift=-.5ex]current bounding box.center)}]
    \draw[gray] (0.86602540378, -1/2)--(0, 0) -- (-0.86602540378, -1/2);
\draw[gray] (0, 0) -- (0, 1);
    \draw[-stealth, red, line width = 0.75 mm] (0, 0) -- (0, 1);
    \node at (0, 0) {\normalsize $2$};
    \end{tikzpicture}
    \quad \begin{tikzpicture}[scale = 0.5, baseline={([yshift=-.5ex]current bounding box.center)}]
    \draw[gray] (0.86602540378, -1/2)--(0, 0) -- (-0.86602540378, -1/2);
\draw[gray] (0, 0) -- (0, 1);
    \draw[-stealth, red, line width = 0.75 mm] (0, 0) -- (0.86602540378, -1/2);
    \node at (0, 0) {\normalsize $3$};
    \end{tikzpicture}
\end{equation}
Similarly, the physical legs of the $B$ tensor will encode: 
\begin{equation}
    d = \quad \begin{tikzpicture}[scale = 0.5, baseline={([yshift=-.5ex]current bounding box.center)}]
    \draw[gray] (0.86602540378, 1 + 1/2)--(0, 1) -- (-0.86602540378, 1+1/2);
\draw[gray] (0, 0) -- (0, 1);
    \node at (0, 1) {\normalsize $0$};
    \end{tikzpicture}
    \quad \begin{tikzpicture}[scale = 0.5, baseline={([yshift=-.5ex]current bounding box.center)}]
    \draw[gray] (0.86602540378, 1 + 1/2)--(0, 1) -- (-0.86602540378, 1+1/2);
\draw[gray] (0, 0) -- (0, 1);
    \draw[stealth-, red, line width = 0.75 mm] (0, 1) -- (0.86602540378, 1+1/2);
    \node at (0, 1) {\normalsize $1$};
    \end{tikzpicture}
    \quad \begin{tikzpicture}[scale = 0.5, baseline={([yshift=-.5ex]current bounding box.center)}]
    \draw[gray] (0.86602540378, 1 + 1/2)--(0, 1) -- (-0.86602540378, 1+1/2);
\draw[gray] (0, 0) -- (0, 1);
    \draw[-stealth, red, line width = 0.75 mm] (0, 0) -- (0, 1);
    \node at (0, 1) {\normalsize $2$};
    \end{tikzpicture}
    \quad \begin{tikzpicture}[scale = 0.5, baseline={([yshift=-.5ex]current bounding box.center)}]
    \draw[gray] (0.86602540378, 1 + 1/2)--(0, 1) -- (-0.86602540378, 1+1/2);
\draw[gray] (0, 0) -- (0, 1);
    \draw[stealth-, red, line width = 0.75 mm] (0, 1) -- (-0.86602540378, 1+1/2);
    \node at (0, 1) {\normalsize $3$};
    \end{tikzpicture}
\end{equation}
Numerical method in hand, we can now test the dynamical preparation of a $U(1)$ spin lake by simulating trotterized time-evolution as in the $\mathbb{Z}_2$ tree case.

\subsection{Dynamical Preparation of $U(1)$ Spin Lake}

We now once again initialize our system in the Higgs phase and start with $\delta= -10$ and $\Omega = 0$ (effectively starting with a product state on the Bethe lattice).
By linearly ramping up the value of $\delta$ to $\delta = 10$ and turning up the value of $\Omega$ in the same way as Fig.~\ref{fig:Z2Tree}(b), we sweep from the ground state of the Higgs phase into the region of the phase diagram with an emergent gauge theory.
Our expectation is that the final state that we prepare will be the initial state with Gauss law violations projected out following Eq.~\eqref{eq-sweeping-proj}.
Similar to the $\mathbb{Z}_2$ case of Sec.~\ref{sec-Summary}~and~\ref{sec-Tree}, we can make the mean-field ansatz for the initial state of the sweep after $\Omega$ has been ramped up to $1$ given by Eq.~\eqref{eq-RydbergMFansatz}.
Subsequently, when we project out Gauss law violations, the resulting state will be the RVB (for the same reasons as Eq.~\eqref{eq-RydbergProj} which is the fixed point state of the $U(1)$ QSL.
On the infinite tree, at any finite rate, we will prepare a finite-size quantum spin lake with the size of the lake getting larger with decreasing rate.
We can diagnose the onset of the $U(1)$ quantum spin lake numerically through three probes.

First, to verify that our final state satisfies the dimer constraint, we plot the density $n_{\alpha} = (1 + Z_{\alpha})/2$ of the three qubits on site $A$ (equivalently $B$ as the qubits in $B$ are ferromagnetically locked to the qubits in $B$) as a function of time [Fig.~\ref{fig:U1QSL}(b)].
We find that the density of each qubit saturates to $1/3$ as we approach the end of the sweep implying a maximal packing of dimers.
As such, we know that the final state satisfies the Gauss law and can conclude that it is rotationally invariant.
This leaves open the possibility of either the $U(1)$ QSL or a rotationally invariant valence bond solid (VBS) state---a symmetry breaking state consisting of a finite superposition of dimer states.

Next, in Fig.~\ref{fig:U1QSL}(c), we plot the entanglement across the central bond of the tree tensor network state as a function of the time along the sweep.
We see that the entanglement saturates to around $\log(2)$ towards the end of the sweep.
This is consistent with the fixed point entanglement of the RVB state of the $U(1)$ QSL and also with a cat state of two VBS configurations.

To distinguish the VBS cat state from the $U(1)$ QSL, we now turn our attention to the behavior of correlation functions in the final state of the sweep.
The prediction is that, since the $U(1)$ QSL is gapless, on a tree lattice, it will decay with the maximal possible correlation length for physical (non-cat) states, $\xi = 1/\log(2)$.
On the other hand, the VBS cat state is predicted to have a correlation length larger than $\xi > 1/\log(2)$.
To distinguish these two cases, we compute the following correlation function for the final state of the dynamical sweep $C(x) = |\langle [\bar{n}(x + 1) - \bar{n}(x)] [\bar{n}(1) - \bar{n}(0)]|$, as a function of the total time of the sweep and distance, where $\bar{n}(x) = \sum_{\alpha} n_{\alpha}(x)/3$ is the average density of dimers at vertex $x$ of the Bethe lattice.
Such a correlation function is numerically convenient because it manifestly tends to a zero value at long distances (due to the translation invariance of our tree tensor network ansatz) and hence enables us to probe the correlation function at long distances without being mired by numerical errors in the value of the one-point function.
The results of this correlation function are shown in Fig.~\ref{fig:U1QSL}(d).
We find that as we increase the time of our sweep, the correlation function goes from decaying faster than $1/\log(2)$ to saturating at a $1/\log(2)$ decay (see Appendix~\ref{app-U1}).
This is strong evidence signaling the onset of the gapless $U(1)$ quantum spin lake!

\section{Outlook}

We have considered systems containing two \textit{emergent} low-energy degrees of freedom---such as the $e$- and $m$-anyons of the $\mathbb Z_2$ spin liquid---whose dynamics are controlled by well-separated energy scales.
In such a setting, we first demonstrated that there exists a dynamical regime wherein one is nearly in equilibrium (quasi-adiabatic) relative to one degree of freedom while out-of-equilibrium relative to the other (sudden).
A remarkably clean observational signature of being in this regime is given by an ``echo'' experiment wherein one sweeps back and forth as in Fig.~\ref{fig-dTC}(e): this displays a revival that is distinct from the purely-adiabatic or purely-sudden regime.

While the existence of such a non-equilibrium regime is already of interest, it is even more interesting that this dynamical regime can be exploited to help realize exotic quantum states.
In particular, consider a protocol wherein one sweeps parameters in the Hamiltonian between two different phases, where the quasi-adiabatic degree of freedom goes from being condensed in the ground state to not being condensed.
Then, dynamics in the aforementioned regime can be used to effectively implement a projection operator (corresponding to adiabatically pushing out only one degree of freedom, whilst leaving the other unaffected).
We demonstrated that if one starts in an initial product state (where an anyonic degree of freedom was condensed), then this novel non-equilibrium regime can prepare a QSL-like state by ``projecting'' into a constrained subspace of the Hilbert space.
We extensively illustrated this for $\mathbb Z_2$ spin liquids, both in the toric code as well as the Rydberg ruby lattice settings (whose constrained spaces are loop and dimer states, respectively).

Equally important as recognizing the existence of the preparation scheme above is understanding its limitations.
In particular, since the preparation scheme involves sweeping parameters in the Hamiltonian between two phases, the separation of energy scales between the two degrees of freedom required for the foregoing non-equilibrium regime can only be guaranteed for a finite range of system sizes.
For larger systems, Kibble-Zurek-type considerations imply that projection only takes place of a finite length scale creating a finite-size ``quantum spin lake'' instead of a full QSL.
Despite this limitation, the current range of available system sizes and coherence times in near-term quantum devices make quantum spin lakes a very promising alternative to traditional ground state preparation.

This mechanism gives a variety of exciting new directions to explore.
We have already highlighted how our protocol even allows us to approximately realize a $U(1)$ spin liquid as a honeycomb dimer model, which is all the more remarkable given that this does not arise as a stable ground state in two spatial dimensions.
We have argued that the dynamical preparation is achievable in Rydberg atom tweezer arrays, using the same ingredients as already demonstrated in the ruby lattice experiment \cite{Semeghini21}.
Moreover, we argued and demonstrated how simulating the dynamical protocol on a tree gives a new handle on matching experimental data.

One striking aspect of our mechanism is that it suggests that \emph{larger plaquettes are better}.
Indeed, these lead to smaller energy scales for the magnetic excitations, making it easier to be sudden with respect to them.
For the two Rydberg-related models we discussed---namely a $\mathbb Z_2$ ($U(1)$) spin liquid for a kagome (honeycomb) dimer model---these plaquettes were already sizable, consisting of six bonds.
However, it would be interesting to explore lattices with larger plaquettes. E.g., instead of placing Rydberg atoms on the bonds of the kagome, they can be placed on the bonds of the Fisher (or star) lattice, which is a decorated version of the former.
In addition, going to 3D naturally suggests the hyperkagome lattice. In fact, the tree geometry we studied is the limit of infinitely large plaquettes.
More generally, one could explore hyperbolic lattices, which allow for arbitrary plaquette size, as captured by the Schl\"afli symbol \cite{coxeter1973regular}.

With regard to further possible generalizations, we note that the broader context of our work is the non-equilibrium dynamics of gauge theories. A special feature here is the interplay of charge and flux dynamics, which can be decoupled to an extent in this non-equilibrium context.
Thus, we are led to our ``two-time'' criteria, representing the independent equilibration times of the charge and the flux. In this language, our mechanism should also apply to other gauge groups, such as $\mathbb Z_3$, of which there exist interesting ground state proposals \cite{Motrunich02,Z3} where our dynamical mechanism might provide a route toward their realization.
Similarly, it would be interesting to explore the potential applicability of our mechanism in the context of deconfined gauge theories obtained by local two-body interactions using combinatorial gauge symmetry \cite{Chamon20,Wu21,Green22}.

A fascinating question for future study is how these results extend to non-Abelian gauge theories. Discrete non-Abelian gauge groups lead to excitations with non-Abelian statistics. Their classical counterparts---non-Abelian defects in ordered media---can lead to glassy dynamics \cite{NelsonDefects}. Are there analogs in the quantum dynamics of non-Abelian gauge theories? Looking further afield, the dynamics of Yang-Mills theories is clearly of prime importance in a variety of situations  including the early universe \cite{mukhanov_2005}. Detailed studies of real time dynamics of gauge fields in concrete models that can be experimentally realized, are likely to make significant contributions to this important area.

Even beyond realizing gauge theories, there is likely a broad range of applications of our protocol for dynamically implementing a projection operator using a non-equilibrium sweep. One tantalizing option is to start with a free-fermion state and dynamically implement the Gutwziller projection. Combining our non-equilibrium protocol with the technical ingredients introduced in Ref.~\cite{Kale22} could lead the way to implementing this idea.

In conclusion, quantum spin lakes present an exciting new interplay between non-equilibrium dynamics, topological states, and NISQ devices. While we have focused here on state preparation, it would be worthwhile to explore the potential quantum-information-theoretic applications of quantum spin lakes. Moreover, this new non-equilibrium regime might be interesting to explore in its own right, offering a new handle on the rich phenomenology of quantum dynamics.

\section{Acknowledgements}

The authors would like to thank Dominic Else, Manuel Endres, Ruihua Fan, Giuliano Giudici, David Huse, Marcin Kalinowski, Misha Lukin, Francisco Machado, Nishad Maskara, Dan Parker, Hannes Pichler, Drew Potter, Saran Prembabu, and Ryan Thorngren for illuminating discussions.
DMRG simulations were performed on the Harvard FASRC facility using the TeNPy Library \cite{Hauschild18}, which was
inspired by a previous library \cite{Kjaell13}.
RS acknowledges support by the U.S. Department of Energy, Office of Science, Office of Advanced Scientific
Computing Research, Department of Energy Computational Science Graduate Fellowship under Award Number
DESC0022158.
RV is supported by the Harvard Quantum Initiative Postdoctoral Fellowship in Science and Engineering, and RV and AV by the
Simons Collaboration on Ultra-Quantum Matter, which is a grant from the Simons Foundation (651440, AV).
A.V. further acknowledges support by NSF-DMR 2220703.

\bibliography{refs.bib}

\appendix

\section{Deformed Toric Code Model Numerics}

In this appendix, we provide some additional details regarding the numerics reported in Section~\ref{sec-DTC} on the deformed toric code model of Eq.~\eqref{eq-TCf_2}.
For convenience, we reiterate the model takes the following form:
\begin{equation} \label{eq-appTCf_2}
    H_{\text{TC + f}} = -K\sum_v \begin{tikzpicture}[scale = 0.5, baseline = {([yshift=-.5ex]current bounding box.center)}]
    \draw[gray] (1.5,0) -- (-1.5, 0);
    \draw[gray] (0,1.5) -- (0,-1.5);
    \node at (0.75, 0) {\normalsize $Z$};
    \node at (-0.75, 0) {\normalsize $Z$};
    \node at (0, 0.75) {\normalsize $Z$};
    \node at (0, -0.75) {\normalsize $Z$};
\end{tikzpicture} - h_x \sum_{\ell} X_{\ell} - h_z \sum_{\ell} Z_{\ell}
\end{equation}
where, in our numerics, we always take $h_z/h_x \ll 1$.
Our numerics for this model are all performed by using infinite cylinder matrix product state algorithms which are implemented in the Tensor Network Python (TeNPy) numerical package \cite{Hauschild18}.

This appendix is organized into two parts.
In Appendix~\ref{app-GSsim}, we provide numerical details involved in extracting the ground state phase diagram of the model above, reported in Fig.~\ref{fig-dTC}(a).
In Appendix~\ref{app-dtcdynamics}, we provide numerical details for the dynamical sweep simulations reported in the main text.

\subsection{Additional Details for Ground State Simulations} \label{app-GSsim}

In Section~\ref{sec-DTC}, we found the ground state phase diagram of the model of Eq.~\ref{eq-appTCf_2}.
This phase diagram was found by simulating the aformentioned model using the infinite cylinder density matrix renormalization group (DMRG).
Unless otherwise reported, the numerics we report for the ground state phase diagram both here and in the main text are performed for a cylinder circumference $L_y = 4$ and $\chi = 64$.
To clarify conventions, since qubits living on the links of the square lattice have a two-site unit-cell:
\begin{align*}
\begin{tikzpicture}[scale = 0.7, baseline={([yshift=-.5ex]current bounding box.center)}]
\foreach \i in {0,...,0} {
    \draw[gray] (\i, -0.5) -- (\i, 1);
    \draw[gray] (-0.5, \i) -- (1, \i);
    \foreach \j in {0,...,0} {
        \filldraw[blue] (\i, \j + 0.5) circle (2 pt);
        \filldraw[red] (\i+ 0.5, \j) circle (2 pt);
    }
}
\end{tikzpicture}
\end{align*}
$L_y = 4$, corresponds to $8$ qubits around the cylinder.
Here, we report how we identify the different phases of the model and how critical points are extracted.
We further report why we expect the qualitative features of the phase diagram not to vary as a function of matrix product state cylinder circumference $L_y$.

\subsubsection{Identification of Phases}\label{app-Phase-and-CP-extraction}

We identify the phases in Fig.~\ref{fig-dTC}(a) by using the FM order parameter [defined in Eq.~\eqref{eq-FMOP}].
The flow of the FM order parameter downwards with increased string length is a precise indicator of a topologically ordered phase.
A large non-zero value of $\langle Z \rangle_{FM}$ ($\langle X \rangle_{FM}$) is a nice heuristic for defining the confined (Higgs) phase but we remark that since these phases can be adiabatically connected to one another, they cannot be sharply distinguished.

\begin{figure}
    \centering
    \includegraphics[width = 240pt]{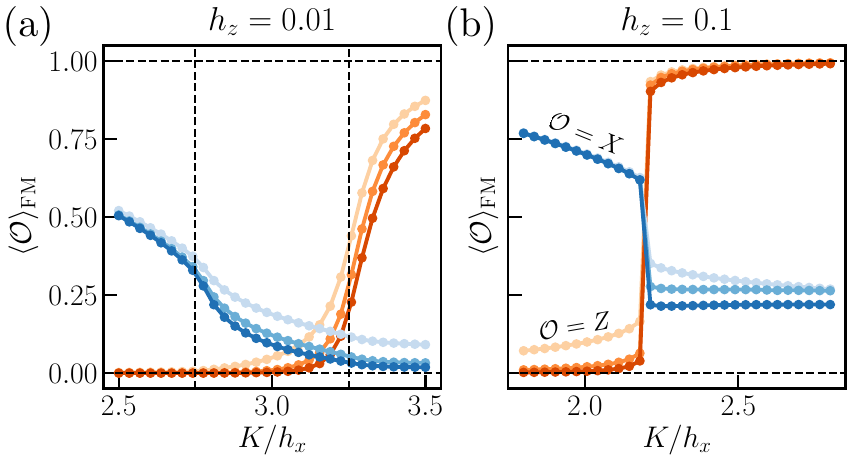}
    \caption{\textbf{Fredenhagen-Marcu Order Parameter for Ground State Phase Detection.}
    Here, we show the FM order parameter for different string lengths with darker colors indicated longer string length (See insets of Fig.~\ref{fig-FMdTC} for precise string configuration).
    In (a), we show the FM order parameter computed as a function of $K$ for $h_z = 0.01$. Here, we find an intermediate regime $2.75 \lesssim K/h_x \lesssim 3.25$ where both $\langle X \rangle_{\text{FM}}$ and $\langle Z \rangle_{\text{FM}}$ are small and decay to zero with increased string length. To the left of this region, $\langle X \rangle_{FM}$ does not decay with distance suggesting a trivial Higgs phase. To the right, $\langle Z \rangle_{FM}$ is large suggesting a trivial confined phase. In (b), we show the same plot but now for $h_z = 0.1$. We find that there is no region where both order parameters decay to zero with increased string length. Instead, we find that when $K/h_x \lesssim 2.2$, $\langle X \rangle_{FM}$ is large and when $K/h_x \gtrsim 2.2$, $\langle X \rangle_{FM}$ is large with the value of these jumping sharply across $K/h_x = 2.2$. This suggests a first-order transition between a Higgs and confined phase. }
    \label{fig:app-FM-GS}
\end{figure}

We report the value for the FM order parameter for two representative cuts through the phase diagram and for the three possible string lengths possible on a cylinder of length $L_y = 4$ (See insets of Fig.~\ref{fig-FMdTC} for specific configurations of string operators).
First at $h_z = 0.01$ [Fig.~\ref{fig:app-FM-GS}(a)]---where we reported the existence of three phases (Higgs, Deconfined, Confined) separated by two second order phase transitions---and then at $h_z = 0.1$ [Fig.~\ref{fig:app-FM-GS}(b)] ---where we reported only two phases (Higgs and Confined) separated by first-order phase transitions.

For $h_z = 0.01$, we find three regimes in the behavior of $\langle Z  \rangle_{\text{FM}}$ and $\langle X \rangle_{\text{FM}}$ consistent with the presence of the Higgs, deconfined, and confined phase
In particular, we find that for $K/h_x \lesssim 2.75$, the value of $\langle X \rangle_{FM}$ does not decay with increased FM string length, suggesting a Higgs phase.
For $2.75 \lesssim K/h_x \lesssim 3.25$, we find that both  FM order parameters are small and decay to zero suggesting a deconfined phase.
Finally, when $ K/h_x \gtrsim 3.25$, the value of $\langle Z \rangle_{FM}$ rapidly increases suggesting the existence of a confined phase.
Although, in this regime $\langle Z \rangle_{FM}$ decreases with increased string length, we expect that if we were able to go to longer string lengths, it would saturate to a non=zero value.
To sharply identify the presence of three phases in this regime of $h_z$, we use the correlation length (see next subsubsection).

For $h_z = 0.1$, we find two regimes in the behavior of the FM order parameters.
Namely, when $K/h_x \lesssim 2.2$, $\langle X \rangle_{FM}$ suggesting a Higgs phase.
Conversely, as we increase $K/h_x$, the values of the FM order parameters jumps discontinuously with $\langle Z \rangle_{FM}$ becoming large.
This suggests a confined phase for $K/h_x \gtrsim 2.2$.

\subsubsection{Extraction of Critical Points}

Having identified the phases in our phase diagram, we now comment on how we extract phase boundaries between such phases.
For the portion of the phase diagram where we identify only two phases, we utilize a jump in the FM order parameter to diagnose the location of the first order phase transition [see Fig.~\ref{fig-FMdTC}(b)].

\begin{figure}
    \centering
    \includegraphics[width=150pt]{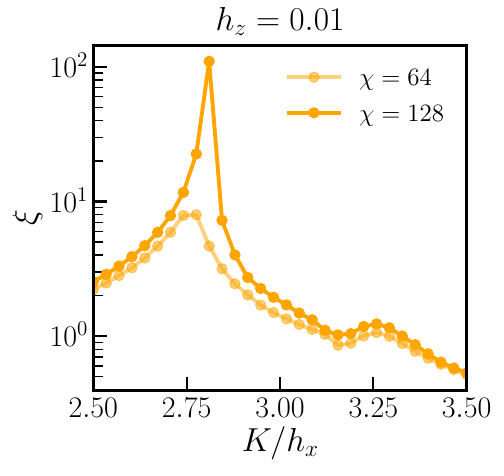}
    \caption{\textbf{Extraction of Critical Points from Correlation Length Peaks.} To extract the location of the two second-order phase transitions at low values of $h_z/h_x$ ($\lesssim 0.025$), we look for peaks in the correlation length. We find two such peaks as we would expect. }
    \label{fig:app_xiLy4}
\end{figure}

For the portion of the phase diagram where we identify three phases, we utilize the correlation length $\xi = \frac{1}{2 L_y} \frac{1}{\log(\lambda_2)}$ computed from the largest non-trivial transfer matrix eigenvalue of the matrix product state (where the inverse factor of $1/(2L_y)$ comes from dividing by the number of qubits around the circumference).
Since the transition between the Higgs and deconfined phase and deconfined and confined phase must be second order, the correlation length, in the infinite bond-dimension limit should, should diverge at the transitions.
At finite bond-dimensions, we expect that the correlation length should increase as $\log(\xi) \propto \log(\chi)$ \cite{Pollmann09} at these critical points.
For our purposes, to simply identify the location of the second-order phase boundaries, we look for locations $K/h_x$ where the correlation displays a peak and increases with bond dimension.
The results for the representative value of $h_z = 0.01$ are shown in Fig.~\ref{fig:app_xiLy4}.

\subsubsection{Phase Boundaries at $L_y = 5$} \label{app-Ly5}

\begin{figure}
    \centering
    \includegraphics[width = 247pt]{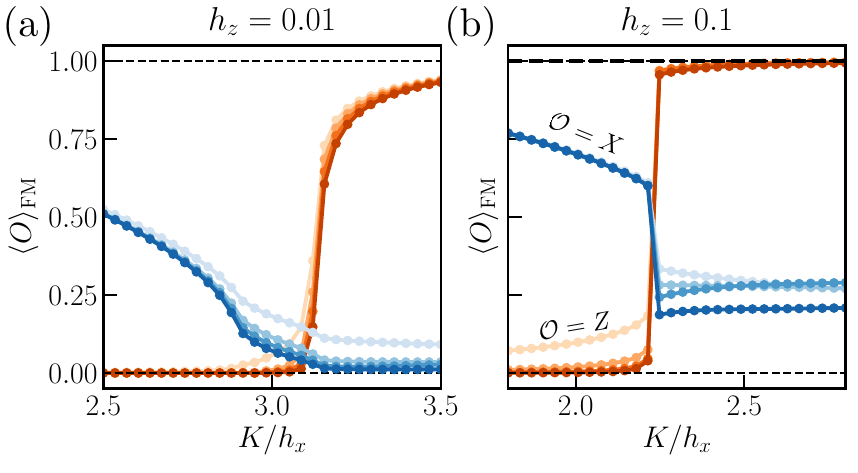}
    \caption{\textbf{Fredenhagen-Marcu Order Parameter at $L_y = 5$.} Here, we show the value of the FM order parameter for two values of $h_z$ at $L_y = 5$. In (a), we show the FM order parameters for $h_z = 0.01$ where we find three distinct regimes of the behaviors of the two FM order parameters corresponding Higgs, deconfined, and confined from left to right. In (b), we show the FM order parameters for $h_z = 0.01$ where we find two distinct behaviors corresponding to Higgs and confined. Both of these cases are analogous to their counterparts at $L_y = 4$.}
    \label{fig:app-FM-Ly5}
\end{figure}

\begin{figure}
    \centering
    \includegraphics[width = 150pt]{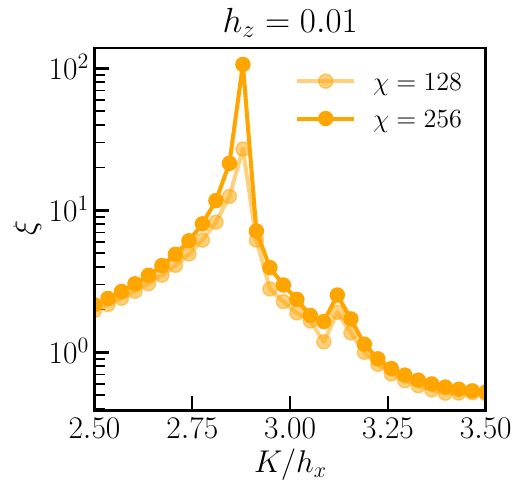}
    \caption{\textbf{Extraction of Critical Points from Correlation Length at $L_y = 5$.} Here, we demonstrate that the correlation length at $L_y = 5$ is able to detect the presence of two transitions at low values of $h_z$ in the same way as $L_y = 4$.}
    \label{fig:app-xi-Ly5}
\end{figure}

We conclude by reporting numerical results for the phase boundaries at $L_y = 5$ to show that the phases we find are qualitatively unchanged.
In particular, we first report the value of the FM order parameter for two values of $h_z/h_x$  ($h_z/h_x = 0.01$ and $h_z/h_x = 0.1$) in Fig.~\ref{fig:app-FM-Ly5}.
For $h_z/h_x = 0.01$ [Fig.~\ref{fig:app-FM-Ly5}(a)], we find three regimes in the order parameter corresponding to Higgs, deconfined, and confined (from left to right) similar to the $L_y = 4$ case.
For $h_z/h_x = 0.1$ [Fig.~\ref{fig:app-FM-Ly5}(b)], we find two regimes corresponding to Higgs and confined, once again qualitatively similar to the $L_y = 4$ case.
Moreover, we show that peaks in the correlation length can once again be used to detect the phase boundaries as was the case at $L_y = 4$.
The results for the correlation length are shown in Fig.~\ref{fig:app-xi-Ly5} where we once again find two peaks in the correlation length.

\subsection{Additional Details for Sweep Dynamics Simulations} \label{app-dtcdynamics}

In Section~\ref{sec-DTC}, we initialized the deformed toric code model of Eq.~\eqref{eq-TCf_2} in its product state ground state at $K = 0$ and small $h_z$ (in the `Higgs phase' of Fig.~\ref{fig-dTC}(a)).
We then linearly increased $K$ at a rate $1/T$ and investigated the nature of the final state.
We found that there was a window of total times $[T_e, T_m]$ wherein the final state of the sweep had high overlap with the initial state of the sweep with Gauss law violations projected out, $\mathcal{P}_G \ket{\psi(0)}$.
In this subsection, we provide additional details for the sweep dynamics numerical simulations performed.
The dynamics numerics performed for Eq.~\ref{eq-appTCf_2} were done by using the so-called ``$W_2$'' method for time-evolving a matrix product state developed in Ref.~\onlinecite{Zaletel15}.
Unless stated otherwise, numerics reported here and in the main text are done at $\chi = 256$ and with a trotter step size of $dt = 0.0025$, with convergence in both checked but not shown here. 
In what follows, we describe how the overlap between the final state of a dynamical sweep and the overlap with the projected state is practically computed in our numerics and demonstrate that at the longest sweep times at $h_z = 0.1$, our final state starts to resemble the ground state.
Finally, we provide some additional numerics for the time dynamics of the Fredenhagen-Marcu order parameter depicted in Fig.~\ref{fig-FMdTC}.

\subsubsection{Applying Gauss Law Projection on Matrix Product State}

In Fig.~\ref{fig-dTC}(d) in Section~\ref{sec-DTC}, we plotted the overlap density between the matrix product state at the end of a dynamical sweep with the initial state with Gauss law violations projected out.
Here, we detail our method for implementing this projection operator in numerics.

First, consider a wavefunction $\ket{\psi}_{\ell} \otimes \bigotimes_v \ket{-}_v$ defined on the Lieb lattice (square lattice with qubits on the links and vertices): 
\begin{align*}
\begin{tikzpicture}[scale = 0.7, baseline={([yshift=-.5ex]current bounding box.center)}]
\foreach \i in {0,...,1} {
    \draw[gray] (\i, -0.5) -- (\i, 1.5);
    \draw[gray] (-0.5, \i) -- (1.5, \i);
    \foreach \j in {0,...,1} {
        \filldraw[red] (\i, \j) circle (2 pt);
        \filldraw[blue] (\i - 0.5, \j) circle (2 pt);
        \filldraw[blue] (\i, \j - 0.5) circle (2 pt);
        \filldraw[blue] (\i, \j + 0.5) circle (2 pt);
        \filldraw[blue] (\i+ 0.5, \j) circle (2 pt);
    }
}
\end{tikzpicture}
\end{align*}
where $\ket{\psi}_{\ell}$ is defined on the link qubits (shown in blue) and the $\bigotimes_v \ket{-}_v$ is defined on the vertex qubits (shown in red).
By evolving the qubits above by the following Hamiltonian:
\begin{align}
H = \sum_v \begin{tikzpicture}[scale = 0.6, baseline={([yshift=-.5ex]current bounding box.center)}] 
\draw[gray] (-1, 0) -- (1, 0);
\draw[gray] (0, -1) -- (0, 1);
\node at (0,0) {\normalsize $Z$};
\node at (0,-0.5) {\normalsize $Z$};
\node at (-0.5,0) {\normalsize $Z$};
\end{tikzpicture}
\ -\ \begin{tikzpicture}[scale = 0.6, baseline={([yshift=-.5ex]current bounding box.center)}] 
\draw[gray] (-1, 0) -- (1, 0);
\draw[gray] (0, -1) -- (0, 1);
\node at (0,0) {\normalsize $Z$};
\node at (0,0.5) {\normalsize $Z$};
\node at (0.5,0) {\normalsize $Z$};
\end{tikzpicture}
\end{align}
for a time $t = \pi/4$ and then projecting the vertex qubits into the $\ket{-}$ state, we will implement $\mathcal{P}_G = \prod_{v} (1 + G_v)/2$.
To see why this is the case, note that for a single vertex:
\begin{align}
&\bra{-} e^{i \frac{\pi}{4} Z_v Z_{\ell_3} Z_{\ell_4}} e^{-i\frac{\pi}{4} Z_v Z_{\ell_1} Z_{\ell_2}} \ket{-}\\
&= \frac{1}{2} \bra{-} ( 1 + i Z_v Z_{\ell_3} Z_{\ell_4}) ( 1 - i Z_v Z_{\ell_1} Z_{\ell_2}) \ket{-}\\
&= \frac{1}{2} \left(1  - Z_{\ell_1} Z_{\ell_2} Z_{\ell_3} Z_{\ell_4}  \right) =  \frac{1 + A_v}{2}
\end{align}
where $\ell_1, \cdots \ell_4$ clockwise label the link qubits that neighbor the qubit at vertex $v$ (with $\ell_1$ labeling the qubit directly below).
Since the Hamiltonian $H$ is fully commuting and the intial vertex qubits are unentangled, time evolving under $H$ for $t = \pi/4$ then projecting vertices to $\ket{-}$ will implement the full projection operator.

Since time evolution and single-site projection is straightforward using a cylinder matrix product state, the above can be used to implement Gauss law projection.

\subsubsection{Recovering the Ground State at Long Times} \label{app-GSrecovery}

\begin{figure}[H]
    \centering
    \includegraphics[width = 220 pt]{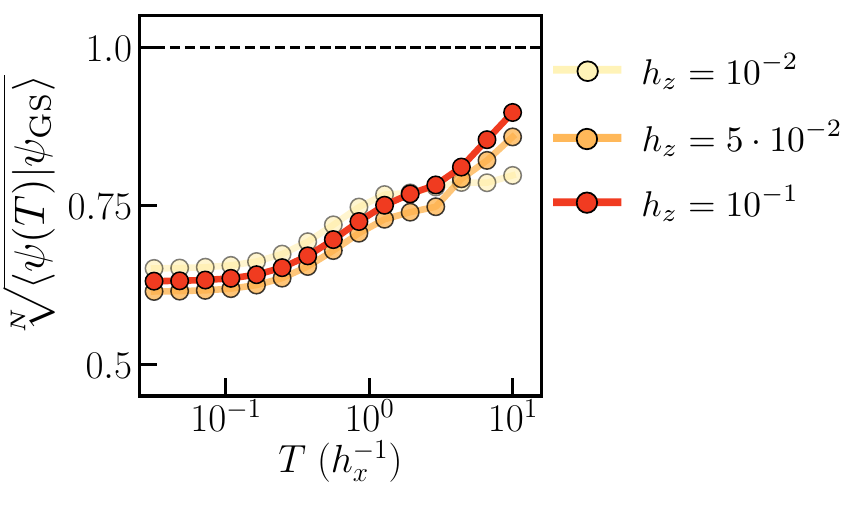}
    \caption{\textbf{Resolving the Ground State with Adiabatic Sweeps.} We depict the overlap density between the final state of the dynamical sweep $\ket{\psi(T)}$ and the ground state of the system at the endpoint of the sweep $\ket{\psi_{GS}}$. We find that for the largest two values of $h_z$ and for long total sweep times $T$, the final state starts to resemble the ground state indicated by a sharp increase in the aforementioned overlap density.}
    \label{fig:supp_GS}
\end{figure}

In Section~\ref{subsec-TCSpinLakes}, we stated that at late time $T > T_m$, our dynamical sweep will discover the ground state via the adiabatic theorem.
Here, we justify that by increasing $h_z$, we indeed approach this regime at long times by plotting the overlap with the final state $\ket{\psi(T)}$ with the ground state of the parameter value of the end of the sweep.
The results are shown in Fig.~\ref{fig:supp_GS}.
As with our other diagnostics, we find three distinct behaviors demarcated by time scales $T_e < T_m$.
When $T < T_e$, the overlap density with the ground state is constant and much less than maximal.
Subsequently, the overlap density increases to another constant value that is still less than maximal.
This region coincides with the window $[T_e, T_m]$ where we dynamically prepare a quantum spin lake.
The size of this window decreases as $h_z$ is increased similar to the main text (for $h_z = 10^{-2}$, we don't probe $T$ large enough to go past it).
Finally, when $T>T_m$, we find that the overlap density increases rapidly signaling that the system resolves the ground state as claimed.

\subsubsection{Additional Numerics for the Fredenhagen-Marcu Order Parameter}\label{app-FM}

\begin{figure}
    \centering
    \includegraphics[width = 150 pt]{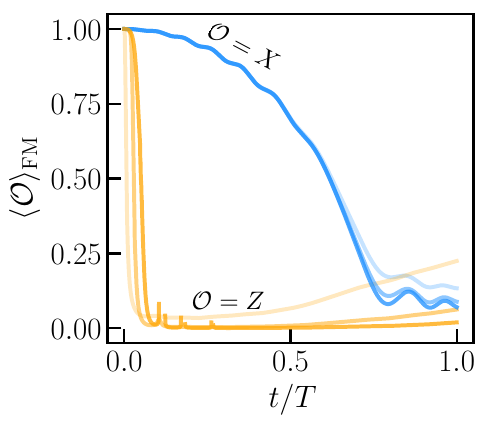}
    \caption{\textbf{Time Trace of Fredenhagen-Marcu Order Parameter.} We depict the time-trace for the Fredenhagen-Marcu order parameter for a sweep taken at $h_z/h_x = 0.1$ [the red arrow in Fig.~\ref{fig-dTC}(a)] for a total time $T = 10\ h_x^{-1}$. We remark that the numerical instabilities in $\langle Z \rangle_{\text{FM}}$ that occur at small $t/T$ are because both the numerator and denominator of the FM order parameter are almost zero to machine precision.}
    \label{fig:app-FMtimetrace}
\end{figure}

In the main text, in Fig.~\ref{fig-FMdTC}(a), we reported the maximum value of the FM order parameter $\langle Z \rangle_{\text{FM}}$ obtained during a dynamical sweep that took a total time $T$, as a function of $T$.
Subsequently, in Fig.~\ref{fig-FMdTC}(b), we reported the value of $\langle X \rangle_{\text{FM}}$ at the time in the sweep where $\langle Z \rangle_{\text{FM}}$ obtained its minimum value.
Our motivation to do this was that, towards the end of the dynamical sweep, we found that the value of both FM order parameters oscillated.
Here, we provide the full numerical time traces for the FM order parameter for the representative value of $T = 10$ to show these oscillations (see Fig.~\ref{fig:app-FMtimetrace}).

\section{Tree Tensor Network Method}

In this appendix, we provide some further details regarding the infinite tree tensor network method we use in the main text.
In particular, we will define the canonical form for the infinite tree tensor network ansatz that makes them numerically efficient to simulate.
Furthermore, we will provide the fixed-point infinite tree tensor network for the RVB state on the tree.

For convenience, we re-iterate that our infinite tree tensor network ansatz takes the following form.
\begin{equation} \label{eq-app-TTN}
    \includegraphics[width = 70pt, valign = c]{TreeTN.pdf}
\end{equation}
where, in full generality, $A$ and $B$ are $d \times \chi_a \times \chi_b \times \chi_c$ tensors that encode the state of the unit cell of the tree lattice we are simulating (with $d$ being the local Hilbert space dimension) and the $\{\chi_{\alpha}\}$ are called the bond dimensions and refers to the ranks of the non-dangling legs of the tensors.
As an example, in the case of the Husimi cactus lattice, the lower legs of rank$-d$ label the states of the right-side-up and up-side-down triangles of Eq.~\eqref{eq-Atriangle}~and~\eqref{eq-Btriangle}.
The small tensors that are on the bonds between $A$ and $B$ are ``singular value tensors'' in a sense that will be clear in the next section.

\subsection{Canonical Form}

Similar to the matrix product state, efficient computations with infinite tree tensor network ansatz are enabled due to the existence of a canonical form.
This canonical form enables computing expectation values, correlation functions, and the entanglement entropy using only local data in the tensor network. 
The canonical form for tree tensor networks are defined as:
\begin{align}
\includegraphics[width = 40 pt, valign = c]{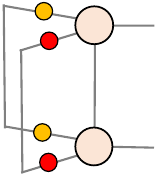} = \includegraphics[width = 40 pt, valign = c]{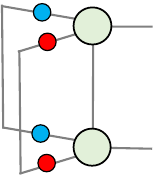} = \includegraphics[width = 40 pt, valign = c]{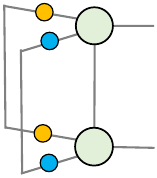} = \includegraphics[width = 20 pt, valign = c]{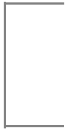}
\end{align}
and also:
\begin{align}
\includegraphics[width = 40 pt, valign = c]{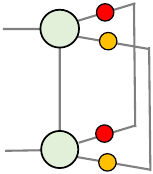} = \includegraphics[width = 40 pt, valign = c]{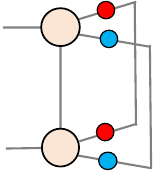} = \includegraphics[width = 40 pt, valign = c]{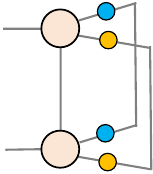} = \includegraphics[width = 20 pt, valign = c]{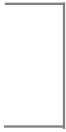}
\end{align}
It was shown in Refs.~\onlinecite{Li_TTN, nagy2012simulating}, that such a canonical form enables efficient computation of local observables and entanglement with trees.

\subsection{Fixed-Point $\mathbb{Z}_2$ Resonating Valence Bond Liquid State} \label{app-fprvb}

We want to construct the tree tensor network for the RVB in canonical form. To do so, consider the following tensor network:
\begin{align}
    \includegraphics[width = 60 pt, valign = c]{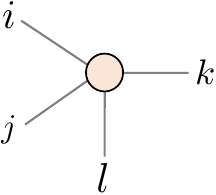}\ =\ \includegraphics[width = 60 pt, valign = c]{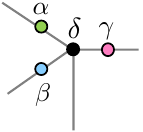}
\end{align}
and
\begin{align}
    \includegraphics[width = 60 pt, valign = c]{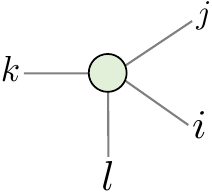}\ =\ \includegraphics[width = 60 pt, valign = c]{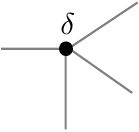}
\end{align}
where $\delta$ is the rank-4, $4 \times 4 \times 4 \times 4$ kronecker delta tensor and $\alpha, \beta, \gamma$ are $4 \times 4$ matrices that can be expressed as follows (with the second index always labeling the index pointing away from the $\delta$ tensor):
\begin{align}
    \alpha_{i' i} &= \left[ \delta_{i'3} (\varepsilon_{i1} - \delta_{i0}) \right. \nonumber \\
    &+ \left.(\delta_{i'1} + \delta_{i'2}) (\delta_{i1} + \delta_{i0})  + \delta_{i'0} \varepsilon_{i0} \varepsilon_{i1}  \right]/\sqrt{2} \\
    \beta_{j' j} &= \left[ \delta_{j'2} (\varepsilon_{j2} - \delta_{j0}) \right. \nonumber \\
    &+ \left.(\delta_{j'1} + \delta_{j'3}) (\delta_{j2} + \delta_{j0})  + \delta_{j'0} \varepsilon_{j0} \varepsilon_{j2}  \right]/\sqrt{2} \\
    \gamma_{k' k} &= \left[ \delta_{k'1} (\varepsilon_{k3} - \delta_{k0}) \right. \nonumber \\
    &+ \left.(\delta_{k'2} + \delta_{k'3}) (\delta_{k3} + \delta_{k0})  + \delta_{k'0} \varepsilon_{k0} \varepsilon_{k3}  \right]/\sqrt{2}
\end{align}
where $\varepsilon_{ij} = (1 - \delta_{ij})$.
Such tensors assign an equal weight and equal phase to states that satisfy the Gauss law constraint of the $\mathbb{Z}_2$ RVB state.
As such, the tensor network above describes the (non-normalized) fixed-point RVB state.
To put the above tensor network in canonical form, we note that:
\begin{align}
\sum_{j} \alpha_{j', j} \alpha^*_{j^{''}, j} = \delta_{j', j^{''}} 
\end{align}
and similarly with $\beta$ and $\gamma$.
This means that the tensor network above is almost in canonical form.
To put it in canonical form, we simply perform an SVD for each of the tensors $\alpha, \beta, \gamma$ with $\alpha = U_{\alpha} \Lambda_{a} V_{\alpha}^{\dagger}$, $\beta = U_{\beta} \Lambda_{b} V_{\beta}^{\dagger}$, and $\gamma = \alpha = U_{\gamma} \Lambda_{c} V_{\gamma}^{\dagger}$.
Graphically, 
\begin{align}
    \includegraphics[width = 50 pt, valign = c]{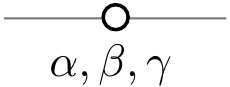}\  = \includegraphics[width = 50 pt, valign = c]{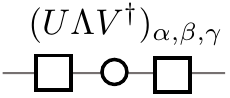}\
\end{align}
Then, by pushing the $U$ tensors into the $B$ tensor and the $V^{\dagger}$ tensors into the $A$ tensor, we can define a new $A$ and $B$ tensors as follows: 
\begin{align}
    \includegraphics[width = 60 pt, valign = c]{A_tensor.pdf}\ =\ \includegraphics[width = 60 pt, valign = c]{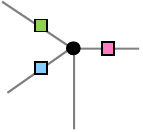}
\end{align}
and
\begin{align}
    \includegraphics[width = 60 pt, valign = c]{B_tensor.pdf}\ =\ \includegraphics[width = 60 pt, valign = c]{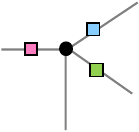}
\end{align}
with the singular values being the $\Lambda$ matrices. 
This tensor network is in canonical form and can be normalized by normalizing the singular values (i.e. $\text{Tr}(\Lambda^2) = 1$) and multiplying the $A$ and $B$ tensor by $4$.

\section{Additional Numerics for $U(1)$ Tree Tensor Networks} \label{app-U1}

\begin{figure}[H]
    \centering
    \includegraphics[width = 150 pt]{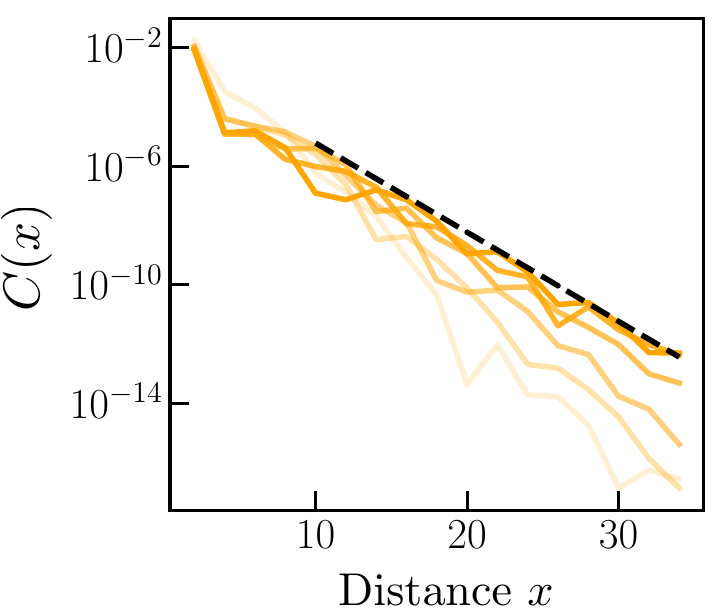}
    \caption{\textbf{Raw Correlation Function Data for $U(1)$ Spin Liquid on Bethe Lattice.} We plot density-density correlations as a function of distance for different total sweep times  (from light to dark, $T \in \{2.5, 3.5, 4.5, 4.75, 5, 7.5\}$) on the tree lattice.
    We find that they fall off as $2^{-x}$ which is the predicted fall off of gapless states on tree lattices \cite{nagy2012simulating, Laumann09}.
    For our numerics, we use a bond dimension of $\chi_{\alpha} = 10$ ($\alpha = a, b, c$) and $dt = 0.005$.  }
    \label{fig:rawU1}
\end{figure}

In Section~\ref{sec-U1} of the main text, we plotted a density correlation function in Fig.~\ref{fig:U1QSL}(d). 
In that figure, we performed a rolling average of $3$ data points in order to smooth out the density correlation functions for ease of viewing the decay profile of the correlation function.
In Fig.~\ref{fig:rawU1}, we show the raw data.

\end{document}